\documentclass[onecolumn]{aa} 

\usepackage{graphicx}
\usepackage{dcolumn}
\usepackage{morefloats}
\usepackage{txfonts}
\usepackage{natbib}
\bibpunct{(}{)}{;}{a}{}{,} 

\def\virgl{``}
\def\virgr{"}

\def\arcsec{\hbox{$^{\prime\prime}$}}
\def\deg{\hbox{$^\circ$}}

\def\gtsim{\; \raise0.3ex\hbox{$>$\kern-0.75em \raise-1.1ex\hbox{$\sim$}}\; }

\def\ltsim{\; \raise0.3ex\hbox{$<$\kern-0.75em \raise-1.1ex\hbox{$\sim$}}\; }

\usepackage{changepage}
\usepackage{caption}
\usepackage{subcaption}
\usepackage{textcomp}
\usepackage{multirow}
\usepackage{geometry}
\usepackage{pdflscape}
\usepackage{bm}

\begin{document} 

\defcitealias{OSullivan2017}{OS17}
\defcitealias{Pasetto2016}{Pa16}

\title{Broadband radio spectro-polarimetric observations of high Faraday rotation measure AGN}
\titlerunning{Broadband spectro-polarimetry study of high RM AGNs}
\authorrunning{Pasetto et al.}


   \author{ Alice Pasetto \inst{1,3}\fnmsep\thanks{
  \email{a.pasetto@crya.unam.mx}\newline},
              	Carlos Carrasco-Gonz\'alez \inst{1},
		Shane O'Sullivan \inst{2},               
		Aritra Basu \inst{3}, 
		Gabriele Bruni \inst{4,3},
                	Alex Kraus \inst{3}, 
               	 Salvador Curiel \inst{5},
	 	Karl-Heinz Mack \inst{6}
                }

   \institute{Instituto de Radioastronom\'ia y Astrof\'isica (IRyA-UNAM), Antigua Carretera a P\'atzcuaro, Morelia, Michoac\'an, M\'exico
              \and
             Hamburger Sternwarte, Universit\"at Hamburg, Gojenbergsweg 112, 21029, Hamburg, Germany
	    \and
		Max-Planck Institut f\"ur Radioastronomie (MPIfR), Auf dem H\"ugel 69, D-53121 Bonn, Germany           
		\and
		Istituto di Astrofisica e Planetologia Spaziali (IAPS-INAF), via Fosso del Cavaliere 100, I-00133 Rome, Italy
		\and
		Instituto de Astronom\'ia, UNAM, Apartado Postal 70-264, 04510 M\'exico, DF, Mexico
             \and
             Istituto di Radioastronomia (IRA-INAF), Via Gobetti 101, I-40129 Bologna, Italy
             }

  \abstract{

We present broadband polarimetric observations of a sample of high Faraday rotation measure (RM) AGN using the Karl. G. Jansky Very Large Array (JVLA) telescope from 1 to 2 GHz, and 4 to 12 GHz.
The sample (14 sources) consists of very compact sources (linear resolution smaller than $\approx$ 5 kpc) that are unpolarized at 1.4 GHz in the NRAO VLA Sky Survey (NVSS). Total intensity data have been modelled using combination of synchrotron components, revealing complex structure in their radio spectra.
Depolarization modelling, through the so called qu-fitting (the modelling of the fractional quantities of the Stokes Q and U parameters), have been performed on the polarized data using an equation that attempts to simplify the process of fitting many different depolarization models that we can divide into two major categories: External Depolarization and Internal Depolarization models.
Understanding which of the two mechanisms are the most representative, would help the qualitative understanding of the AGN jet environment, whether it is embedded in a dense external magnetoionic medium or if it is the jet-wind that causes the high RM and strong depolarization.
This could help to probe the jet magnetic field geometry (e.g. helical or otherwise). 
This new high-sensitivity data, shows a complicated behaviour in the total intensity and polarization radio spectrum of individual sources. We observed the presence of several synchrotron components and Faraday components in their total intensity and polarized spectra. For the majority of our targets, (12 sources) the depolarization seems to be caused by a turbulent magnetic field. Thus, our main selection criteria (lack of polarization at 1.4 GHz in the NVSS), results in a sample of sources with very large RMs and depolarization due to turbulent magnetic fields local to the source.
These broadband JVLA data reveal the complexity of the polarization properties of this class of radio sources. We show how the new qu-fitting technique can be used to probe the magnetised radio source environment and to spectrally resolve the polarized components of unresolved radio sources.
}


  \maketitle
%

\section{Introduction}

Active Galactic Nuclei (AGN) are powered by the supermassive black hole (SMBH) at their centres. They can emit across the full electromagnetic spectrum, from low radio frequencies to high energy X-ray and $\gamma$-ray. The radio emission of radio loud AGN is dominated by synchrotron radiation from their core, jets and lobes. Understanding how these objects interact with the surrounding medium is important for studying the evolution and feedback of the radio sources and on the formation of the radio jets. Important information can be extracted not only through studying the total intensity of the synchrotron radiation but also the polarization information.
During in the last years polarization study is playing an importan role in the understanding of AGN jets and their surrounding \citep[e.g the recent works done by][]{Zavala2003,OSullivan2009,Hovatta2012,OSullivan2012,Farnes14,Pasetto2016,Kravchenko2017}.
In the radio regime, the polarization state of the electromagnetic radiation of extragalactic sources is described using the Stokes parameters I,Q, U and V. 
Following \cite{Sokoloff98}, the complex linear polarization is:

	\begin{equation}
	P = Q+iU = pIe^{2i\chi}
	\label{simple}
	\end{equation}

where I, Q and U are the measured Stokes parameters, $\textit{p}$ and $\chi$ are the fractional polarization and the polarization angle of the polarized wave described as:
 
	\begin{equation}
	p = \frac{S_{pol}}{I}=\sqrt{q^2+u^2}
	\label{degpol}
	\end{equation}
	
\noindent
where we consider q=Q/I and u=U/I as the fractional values of the Stokes parameters Q and U respectively, as used in \cite{Farnsworth11}, and  
	
	\begin{equation}
	\chi = \frac{1}{2} \arctan\frac{u}{q}
	\label{polang}
	\end{equation}

\noindent
The study of the polarization state of radio sources provides the opportunity to analyze two important effects: Faraday rotation and Faraday depolarization.
Characterizing the Faraday structure and studying the effect of Faraday depolarization in radio sources allows one to probe the properties of the magneto-ionic medium, such as the strength, degree of order and orientation of the magnetic field, and the distribution of the relativistic and thermal electron populations. 
Faraday rotation is the rotation of the plane of an electromagnetic wave as it passes through a magnetized medium; which for a particular region can be described by the Faraday depth \citep{Burn66}:

	\begin{equation}
	\phi = 0.81\int^{telescope}_{emission} nB_{\parallel}\cdot dl  \qquad [rad \quad m^{-2}]
	\label{phi}
	\end{equation}
\noindent
where $n$ is the free electron density (in units of cm$^{-3}$), $B_{\parallel}$ is the parallel component of the magnetic field (in $\mu$G) and $\textit{l}$ is the distance along the line of sight (in parsec).
In the simplest scenario, in which there is a background source emitting synchrotron radiation and only one single uniform Faraday screen in the foreground, the Faraday depth $\Phi$ is identical to the rotation measure (RM). Traditionally, the RM has been determined by fitting the equation:

	\begin{equation}
	\chi (\lambda^2)=\chi_{0}+RM\lambda^2, 
	\label{RM}
	\end{equation}
	
\noindent
where $\chi_{0}$ is the intrinsic polarization angle.
This equation is true when one restricts the RM fitting to regions of $\lambda^{2}$ space, where the fractional polarization, p($\lambda^2$), is constant or decreases monotonically \citep[e.g., ][]{Simard-Normandin1981}, or when one restricts the fitting to $\lambda<\lambda_{1/2}$ (therefore, p($\lambda_{1/2}$)/p(0)=0.5), beyond which \cite{Burn66} suggests a non-linear behaviour of the polarization angle, $\chi(\lambda^2)$.
In more realistic astrophysical cases, a non-linear behaviour of the polarization angle, therefore a complex Faraday structure, occurs \citep[e.g., ][]{Roy2005,Anderson2015,Anderson2016} \citep[][hereafter Pa16]{Pasetto2016}.
The presence of a magnetized and dense medium surrounding and/or interacting with the emitting radio source can depolarize the emission. This effect is generally seen as a decrease of the degree of polarization with increasing wavelength, and has been studied for many years \citep[e.g.,][]{Burn66,Tribble91,Sokoloff98}. The depolarization can be caused internally, where the synchrotron emitting and the Faraday rotating regions are spatially coincident, or externally, where the synchrotron emitting and the Faraday rotation regions are different (see Section \ref{JVLAdepolmodels} for a more exhaustive explanation).
To extract information about the magnet-ionic medium, it is necessary to study and model both the fractional polarization, p($\lambda^2$), and the polarization angle, $\chi(\lambda^2)$, as suggested by \cite{Farnsworth11}. However, until now the lack of wide bandwidth coverage made this kind of study difficult. 
To study the complex bahaviour of the several Faraday structures, well sampled polarized data and more sophisticated modelling are required.
Very recently, thanks to new broadband receivers at radio facilities, spectropolarimetric studies have been performed by \cite{OSullivan2012, Anderson2015, Anderson2016, OSullivan2017} (hereafter OS17) on discrete sources. They selected polarized sources at 1.4 GHz and analyzed the complex behaviour by studying and modelling the polarization signal. 

In this work we present well sampled, wide-band polarimetric data of a sample of AGN observed with the Karl. G. Jansky Very Large Array (JVLA) of the National Radioastronomy Observatory (NRAO)\footnote{The NRAO is a facility of the National Science Foundation operated under cooperative agreement by Associated Universities, Inc.} at L, C and X bands (1 to 2 GHz bandwidth is L band; 4 to 8 GHz bandwidth is C band and 8 to 12 GHz bandwidth is X band). The sources have been selected from previous work, a polarimetric Effelsberg single dish study, done by \citetalias{Pasetto2016}. The sources in \citetalias{Pasetto2016} have been selected to show strong depolarization in the 1.4 GHz NRAO VLA Sky Survey (NVSS, 45$\arcsec$ FWHM resolution, effective continuum IF bandwidth of $\Delta\nu$ $\sim$ 42 MHz) with $\textit{p}$ $\le$ 0.3$\times$10$^{-2}$ and sources with no polarization information (blanked sources in the NVSS, i.e., pixels with lower weight blanked to eliminate regions with inadequate coverage and poor sensitivity). 
\citetalias{Pasetto2016} turned out that the strong depolarization of these sources is most likely related to the large values of RM derived from linear $\chi(\lambda^2)$ fits (with RM measured with single dish observations, RM$_{dish}$ $>$ 500 rad/m$^2$).  
In this work, we present and analyze their wide-band JVLA total intensity data and their polarimetric data (at L, C and X bands) by modelling the total intensity radio spectra using combination of several synchrotron components and by modelling the polarization information following the approach in \cite{OSullivan2012}. We propose a general interpretation of the complex medium of these sources.
In section \ref{sample} we present the sample, in section \ref{JVLAobs} we describe the observations and calibration process. In section \ref{JVLAdepolmodels} we describe the polarization models and in section \ref{fitting} we describe the fitting of the depolarization modelling and how to extract the polarization and Faraday rotation parameters. Section \ref{JVLAresults} presents our results for all sources, with comments on individual sources in section \ref{Commentsindividualsources}. In Section \ref{discussion} we discuss possible implications and our summary and conclusion is in Section \ref{JVLAsummary}. 
Throughout this paper, we assume a cosmology with H$_{0}$ = 71 km s$^{-1}$ Mpc$^{-1}$, $\Omega_{M}$ = 0.27 and $\Omega_{\Lambda}$ = 0.73, and define the spectral index, $\alpha$, such that the observed flux density (S) at frequency $\nu$ follows the relation S$_{\nu}$$\propto$ $\nu^{\alpha}$.


\section{The sample}
\label{sample}

The sources have been selected from \citetalias{Pasetto2016} where single dish observations with the Effelsberg 100-m radio telescope of a sample of more than 500 AGN have been performed in order to search for sources with high RM. The two principal characteristics of \citetalias{Pasetto2016} sources are (1) their lack of sufficient polarized flux density at 1.4 GHz in the NVSS (i.e., blanked polarized images and low polarization flux density detection) with S$_{pol}^{1.4}$ $\le$ 0.87 mJy which represents the 3$\sigma$$_{pol}^{1.4}$ \citep[the rms fluctuation level $\sigma$$^{pol}_{1.4}$ = 0.29 mJy/beam, for the NVSS survey, ][]{Condon98} and (2) their compactness at arcsec scale, they are unresolved in the FIRST catalogue \citep[5$\arcsec$ FWHM, ][]{White97} after subsequent cross correlation with the NVSS catalogue.
The first characteristic was important for the previous project because of its possible relation with strong depolarization (with $\textit{p}$ $\le$ 0.3$\times$10$^{-2}$ ) due to high RM. The second characteristic allowed us to select possibly compact/or high-redshift candidates and avoid extended structure sources that might be affected by beam depolarization (cancellation of the polarized vectors within the telescope beam) at the NVSS and FIRST resolution.
High frequency (at 10 GHz) single dish observations of the initial big sample of more than 500 AGN, resulted in a list of 30 targets with detectable polarization. Subsequent follow-up single dish observations (from 1 to 10 GHz) have been performed in order to study the total intensity and polarized spectra. \citetalias{Pasetto2016} found that almost half of the sample have high RM values with RM$_{dish}$ $\ge$ 500 rad/m$^{2}$. Moreover, they noticed first signs of complexity of the polarization information (i.e., fractional polarization and polarization angle) and deviations from the standard linear-$\lambda^2$ behaviour expected for simple RM structure.
 
Here we present a study of this polarization complexity, performing high sensitivity, broadband JVLA observations on 14 sources, all Quasar (QSO) type, with RM$_{dish}$ $\ge$ 500 rad/m$^2$ (see Tab.\ \ref{TabHighRMsources} for the complete list of the sources). Note that the source 0239--0234 shows a RM$_{dish}$ value that is lower than 500 rad/m$^2$. However, we decided to include this source to test if the broadband spectropolarimetry technique could reveal Faraday components with higher RM values previously hidden because of the lack of data coverage.
According to our previous single dish work and based on the characteristics of the new JVLA observations, the sources have the following selection criteria:
	\begin{itemize}

	\item Galactic latitude l $>$ 30$\deg$: in order to avoid the contribution of the galactic plane

	\item Flux density at 1.4 GHz in the NVSS, S$_{1.4}$ $>$ 300 mJy

	\item Unpolarized at 1.4 GHz in the NVSS (S$_{pol}^{1.4}$ $\le$ 0.87 mJy or blank polarized images)

	\item RM$_{dish}$ $\ge$ 500 rad/m$^2$

	\item Unresolved in the FIRST catalogue and at all the JVLA configurations used (reaching a resolution of 0.6$\arcsec$ at X band for the B configuration)

	\end{itemize}
	
This source selection resulted in a sample of 14 unresolved sources (see Tab.\ \ref{TabHighRMsources}). We observed this sample at L, C and X bands with the JVLA with the objective to study their total intensity spectra and their magneto-ionic media. Considering the highest angular resolution reached with this JVLA observations (0.6$\arcsec$ at X band for the B configuration) and the redshift of the sources, we can give an upper limit on their linear sizes. Our observations are sensitive to regions smaller than $\sim$5 kpc (see Linear Resolution - Lin.Res.- in table \ref{TabHighRMsources}). Therefore, the sample is composed by sources much more compact than those polarized at 1.4 GHz selected by \citetalias{OSullivan2017}, for which the median linear size is of the order of $\sim$ 100 kpc.

\begin{table*}[h]
\caption{List of the high-RM sources}
\scriptsize
\begin{center}
\begin{tabular}{llllllllllll}
\hline\hline                                                    
 Source		 & Other	   	& Optical			&	RA			&   DEC			&  RM$_{dish}$~~~      		& z						& Ref.				& scale		& Lin.Res.		& Mag		& NVSS$_{p.info}$	\\
 
 name		 & name$^*$	   	&   ID				&	[J2000]		& [J2000]		& [rad/m$^{2}$] ~ 			&						&	 z 				& [kpc/"]	&  [kpc]	& filter 	&	\\                   
 \hline                                     	        	
  0239--0234 & PKS 0237--02	& 	QSO$^{(1)}$			& 02:39:45.480  &  -02:34:40.98 &    --40    $\pm$   10~~   & 1.116				& $^{(a)}$		& 7.2		& 4.3			& 19.90V$^{(1)}$	& Unpol. \\    
  0243--0550 & 0240--060	& 	QSO$^{(1)}$			& 02:43:12.464  &  -05:50:55.36 &    600     $\pm$   100   	& 1.800				& $^{(b)}$		& 7.7		& 4.6			& 19.90V$^{(1)}$	& Unpol. \\ 
  0751+2716  & B2 0748+27	& 	QSO$^{(1)}$			& 07:51:41.492  &  +27:16:31.65 &    500     $\pm$   100   	& 3.200				& $^{(c)}$		& 7.4		& 4.4			& 21.20I$^{(*)}$	& Unpol. \\  
  0845+0439  &				& 	QSO$^{(2)}$			& 08:45:17.151  &  +04:39:46.64 &    1920    $\pm$   20~~   & 0.800				& $^{(d)}$		& 6.7		& 4.0			& 21.14B$^{(2)}$	& Unpol. \\ 
  0958+3224  & 3C 232		& 	QSO$^{(1)}$			& 09:58:20.939  &  +32:24:02.16 &    2200    $\pm$   100   	& 0.530				& $^{(e)}$ 		& 5.5		& 3.3			& 15.78V$^{(1)}$	& Unpol. \\ 
  1048+0141  & 1045+019		& 	QSO$^{(2)}$ 		& 10:48:22.850  &  +01:41:47.46 &    --2510  $\pm$   30~~   & 0.689				& $^{(f)}$		& 6.3		& 3.8			& 21.86 -- $^{(*)}$ 	& Unpol. \\ 
  1146+5356  &				& 	QSO$^{(1)}$			& 11:46:44.186  &  +53:56:43.36 &    --450   $\pm$   10~~   & 2.201				& $^{(g)}$		& 7.7		& 4.6			& 20.50V$^{(1)}$	& Unpol. \\ 
  1246--0730 & 1243-072		& 	QSO$^{(1)}$			& 12:46:04.231  &  -07:30:46.63 &     880    $\pm$   10~~   & 1.286				& $^{(h)}$		& 7.4		& 4.4			& 18.90V$^{(1)}$	& Blank \\ 
  1311+1417  & 1308+145		& 	QSO$^{(1)}$			& 13:11:07.835  &  +14:17:46.69 &    570     $\pm$   10~~   & 1.952				& $^{(i)}$		& 7.7		& 4.6			& 20.00V$^{(1)}$	& Unpol. \\ 
  1312+5548  &				& 	Cand. QSO$^{(3)}$	& 13:12:53.193 	&  +55:48:13.21 &    --1000  $\pm$   200	& 0.975				& $^{(j)}$ 		& 7.1		& 4.2			& 19.37g$^{(3)}$	& Unpol. \\ 
  1405+0415  & 1402+044		& 	QSO$^{(1)}$			& 14:05:01.113  &  +04:15:35.87 &     1153   $\pm$   4~~~~ 	& 3.211				& $^{(k)}$		& 7.4		& 4.4			& 19.56V$^{(1)}$	& Blank \\ 
  1549+5038  &				& 	QSO$^{(1)}$			& 15:49:17.447  &  +50:38:05.87 &    100     $\pm$   100	& 2.169				& $^{(l)}$		& 7.7		& 4.6			& 18.77V$^{(1)}$	& Unpol. \\  
  1616+0459  & 1614+051		& 	QSO$^{(1)}$			& 16:16:37.530  &  +04:59:31.96 &    2530    $\pm$   40~~	& 3.217				& $^{(k)}$		& 7.4		& 4.4			& 19.60V$^{(1)}$	& Unpol. \\ 
  2245+0324  &				& 	QSO$^{(1)}$			& 22:45:28.284  &  +03:24:08.71 &    --800   $\pm$   100	& 1.340				& $^{(m)}$		& 7.5		& 4.5			& 18.00V$^{(1)}$	& Unpol. \\ 
\hline\hline

\end{tabular}
\end{center}
 	\begin{flushleft}
	NOTE -- 
	The table show in column 1: the source name used in this work; column 2: other name associated to the source; column 3: the optical idenitification; the coordinates Right Ascension and Declination in J2000 are reported in column 4 and 5; column 6: the RM value measured with the Effelsberg single dish RM$_{dish}$; column 7 and 8: the redshift and its relative literature reference; column 9 and 10: estimation of the scale [kpc/$\arcsec$] and the linear size [kpc] of the sources; column 11: magnitude information and the used filter ; column 12: NVSS polarization information.\\
	$^{(*)}$ from Nasa Extragalactic Database (NED), 
	$^{(1)}$ from \cite{Veron-Cetty2006} catalogue, 
	$^{(2)}$ from \cite{Souchay2012} catalogue, 
	$^{(3)}$ from \cite{Richards2009}, $^{(a)}$: \cite{Fricke1983}, $^{(b)}$: \cite{Baldwin1981}, 	$^{(c)}$:\cite{Tonry1999}, $^{(d)}$: \citetalias{Pasetto2016}, $^{(e)}$:\cite{Brotherton1996}, $^{(f)}$:\cite{Labiano07}, $^{(g)}$: \cite{Xu1994}, $^{(h)}$: \cite{Wilkes1986}, $^{(i)}$: \cite{Peck2000}, $^{(j)}$: \cite{Richards2009},  $^{(k)}$: \cite{Tytler1992}, $^{(l)}$: \cite{Stickel1994}, $^{(m)}$: \cite{Wolter97}. \textit{Lin.Res.} is the Linear Resolution estimated considering the highest resolution reached (0.6 $\arcsec$) and the redshifts of the sources. \textit{Unpol.} are the sources with low polarization flux density in the NVSS and \textit{Blank} are the sources for which the NVSS polarization images have been blanked because of technical issue.	 
	\end{flushleft}

\label{TabHighRMsources}
\end{table*}%


\section{Observations and data reduction}
\label{JVLAobs}
We observed our sample of 14 sources with the JVLA in full polarization mode. Observational details are summarized in Tab.s \ref{tab1} and \ref{tab2}. Observations were made in different epochs during semesters 2013B (project code: 13B-236), 2014B and 2015A (project code: 14B-184), and used different configurations of the JVLA (see Table \ref{tab1}). Different observational setups were used for C and X bands (2 GHz bandwidth in 13B-236, and 4 GHz bandwidth in 14B-184; see Table \ref{tab1}). For L band a bandwidth of 1 GHz was used.

\begin{table*}[]
\small
\caption{General information of the projects.}
\begin{center}
\begin{tabular}{ccrccccc}
\hline
Project & Frequency  &$\lambda$ ~ & $\nu$-Range     & Nspw  & $\Delta\nu$   &  Chan    & $\Delta\nu$  \\  
    ID    &  Band     & (cm)       &    (GHz)        &      & per spw   	&  per spw & per chan \\   
\hline\hline
13B-236 & L     & 20 cm      & 1.0~ -- ~2.0    & 16    &  64  MHz   &  64   & 1 MHz    \\
13B-236 & C     &  6 cm      & 4.0~ -- ~6.0    & 16    &  128 MHz  	&  64   & 2 MHz    \\
13B-236 & X     &  3 cm      & 8.0  -- 10.0    & 16    &  128 MHz  	&  64   & 2 MHz    \\
\hline
14B-184 & L    & 20 cm       &  1.0~ -- ~2.0   & 16    &  64  MHz	&  64   & 1 MHz    \\
14B-184 & C    &  6 cm       &  4.0 ~--~ 8.0   &  32   &  128 MHz  	&  64   & 2 MHz    \\
14B-184 & X    &  3 cm       &  8.0 -- 12.0    &  32   &  128 MHz  	&  64   & 2 MHz    \\
\hline
\end{tabular}
\end{center}
\label{tab1}
NOTE:\\
Column 1 is the \textit{Project ID}; column 2 is the frequency band with the corresponding cm wavelength in column 3. Column 4 reports the frequency range used. \textit{Nspw} (column 5) represents the number of spectral windows available in the frequency range and \textit{$\Delta\nu$ per spw} (column 6) is the frequency range available for each spw. In the last two columns (column 7 and 8): \textit{Chan} is the number of channels for each spw and \textit{$\Delta\nu$ per chan} is the frequency interval for each channel. 

\end{table*}%

\begin{table*}[h]
\small
\caption{Log file of the projects.}
\begin{center}
\begin{tabular}{lccccc}

  Sources    &      Date      &   Conf  &    Bands    & Flux/Pol.Ang. cal. &    Leakage cal. \\

\hline
\multicolumn{6}{c}{Project Code 13B-236} \\ 
\hline
0243--0550   &   19th-Nov-13  &    B    &   L C X   &        3C48         &   J0319+4130   \\
0239--0234   &   19th-Nov-13  &    B    &   L C X   &        3C48         &   J0319+4130   \\ \hline  

\multicolumn{6}{c}{Project Code 14B-184} \\ 
\hline

0751+2716   &   28th-Nov-14  &    C     &   C X      &        3C138         &   J0713+4349    \\
0751+2716   &   16th-Jan-15  &    CnB   &   L        &        3C286         &   J1407+2827    \\

0845+0439   &   28th-Nov-14  &    C     &   C X      &        3C138         &   J0713+4349    \\
0845+0439   &   16th-Jan-15  &    CnB   &   L        &        3C286         &   J1407+2827    \\
0958+3224   &   8th-Nov-14   &    C     &   C X      &        3C138         &   J0713+4349    \\
0958+3224   &   16th-Jan-15  &    CnB   &   L        &        3C286         &   J1407+2827    \\

1048+0141   &   9th-Jan-15   &    CnB   &   C X      &        3C286         &   J1407+2827    \\
1048+0141   &   16th-Jan-15  &    CnB   &   L        &        3C286         &   J1407+2827    \\

1146+5356   &   8th-Jan-15   &    CnB   &   C X      &        3C286         &   J1407+2827    \\

1246--0730  &   9th-Jan-15   &    CnB   &   C X      &        3C286         &   J1407+2827    \\
1246--0730  &   16th-Jan-15  &    CnB   &   L        &        3C286         &   J1407+2827    \\

1311+1417   &   9th-Jan-15   &    CnB   &   C X      &        3C286         &   J1407+2827    \\
1311+1417   &   16th-Jan-15  &    CnB   &   L        &        3C286         &   J1407+2827    \\

1312+5548   &   8th-Jan-15   &    CnB   &   C X      &        3C286         &   J1407+2827    \\
1312+5548   &   16th-Jan-15  &    CnB   &   L        &        3C286         &   J1407+2827    \\

1405+0415   &   8th-Jan-15   &    CnB   &   C X      &        3C286         &   J1407+2827    \\
1405+0415   &   16th-Jan-15  &    CnB   &   L        &        3C286         &   J1407+2827    \\

1549+5038   &   17th-Apr-15  &    B     &   L C X    &        3C286         &   J1407+2827    \\
1616+0459   &   17th-Apr-15  &    B     &   L C X    &        3C286         &   J1407+2827    \\
2245+0324   &   10th-Oct-14  &    C     &   L C X    &         3C48         &   J2355+4950    \\
\hline

\end{tabular}
\end{center}
\label{tab2}
NOTE:\\
Column 1: source name; column 2: date of the observations; column 3: JVLA configuration associated to the project; column 4: the observational bands; flux density and polarization angle calibrators and the Leackage calibrators used during the observations are reported in column 5 and 6.

\end{table*}%

\begin{figure}[h]
\begin{center}
		\includegraphics[width=0.5\textwidth]{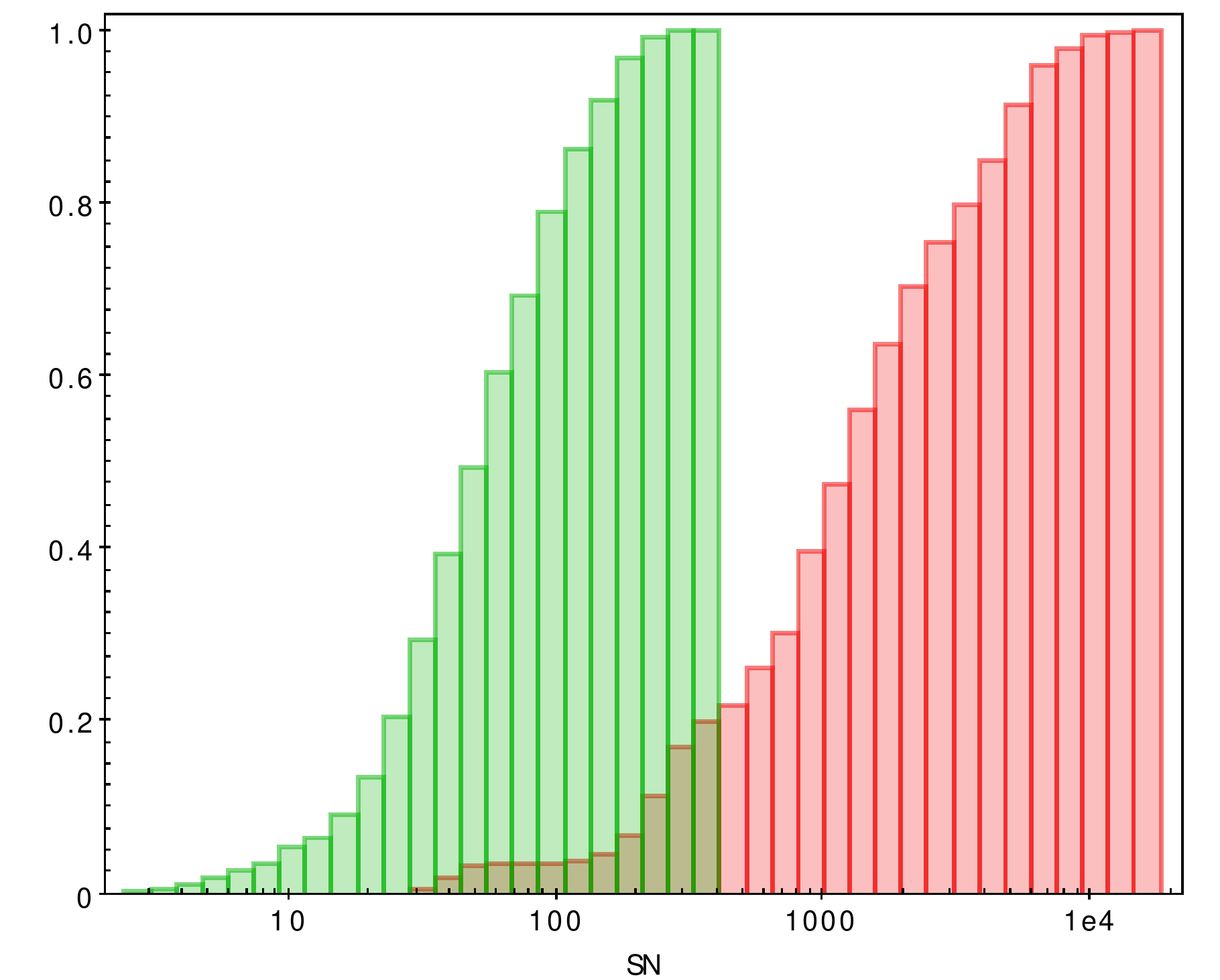}

\caption{Cumulative plot of the Signal-to-Noise distribution of the total intensity (red) and the polarized intensity (green) data points considering all the observed frequencies.}
\label{histoSN}
\end{center}
\end{figure}

The sources in our sample are catalogued as phase calibrators in the VLA Calibrator Manual. This means that all of them are bright and their emission is dominated by a point-like component. Therefore, for the phase calibration of the sources in our sample, we initially self-calibrated them by assuming a point-like model. After calibration, we checked for the possible presence of diffuse extended emission in high sensitivity  tapered images (using the full 4 GHz bandwidth and weighting the visibilities given more weight to short baselines). We did not detect any evidence of an additional extended emission component in any of the sources, confirming that their emission is dominated by only a point-like component at the sensitivity and resolutions of our observations. Therefore, additional phase calibration on the sample sources were not necessary. On-source times were fixed around 1 minute per source/band. This integration time was enough to have signal to noise ratios, at all the observed frequencies, larger than $\sim$ 30 for the total intensity, with 80\% of the data points having SNR $\sim$ 3500 (see Fig.\ \ref{histoSN} red histogram) and signal to noise ratios larger than $\sim$ 3, with 80\% of the data points having SNR $\sim$ 100 (see Fig.\ \ref{histoSN} green histogram) for the polarized intensity. In each session, we included observations of a standard flux/polarization angle calibrator, as well as a leakage calibrator (see Table \ref{tab2}).

Data editing and calibration were done using the data reduction package CASA (Common Astronomy Software Applications\footnote{https://science.nrao.edu/facilities/vla/data-processing}; version 4.4.0) following standard JVLA procedures. We wrote scripts in order to perform calibration in a quasi-automatic way. Our calibration scripts use prior known corrections of the data provided by the NRAO (antenna positions, antenna gain curves, atmosphere opacity corrections, and requantizer gains). Then, it performs bandpass and delays calibrations (using the flux calibrator of each run), and complex gain calibration (by self-calibrating each source) and polarization calibration. The procedure checks the calibrated data, makes additional flags when necessary after a visual inspection, and re-runs the calibration scripts. 

\begin{figure*}[h!]
	\begin{center}
		\includegraphics[width=0.49\linewidth]{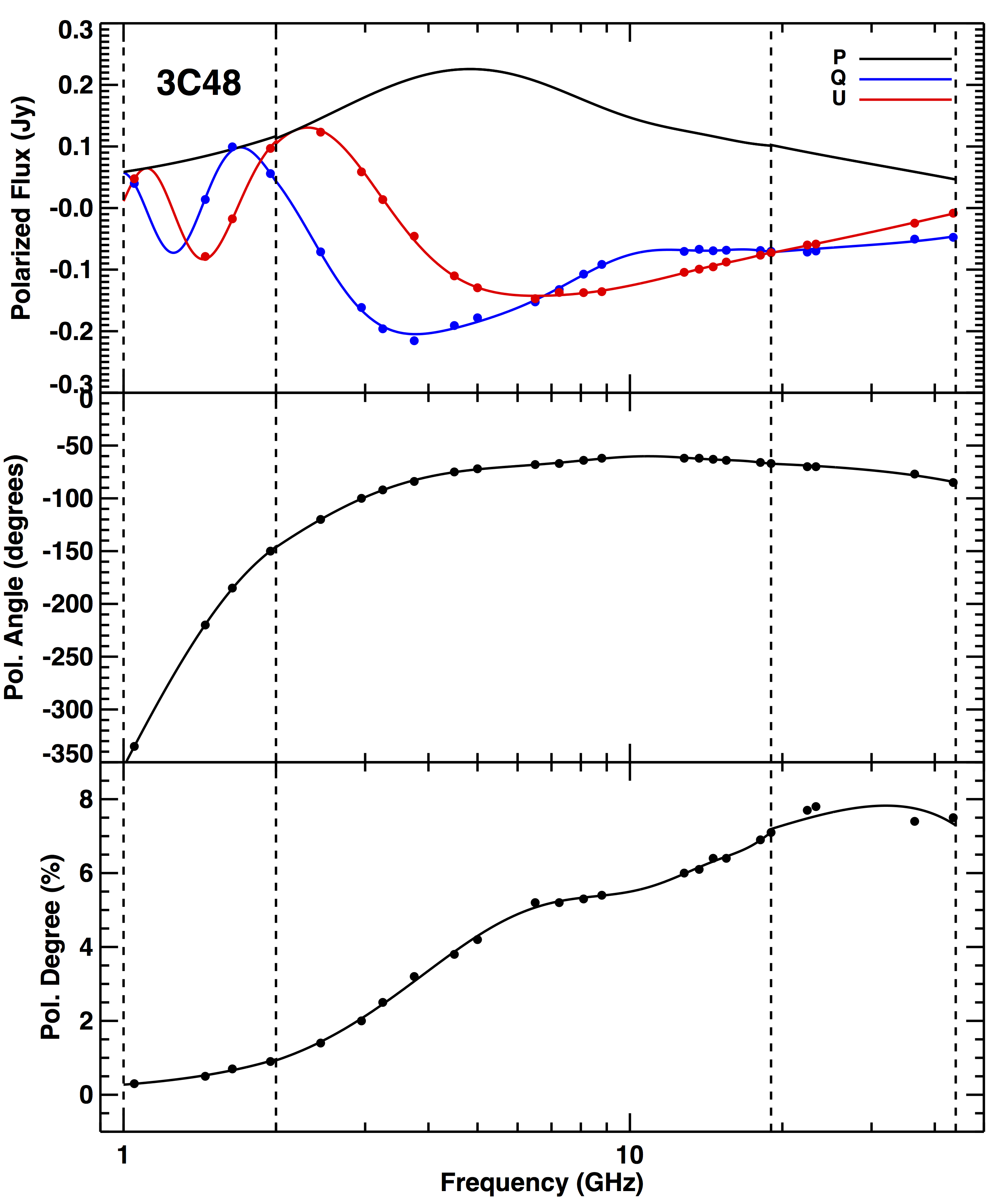}
		\includegraphics[width=0.49\linewidth]{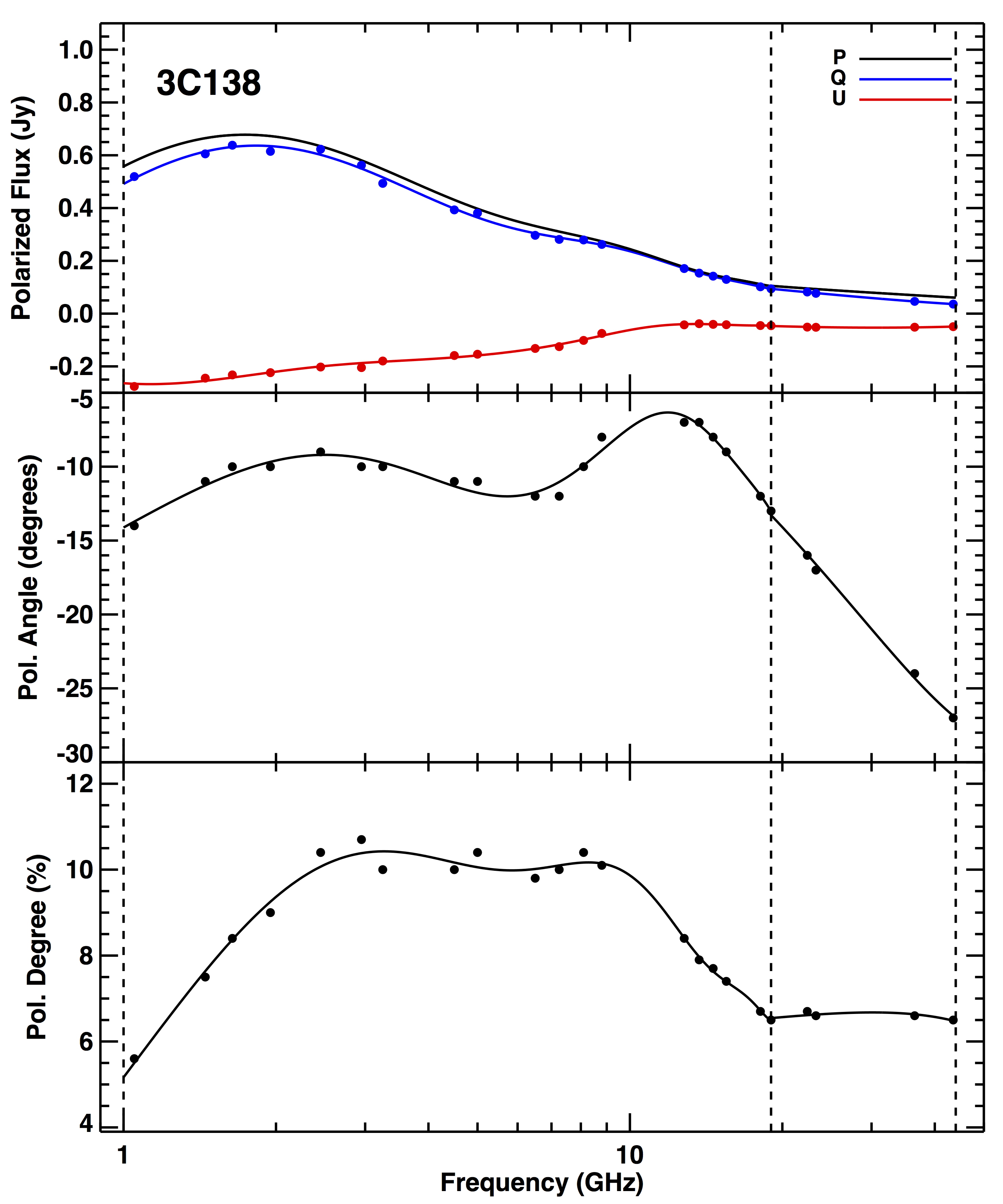}
		\includegraphics[width=0.49\linewidth]{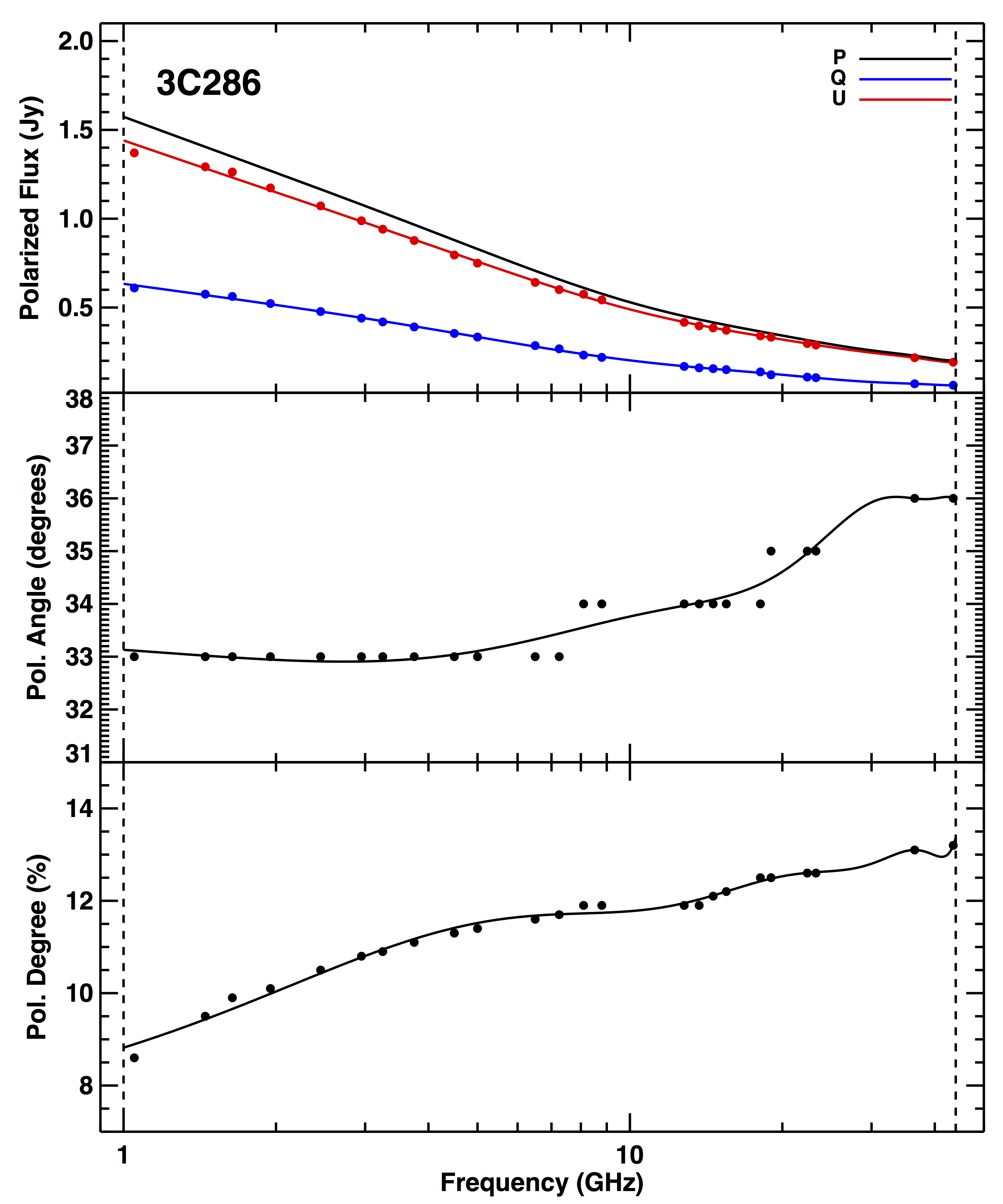}
		\caption{modelling of the polarization parameters for the calibrators: 3C48, 3C138 and 3C286. Frequency coverage: from 1 GHz to 50 GHz. For each of the sources: upper panel shows the polarized flux density (black line), the Stokes parameters Q (blue line) and U (red line); the middle panel shows the polarization angle variation; the last panel shows the polarization degree in percentage variation. The dashed vertical lines mark the frequency ranges used for the modelling. }
	\label{ModelCalib}
	\end{center}
	\end{figure*}

For the flux calibration of the Stokes I, we used resolved models of the flux calibrators provided by the NRAO. Total flux density at each frequency is set by using their known spectrum \citep{Perley13}. The polarization angle was calibrated by using known polarization parameters of the flux calibrators 3C286, 3C138 and 3C48. However, there are not yet models of the fractional polarization and the polarization angle at different frequencies for these calibrators. Therefore, in order to calibrate the linear polarization for our wide band observations, we used the known values of the fractional polarization (\textit{p}) and polarization angle ($\chi$) at different frequencies reported by \cite{Perley13Pol} and performed, within the 1 to 50 GHz frequency range, the following polynomial functions to the data:

\begin{equation}
 \chi=\chi_{0}+\sum_{1}^{n}\chi_{n}\left((\nu-\nu_{c})/\nu_{c}\right)^{n}, 
\end{equation}
\label{polichi}
\begin{equation}
p=p_{0}+\sum_{1}^{n}p_{n}\left((\nu-\nu_{c})/\nu_{c}\right)^{n},
\end{equation}
\label{polip}

\noindent 
where $\nu_{c}$ is the central frequency of the fitted frequency ranges. The result of these fittings are shown in Figures \ref{ModelCalib} and the coefficients used for the fitting are listed in Tab.\ \ref{politab}. For the calibrator 3C286 we performed a single model for the entire frequency range (1 to 50 GHz) while, for 3C138 and 3C48, because of the more complicated behaviour of the polarized parameters, we divided their spectra into several frequency ranges, see Tab.\ \ref{politab}. We therefore set the model for each spectral window (spw, 128 MHz bandwidth) using the standard "setjy" CASA task. For each spw we calculated the values of Stokes I \citep[using the fit performed by][]{Perley13}, the fractional polarization (\textit{p}) and the polarization angle ($\chi$) (using our fitting models) at the borders (I$_{0}$, I$_{1}$, \textit{p}$_{0}$, \textit{p}$_{1}$ and $\chi$$_{0}$, $\chi$$_{1}$) and at the center (I$_{c}$,  \textit{p}$_{c}$ and $\chi$$_{c}$) of each spw. Then, we assigned to each of the spw the central value of the Stokes I$_{c}$, \textit{p$_{c}$} and $\chi_{c}$. We finally let them varying linearly with frequency within each spw. 

We obtained solutions for the polarization angle and the D-terms, the instrumental polarization, for each channel. 
For the polarization angle calibration we constrained the solutions to short baselines (following the suggestions given by NRAO) in order to avoid effects because of possible extended structures emission of the polarization angle calibrators.
To correct for the D-terms we corrected by using an unpolarized calibrator for each observational session.

From the calibrated data, wide band images of Stokes I have been made for all the targets at L, C and X bands.
We run the task CLEAN using the parameter nterms=2 (that takes into account the spectral index of the source) and different weight values of the robust parameter \citep{Briggs95}. All the sources appear unresolved at the highest angular resolution of 0.6$\arcsec$, as assumed in the calibration.
Images of Stokes parameters I, Q and U were also made for each 64 MHz spw at L band and for each 128 MHz spw for C and X bands (using nterms=1 in the task CLEAN). On the individual I, Q and U spectral window images we perform a Gaussian fit to the source, considering a circular region with a diameter 2 times the deconvolved beam size of the image. In this way, information on the Stokes parameters for each 64 MHz and 128 MHz subbands have been collected in order to study the total intensity spectra (at L, C and X bands) and the polarized spectra (at C and X bands) of the sources in our sample.

\begin{figure}[h]
\begin{center}
		\includegraphics[width=0.49\textwidth]{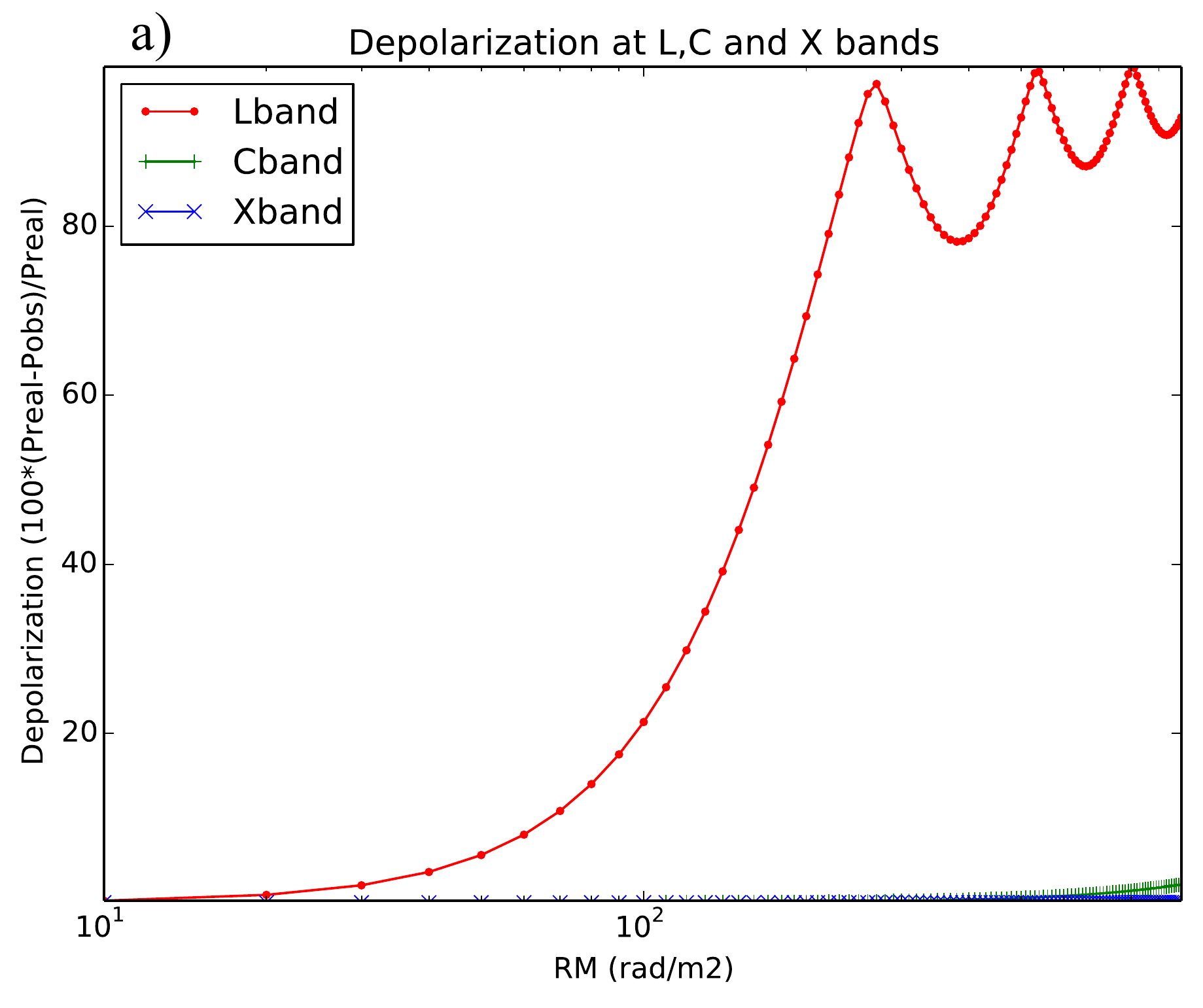}
		\includegraphics[width=0.49\textwidth]{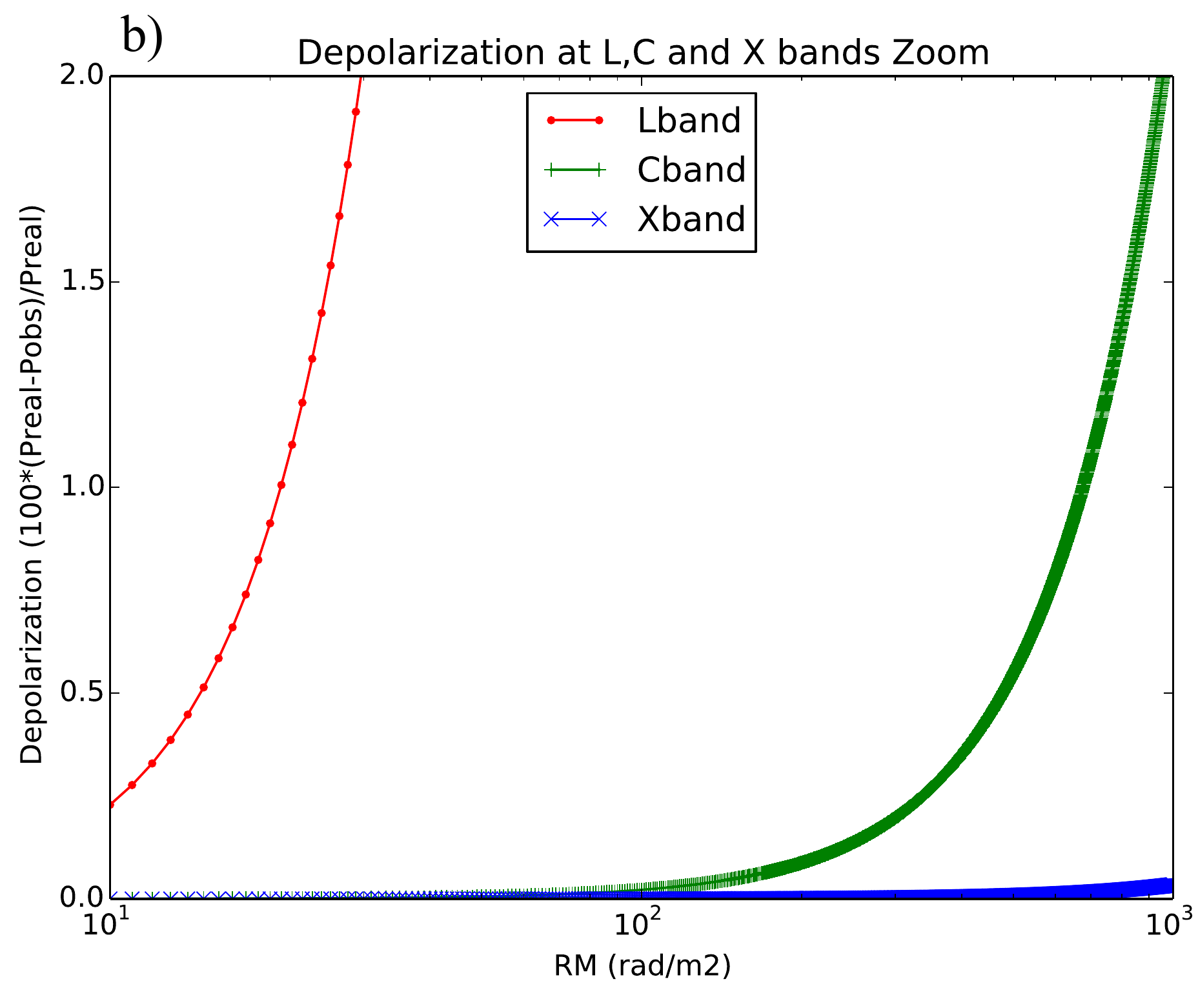}

\caption{In-band depolarization at all the frequency bands $\textit{a)}$: L band(red), C band (green) and X band(blue); and a zoom to only 5\% in-band depolarization $\textit{b)}$ for the three bands.}
\label{inbanddepol}
\end{center}
\end{figure}

The depolarization effects due to in-band depolarization have been estimated, i.e., the depolarization due to the rotation of the polarization angle within a considered bandwidth because of high RM. We estimated the in-band depolarization effect considering the lowest frequency for each observed band (i.e., 1 GHz for L band, 4 GHz for C band and 8 GHz for X band) and their respective spw bandwidth (64 MHz for L band and 128 MHz for C and X bands) (see Fig.\ \ref{inbanddepol} a) and its zoom to low percentage b)).
Considering the lowest frequency at L band (i.e.,1 GHz), the in-band depolarization reaches values $>80\%$ for a RM value of $\sim$ 1000 rad/$m^2$ however, in-band depolarization of $\sim$ 50\% can be reached already for RM value of $\sim$ 200 rad/$m^2$.
For the lowest frequency at C band (i.e., 4 GHz), we estimated an in-band depolarization of $\sim$ 2 \% for RM of 1000 rad/$m^2$, while, for the same RM value, the in-band depolarization at the lowest frequency at X band (i.e., 8 GHz), is of the order of $\sim$ 0.03\%. With these calculations we want to show that the polarization information extracted at C and X band are not affected by observational effects such as the bandwidth of these receivers, therefore in-band depolarization is negligible at these bands and they are the best bands to perform a modelling of the polarization behaviour. In contrast, at L band, even with low RM values of 200 rad/$m^2$, we expect large in-band depolarization for a 64 MHz bandwidth. In order to relate the large RMs and the in-band depolarization at L band, one would extract polarization information within shorter bandwidths (i.e., BW$<$ 64 MHz), but with an increase of the rms noise. We explored this in our sample and made total intensity and polarization intensity L band images considering smaller bandwidths of 30 MHz and 15 MHz. At the sensitivity we got at L band, we could detect polarized signal for 4 sources: 0239-0234, 0243-0550, 1246-0730 and 1405+0415, the overall total intensity and polarized behaviour is shown in Fig.\ $\ref{pol3}$. For the last three sources (0243-0550, 1246-0730 and 1405+0415 ) we had enough data points to perform a depolarization modelling analysis at L band, after excluding a possible depolarization modelling considering the three bands together (see Section $\ref{DepolLband}$ for more details). Nevertheless, as mentioned, a proper polarization modelling could be also properly performed using the C and X bands.

\begin{table*}[h]
\small
\caption{Coefficients used for the polynomials fits for the calibrators 3C286, 3C138 and 3C48. The $\textit{an}$$_{i}$ are the coefficients taken from \cite{Perley13} to determine the Stokes I behaviour. The $\nu_c$ [GHz] is the central frequency used for the determination of the polynomial functions for the polarized data; the $\chi{_Coeff}$ [rad] are the coefficients used for the parametrization of the polarization angle and the p${_Coeff}$ are the coefficients used for the parametrization of the fractional polarization. UV$_{restrict}$ is a restriction in the UV plane, as suggested by NRAO.}
\begin{center}
\begin{tabular}{|l|l|l|l|l|l|}
\hline

\bf{Source} &	\bf{an$_{1}$} &	\bf{an$_{2}$} &	\bf{an$_{3}$} &	\bf{an$_{4}$} &	\bf{UV$_{restric}$} \\
3C286 &	1.2515 &	--0.4605 &	--0.1715 &	0.0336 &	<400k$\lambda$\\
\hline
\bf{Range [GHz]} 	&	\bm{$\nu_{c}$ }\bf{[GHz]} & 	\bm{$\chi_{Coeff}$} &	\bf{p{$_{Coeff}$} }& & \\
1.0 - 50.0 &	25.5 &	\tiny{0.62, 0.06 ,--0.06, --0.21, 0.16, 0.47,--0.17, --0.37} &	\tiny{0.13, 0.004, 0.02, 0.09, --0.13, --0.34, 0.19, 0.37} & & \\
\hline\hline
\bf{Source} &	\bf{an$_{1}$} &	\bf{an$_{2}$} &	\bf{an$_{3}$} &	\bf{an$_{4}$} &	\bf{UV$_{restric}$} \\
3C138 &	1.0332 &	--0.5608 &	--0.1197 &	0.0410 &	<40k$\lambda$ \\
\hline
\bf{Range [GHz]} 	&	\bm{$\nu_{c}$} \bf{[GHz]} &	\bm{$\chi_{Coeff}$} &	\bf{p{$_{Coeff}$} } & & \\
1.0 - 19.0 & 10.0 & \tiny{-0.13, 0.18, --0.41, --0.58, 1.08, 0.18,--0.92, 0.33} & \tiny{0.10, --0.04, --0.10, 0.10, 0.30, --0.29, --0.30, 0.29} & & \\
19.0 - 50.0 & 34.5 & \tiny{--0.41, --0.29, 0.23} & \tiny{0.07, --0.003, --0.01} & & \\
\hline\hline
\bf{Source} &	\bf{an$_{1}$} &	\bf{an$_{2}$} &	\bf{an$_{3}$} &	\bf{an$_{4}$} &	\bf{UV$_{restric}$} \\
3C48 & 1.3324 &	--0.7690 &	 --0.1950 &	 0.0590 &	<20k$\lambda$ \\
\hline
\bf{Range [GHz]}  	&	\bm{$\nu_{c}$} \bf{[GHz]} &	\bm{$\chi_{Coeff}$} &	\bf{p{$_{Coeff}$} } & & \\
1.0 - 2.0    &	1.5	 &	\tiny{--3.66, 5.066, --6.37, 3.59} &	\tiny{0.0056, 0.010, 0.0049, --0.00066} & & \\
2.0 - 19.0   &	10.5 &	\tiny{--1.05, 0.07, --1.05, 0.61, 3.58, --2.76, --5.90, 5.60} &	\tiny{0.06, 0.01, 0.03, --0.01, --0.16, 0.17, 0.12, --0.15} & & \\
19.0 - 50.0  &	34.5 &	\tiny{--1.34, --0.46, --0.21} &	\tiny{0.08, --0.01, --0.04} & & \\
\hline

\end{tabular}
\end{center}
\label{politab}
\end{table*}%

\section{Faraday screen and depolarization models}
\label{JVLAdepolmodels}

\begin{figure*}[!h]
	\centering
	\caption{Faraday Rotation sketch. The radiation (red arrow) coming from a synchrotron emitting region (yellow circle with non thermal electrons) passes through a magneto-ionic region (blue cloud with thermal electrons) with regular magnetic field (black arrows) that rotates the polarized vector of the radiation (blue arrow). In this scenario the fractional polarization remains constant. }
	\includegraphics[width=0.4\textwidth]{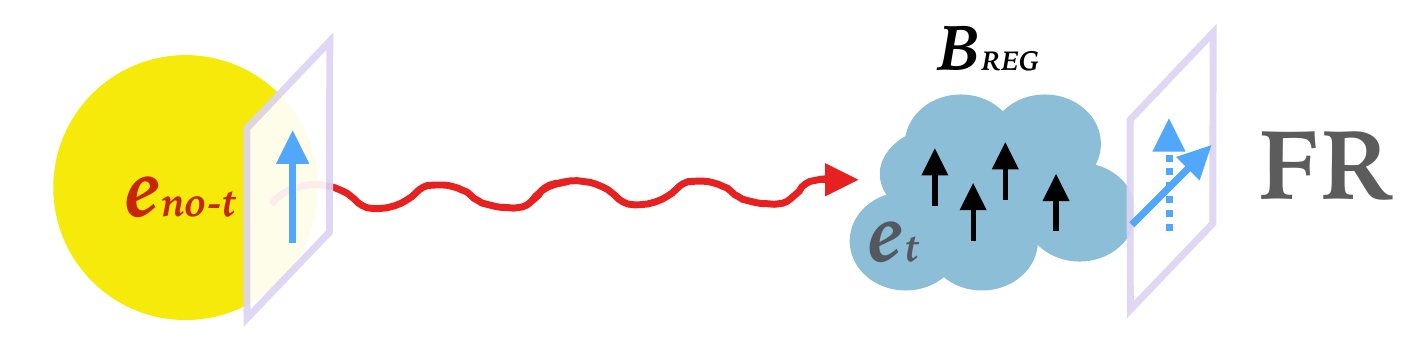}
	\label{FD}
\end{figure*}

\begin{figure*}[!h]
	\centering
	\caption{Depolarization mechanisms. The red undulated arrows represent the synchrotron radiation; the blue arrows represent the polarization vector of the radiation; its length decreases when depolarization occurs. The black arrows represent the magnetic field direction and its intensity is represented by the thickness of the arrows.
	The pictures (a), (b) and (c) represent the cases of internal depolarization where the synchrotron emitting region and the Faraday rotating regions coexist together (dark yellow with thermal, $\textit{e$_{t}$}$ and non thermal electrons $\textit{e$_{no-t}$}$). (a) represents the Differential Faraday Rotation (equation \ref{eqnDFR}) where a regular magnetic field is present. (b) represents the Internal Faraday Dispersion (equation \ref{eqnIFD}) where a turbulent and a regular magnetic field is present. (c) represents the depolarization due to an internal gradient of RM ($\Delta$RM${int}$, equation \ref{eqnDelta2}); the magnetic field is uniform and throughout the region the radiation undergoes to a smooth change of the polarization angle and of the fractional polarization.  
	 The picture (d) represent the case of external depolarization where the synchrotron emitting region (yellow circle containing $\textit{e$_{no-t}$}$) and the Faraday rotating region (blue cloud containing $\textit{e$_{t}$}$) are separated (equation \ref{eqnEFD}). Here the depolarization is due because of the presence of a turbulent magnetic field or because the polarized vectors cancel out within the telescope beam (Beam depolarization, Bd).
	(e) represents the depolarization due to a gradient of RM because of the presence of a foreground region (equation \ref{eqnDelta3}). }
	\includegraphics[width=0.8\textwidth]{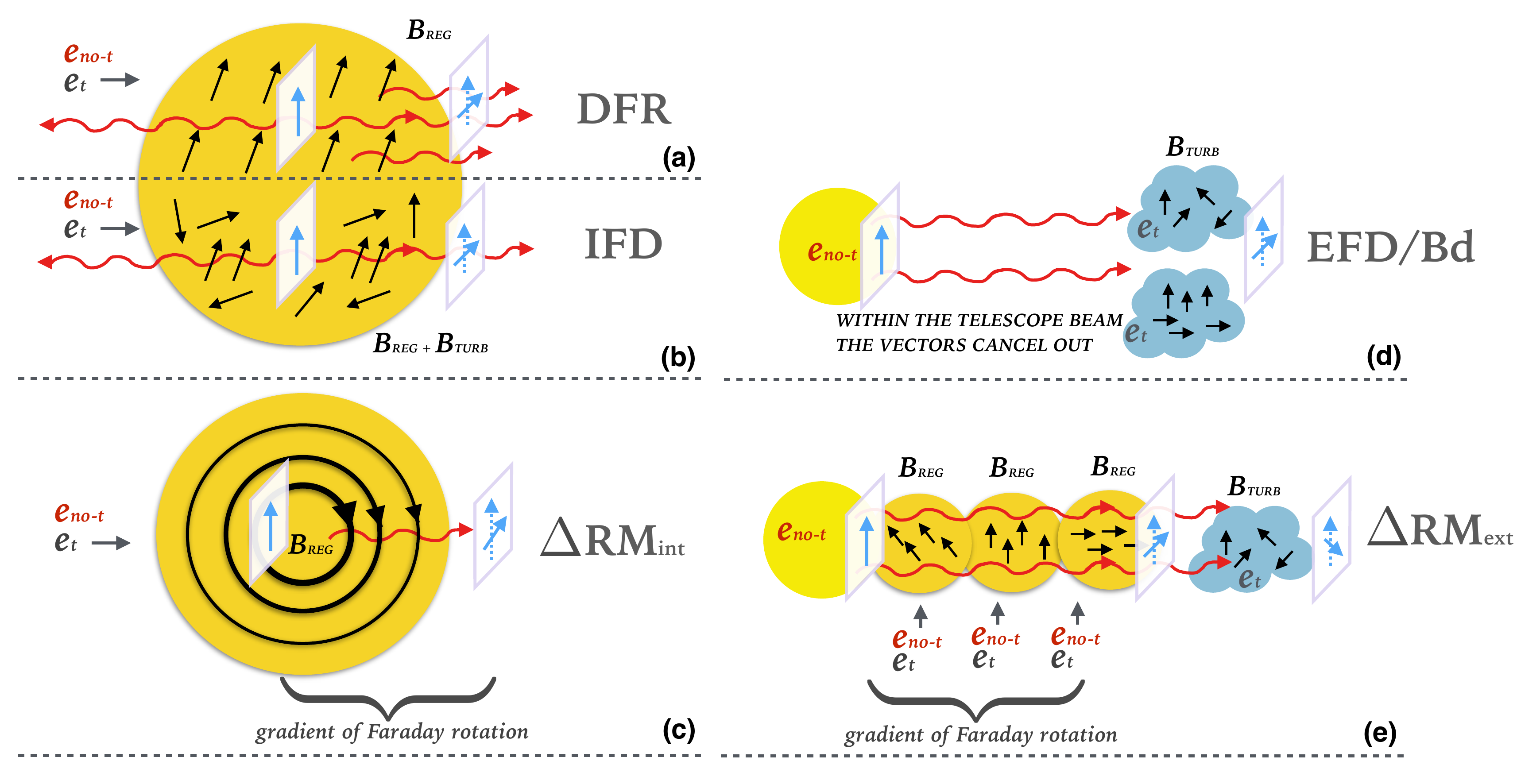}
	\label{FigSumDepol}
\end{figure*}

When synchrotron radiation passes through a magneto-ionic medium (Faraday screen), changes in the direction of its polarization angle (Faraday rotation) and a reduction of its polarization flux density (depolarization) can occur. When depolarization occurs, we can divide this effect into two main categories: \textit{(a) External Depolarization} and \textit{(b) Internal Depolarization} (see Tab.\ \ref{TabSumDepol} and Fig.\ \ref{FigSumDepol} for a schematic and visual explanation of the depolarization equations). Several authors described the two possible scenarios considering the presence in the two depolarization families, of a \textit{(1) uniform} and a \textit{(2) uniform and turbulent} magnetic field \citep[for more details on the different depolarization equations see: ][]{Burn66, Tribble91, Sokoloff98, Rossetti08}. To study the depolarization mechanism we will use the complex representation of the polarized signal in the presence of a Faraday rotation (FR, that does not depolarize) \citep[see Fig.\ \ref{FD}; ][]{Burn66}:
	\begin{equation}
		P = p_{0}e^{2i(\chi_0+RM\lambda^2)},
	\label{simple}
	\end{equation}

\noindent
where p$_{0}$ and $\chi_{0}$ are the intrinsic fractional polarization and the intrinsic polarization angle, respectively. This equation describes a constant behaviour of $\textit{p}$ with $\lambda^2$ and a linear behaviour of $\chi$ with $\lambda^2$. To this equation we add the depolarization contributions. 

When the synchrotron radiation is passing through a magneto-ionic medium that now contains a turbulent magnetic field, the depolarization is called External Faraday Dispersion/Beam depolarization (EFD/Bd) and it is represented by the equation:

	\begin{equation}
		P = p_0 e^{-2\sigma_{RM}^2 \lambda^4} e^{2i(\chi_0 + RM \lambda^2)} ,
	\label{eqnEFD}
	\end{equation}
	
\noindent where $\sigma_{RM}$ is the Faraday dispersion of the random field within the volume traced by the telescope beam \citep{Farnsworth11} and RM is the mean RM across the source on the sky. In this scenario, the depolarization occurs because of the presence of random magnetic cells or, in the case of regular magnetic field, because of the variation in the strength or direction of the field, both cases within the telescope beam.

The synchrotron emitting and the Faraday rotating regions may also be mixed together. In this case the depolarization is internal. The main equations that describe this depolarization scenarios are: (1) Internal Faraday Dispersion (IFD) and the (2) Differential Faraday Rotation \citep[DFR ][]{Gardner1966} with its extension to the case of a RM gradient \citep{Berkhuijsen1990,Sokoloff98}. In these equations, the contribution from a single foreground magneto-ionic material (equation \ref{simple}) is also present. 

In the case in which the synchrotron emitting and Faraday rotating region contains a turbulent and regular magnetic field together, i.e.\ the IFD, the degree of polarization is then given by,

	\begin{equation}
		P = p_0 e^{2i\chi_0}\left(\frac{1 - e^{-S}}{S} \right) ,
	\label{eqnIFD}
	\end{equation}

\noindent where $S=2\sigma_{RM}^2 \lambda^4 - 2i\phi\lambda^2$. In this case, the depolarization occurs because of the combination of both the presence of a regular magnetic field and the turbulent magnetic field. In this scenario, a random walk of the plane of polarization through the region occurs. In this equation $\sigma_{RM}$ is the internal Faraday dispersion of the random field and $\phi$ is the Faraday depth through the region.

When $\sigma_{RM}$ = 0 (therefore, no turbulent magnetic field component), the emitting and rotating regions are co-spatial in the presence only of a regular magnetic field, i.e. the DFR. The complex degree of polarization is given by,

	\begin{equation}
		P = p_0 \frac{\sin \phi \lambda^2}{\phi \lambda^2} e^{2i\left(\chi_0 + \frac{1}{2} \phi \lambda^2\right)} ,
	\label{eqnDFR}
	\end{equation}

\noindent where, $\phi$ is the Faraday depth through the region.
In this case the radiation coming from the most distant part of the region (with respect the observer) undergoes to a different amount of Faraday rotation with respect to the radiation coming from the nearest part of that region. 

Very interesting is the case of a gradient in RM across the beam discussed in the work by \cite{Berkhuijsen1990} and studied in detail in \cite{Sokoloff98}. This special case of DFR occurs when the RM varies systematically across the beam therefore, its magnetic field could be considered uniform. The RM gradient can originate in the synchrotron source and/or in a foreground screen but local to the radio emitting region. 
Following the detailed description in \cite{Sokoloff98}, the internal depolarization due to a smooth change in the RM across the beam (considering a flat beam profile) is given by:

	\begin{equation}
		P \propto  \frac{\sin (2\Delta RM \lambda^2)}{2\Delta RM\lambda^2} \left(e^{4iRM_0 \lambda^2}\right),
	\label{eqnDelta2}
	\end{equation}
	
\noindent 
where $\Delta$RM is the variation in RM across the beam and RM$_0$ is the initial value of RM within the region. If the gradient of RM is originated in a foreground Faraday screen, e.g. for a radio lobes embedded in a intra-cluster medium, the new equation is similar to \ref{eqnDelta2} but with the addition of the random field effect, $\sigma_{RM}$. Therefore, we get (for a flat beam profile):

	\begin{equation}
		P = p_{int}  \frac{\sin (\Delta RM \lambda^2)}{\Delta RM\lambda^2} \left(e^{2iRM_0 \lambda^2 - 2\sigma^2_{RM}\lambda^4}\right).
	\label{eqnDelta3}
	\end{equation}

\noindent However, to describe the more realistic case of a Gaussian beam profile, the $\Delta$RM has to be divided by a factor of 1.35 \citep{Sokoloff98}.

Our JVLA observations revealed complex behaviour both in the total intensity radio spectra and in polarization (see section \ref{JVLAresults}). In fact, the radio spectra were fitted with multiple synchrotron components. We therefore might expect a complex behaviour also in the polarization information with the presence of multiple interfering RM components. To investigate this, we fitted the wide band Stokes Q/I (\textit{q}) and U/I (\textit{u}) spectra following the procedure proposed by \cite{Farnsworth11,OSullivan2012}. The above equations can not explain completely the complex scenarios revealed from our data. The scenario that seems to better represent them is when multiple emitting and/or rotating components exist and they are unresolved within the telescope beam.
We proceed by fitting simultaneously both q($\lambda^2$) and u($\lambda^2$) using, first, the simplest equation of one component RM model (eq.\ \ref{simple}). It can not describe any of the polarized behaviour of the fractional polarization and polarization angle. Therefore, we tried using one component of the depolarization equations listed above. This approach did not give us good results thus, we finally tried multiple RM-component models. The multiple component models are simply constructed as $P=P_1+P_2 +...+P_N$ \citep{OSullivan2012}.

The model fitting used in this study (presented in section \ref{fitting}), attempts to describe the data using the simplest possible parameterization of Faraday depolarization from uniform and random fields (eqn.\ \ref{fitmath}). 
We implicitly assume that the polarized emission comes from optically thin regions with similar spectral index values.

\begin{table}[h]
\caption{Summary of the depolarization equations.}
\begin{center}
\begin{tabular}{|c|c|c|}
\hline
& External & Internal \\
&Depol. & Depol. \\
\hline
B$_{regular}$ 	& FR (eq.\ \ref{simple}, )	& DFR (eq.\ \ref{eqnDFR}) \\
 			& no depol.	$^*$			& 					 \\
			& 					& $\Delta RM$ internal (eq.\ \ref{eqnDelta2}) \\
\hline
B$_{regular}$ 	& EFD (eq.\ \ref{eqnEFD}) 	& IFD (eq.\ \ref{eqnIFD}) \\\cline{2-3}

+& \multicolumn{2}{c|}{$\Delta RM$ foreground (eq.\ \ref{eqnDelta3}) }  \\ 

B$_{turbulent}$ & \multicolumn{2}{c|}{ }  \\ 

\hline
\end{tabular}
 \begin{flushleft}
NOTE. - When the electromagnetic wave passes through an external magneto-ionic medium with a regular magnetic field, the polarization plane suffers for a rotation and its intensity remains constant; therefore, no depolarization occurs.
\end{flushleft}
\end{center}

\label{TabSumDepol}
\end{table}%

\section{Fitting the depolarization models and extracting the polarization parameters}
\label{fitting}


Assuming that all the emission components have the same spectral index, the depolarization models (described in section \ref{JVLAdepolmodels}) can be broadly represented by the following complex polarization equation:

	\begin{equation}
		P = \sum_{j=1,n} p_{0j} e^{2i(\chi_{0j}+RM_j \lambda^2)} sinc(\Delta RM_j \lambda^2) e^{(-2 \sigma_{RMj}^2 \lambda^4)},
	\label{fitmath}
	\end{equation}

\noindent where \textit{j} represents the several Faraday components used for the fitting, $p_{0j}$ is the intrinsic fractional polarization, $\chi_{0j}$ is the intrinsic polarization angle, $\Delta RM_j$ and $\sigma_{RMj}$ describe the variation of the RM in a regular and turbulent magnetic field respectively.
This is the only equation that we used to fit our data. We need to clarify that equation \ref{eqnIFD} is not exactly represented by equation \ref{fitmath}. In fact, the depolarization behaviour pretty closely matched but the polarization angle behaviour often is not. Nevertheless, equation \ref{fitmath} is an attempt to simplify the process of fitting many different models.

We limit the number of the Faraday components to 3 therefore, $P=P_1+P_2+P_3$. For all our sources first we fit a $\virgl$Faraday thin$\virgr$, where no contribution of $\Delta RM$ and $\sigma_{RM}$ are present, then we fit the equation \ref{fitmath} with the contribution of $\Delta RM$ only, then with $\sigma_{RM}$ only and finally with $\Delta RM$ and $\sigma_{RM}$ together. This approach results in a total of twelve equations used for the qu-fitting. Reduced $\chi^2$, standard deviation $\sigma^2$, the Bayesian Information Criterion (BIC) and the Akaike Information Criterion (AIC) have been performed in order to evaluate the quality of the fit.
Visual inspection, together with the combination of the lowest values of all the used statistics methods, help us to choose which equation, therefore which depolarization scenario, well represents the collected polarized data.
In order to compare the several polarization properties (RM, $\Delta RM$ and $\sigma_{RM}$) for each source, we weighted the respective values to the number of components $\textit{j}$. Therefore, we defined the polarization-weighted RM dispersion as:
		
		\begin{equation}
			\sigma_{RM,wdt} = \sum_{j} p_{0j} \sigma_{RM,j} \bigg/  \sum_{j} p_{0j},
		\label{sigmaw}
		\end{equation}
	
\noindent and the polarization-weighted RM gradient:
 
		\begin{equation}
			\Delta RM_{wdt} = \sum_{j} p_{0j} \Delta RM_{j} \bigg/  \sum_{j} p_{0j},
		\label{deltaw}
		\end{equation}

We corrected the measured RM for the Galactic foreground RM contribution (GRM) using the galactic RM map performed by \cite{Oppermann15}. Even though our sources are unpolarized, the \cite{Oppermann15} galactic RM map allowed us to estimate the GRM. 
We obtained the residual rotation measure (RRM) subtracting the polarization-weighted RM (RMwtd) and the GRM correspondent at the position of the source (i.e. RRM = RM$_{wtd}$ -- GRM). The polarization-weighted RM is calculated as

		\begin{equation}
			RM_{wdt} = \sum_{j} p_{0j} RM_{j} \bigg/  \sum_{j} p_{0j},
		\label{RMw}
		\end{equation}


\section{Results}
\label{JVLAresults}
\subsection{Radio spectrum}

\begin{figure*}[!h]
	\centering
	\caption{Radio spectra using L, C and X bands and literature. }
	\includegraphics[width=0.85\textwidth]{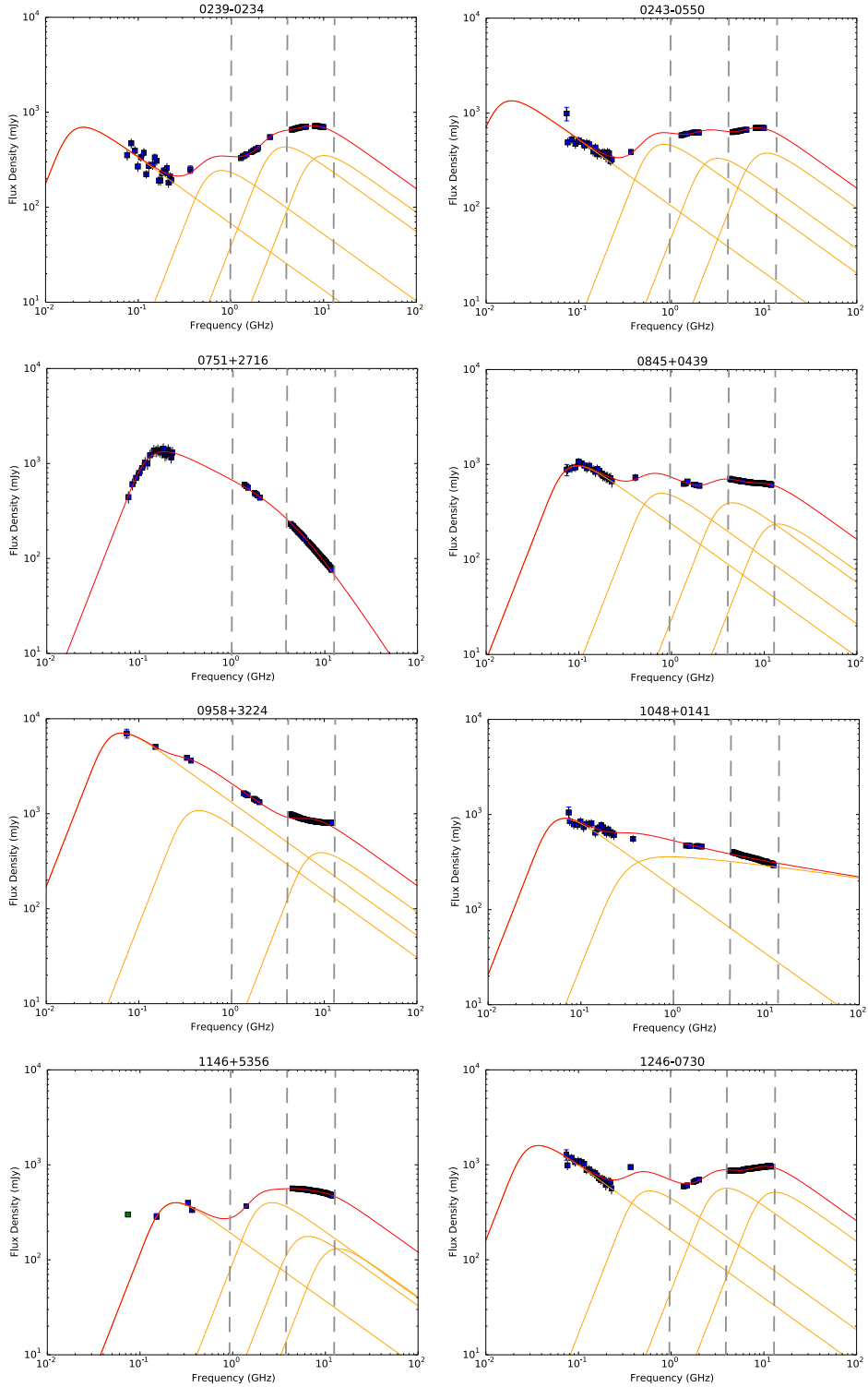}

	\begin{flushleft}
	NOTE:
	Total flux density is expressed in [mJy] and the frequency in [GHz]. Blue points are the JVLA and literature data and the green points are upper limits. Orange lines are the individual synchrotron components used for the radio spectra fit and the red line is their sum. The grey vertical dashed lines indicate the L to C band range and the C to X band range. 
	\end{flushleft}
	\label{sed1}

\end{figure*}

\begin{figure*}[!h]
	\centering
	\caption{Radio spectra using L, C and X bands and literature. }
	\includegraphics[width=0.7\textwidth]{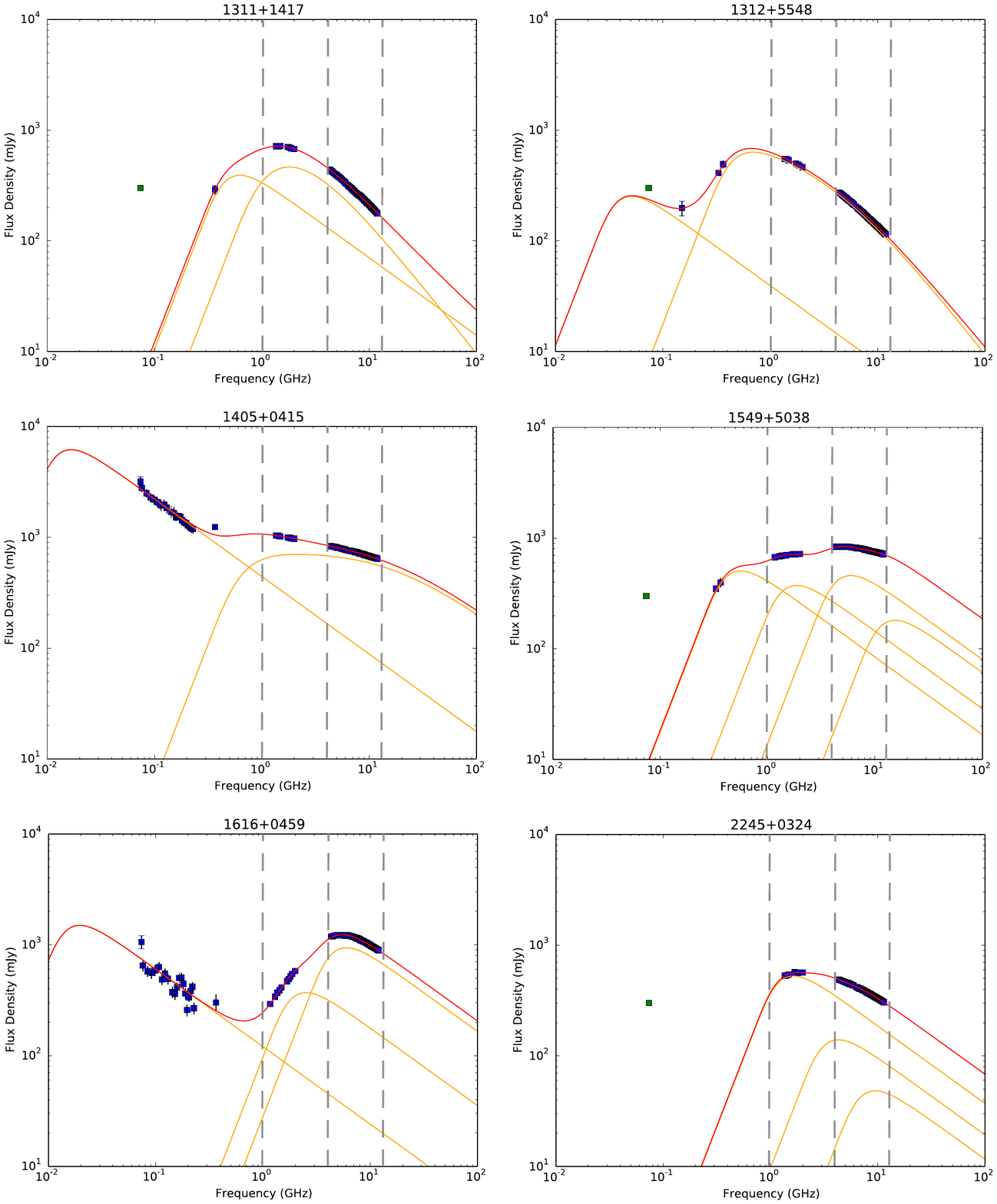}
	\begin{flushleft}
	NOTE:
	Total flux density is expressed in [mJy] and the frequency in [GHz]. Blue points are the JVLA and literature data and green points are upper limits. Orange lines are the individual synchrotron components used for the radio spectra fit and the red line is their sum. The grey vertical dashed lines indicate the L to C band range and C to X band range.
	\end{flushleft}
	\label{sed2}
\end{figure*}

The radio spectra (Fig.\ \ref{sed1} and Fig.\ \ref{sed2}) were made by using our JVLA data at L, C and X bands, as well as data at lower frequencies reported in several surveys, i.e.,\ the VLSS at 74 MHz \citep{Cohen07}, the 7C at 151 MHz \citep{Hales07}, the WENSS at 325 MHz \citep{Rengelink97} and the TEXAS at 365 MHz \citep{Douglas96}. We have also add recent low frequency data from the new broadband total intensity GLEAM survey \citep{Hurley2017}.
The sources in our sample have been previously observed with the Effelsberg 100-m single dish telescope at several frequencies (from 2 to 10 GHz). The radio single dish spectra fitting have been reported in \citetalias{Pasetto2016}. In the previous work we fitted the data using a power law (representing pure optically thin synchrotron emission), one to three synchrotron self absorption components, and a combination of power law models and synchrotron self absorption components \citepalias[see previous paper for more details][]{Pasetto2016}. These new broadband data have been fitted following a similar approach where the total intensity spectra are a composition of multiple synchrotron emitting volumes, considering homogeneous self-absorbed sources with power-law electron energy distributions with spectral index in the optically thick part of the spectrum $\alpha_{thick}$=2.5 and spectral index in the optically thin part of the spectrum $\alpha_{thin}$=--0.7. Because of the better sampled data at our disposal, we were forced to use more synchrotron emitting components to fit the total intensity data. Indeed, to well represent the radio spectra we used mainly a combination of synchrotron self-absorption components (using a maximum of 5 synchrotron components) and a combination of a single synchrotron self-absorption component with a synchrotron component with a break (symptomatic of ageing of the radio source).
The equations used for the radio spectra fitting are the following:
\begin{itemize}
\item a combination of several synchrotron self-absorption components ($S^{ssa}_{\nu} $):

	\begin{equation}
S^{ssa}_{\nu} \propto \sum^{5}_{1} \nu^{2.5}   \left( 1 - \exp\left(- \left(\frac{\nu}{\nu_{0}}\right)^{\alpha_{thin}-2.5}\right)\right),
	\end{equation}

where $\nu_0$ is the frequency where the emission changes from optically thick, with a spectral index of 2.5, to optically thin with a spectral index $\alpha_{thin}$ = --0.7;

\item a single synchrotron component with a break at frequency $\nu_{b}$ ($S^{b}_{\nu} $):
	\begin{equation}
S^{b}_{\nu} = S^{ssa}_{\nu} \left(1-\exp\left(-\left(\frac{\nu}{\nu_b}\right)^{\alpha_{break}-\alpha_{thin}}\right)\right) ;
	\end{equation}

where $\alpha_{thin}$ is the spectral index at frequencies lower than $\nu_{b}$ and $\alpha_{break}$ is the spectral index after the frequency break;

\item a combination of a synchrotron self-absorption component with a synchrotron component with a break at frequency $\nu_{b}$ ($S^{ssa+b}_{\nu} $).

	\begin{equation}
S^{ssa+b}_{\nu} \propto \nu^{2.5}   \left( 1 - \exp\left(- \left(\frac{\nu}{\nu_{0}}\right)^{\alpha_{thin}-2.5}\right)\right)+S^{ssa}_{\nu} \left(1-\exp\left(-\left(\frac{\nu}{\nu_b}\right)^{\alpha_{break}-\alpha_{thin}}\right)\right),
	\end{equation}

\end{itemize}

\noindent Although we used more synchrotron components with respect to the previous single dish radio spectra fitting, these results are consistent with those of the previous single dish observation campaign. Moreover, comparing the single dish and the interferometer fluxes, they are consistent. Therefore, no significant variability has been detected within the two observational sessions (time between the two sessions is roughly 2 years).
We note that this modelling is not unique, there are other possibilities to model the data, e.g, considering a thermal electrons contribution with $\alpha _{thick}$=2.0. In the practice, this is not too different from a 2.5 slope; the data at our disposal can not help us to discern which are the best total intensity components. Moreover, adding other different kind of total intensity components in the radio SED analysis, would complicate too much the overall interpretation. The procedure we decided to follow is a simple one that gives us an idea of the complexity of the sources.


\subsection{Polarization information}

\begin{figure*}[!h]
	\centering
	\caption{Polarization information of the sources at C and X bands. }
\includegraphics[width=0.9\textwidth]{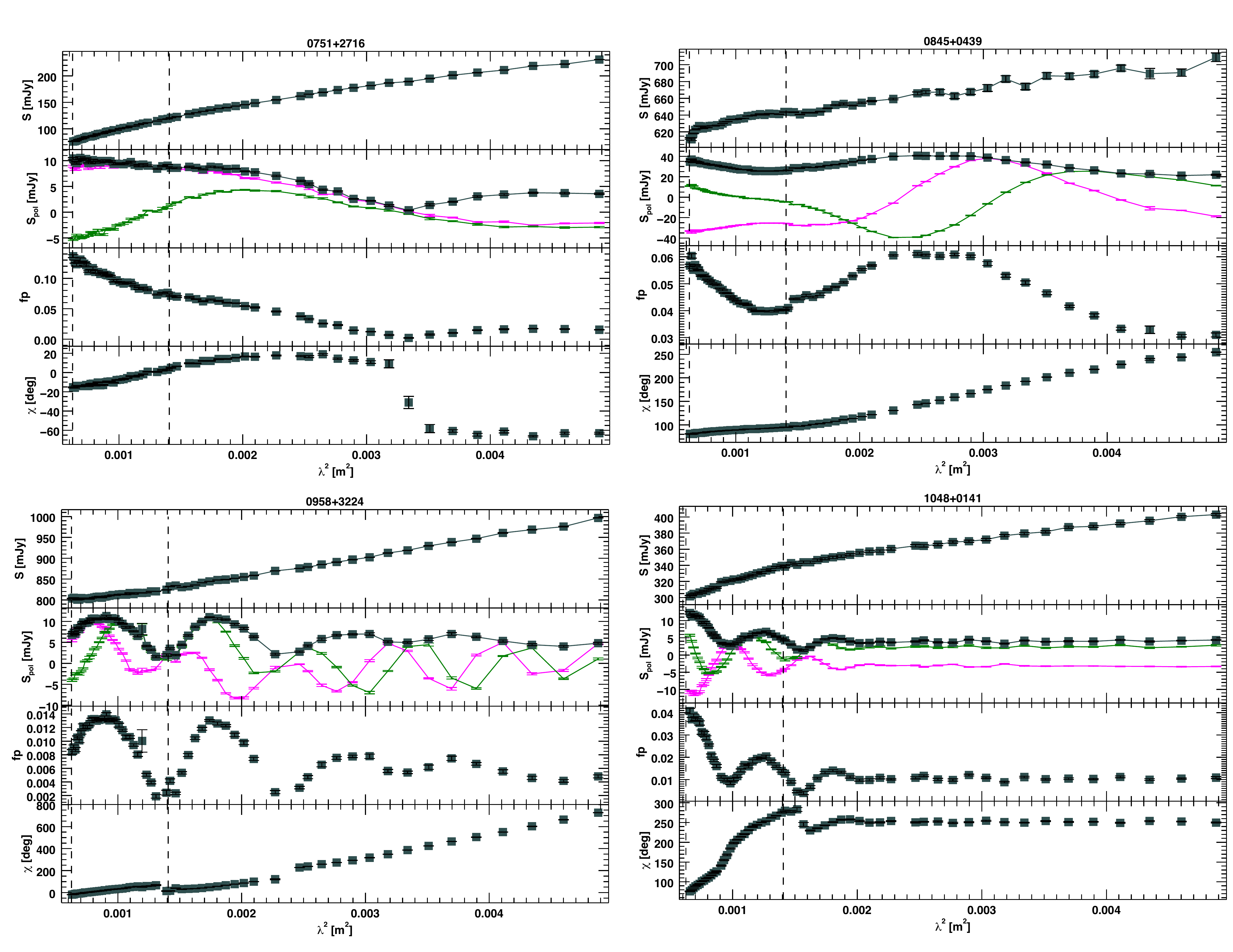}
\begin{flushleft}
\footnotesize{NOTE: For each plot: the first upper panel shows the total intensity \textit{S} expressed in [mJy] (the black line just connects the points), the second panel shows the polarized flux density \textit{S$_{pol}$} expressed in [mJy], the Stokes parameters Q (magenta dots) and U (green dots) expressed in [mJy] (the black, the magenta and the green lines just connect the points); the third middle window reports the fractional polarization; the last bottom window shows the polarization angle $\chi$ expressed in [deg]. All these information are represented in the $\lambda^2$ domain. The vertical dashed lines indicate the C to X band range.}
\end{flushleft}

	\label{pol1}
\end{figure*}

\begin{figure*}[!h]
	\centering
	\caption{Polarization information of the sources at C and X bands. Continued. }

\includegraphics[width=0.9\textwidth]{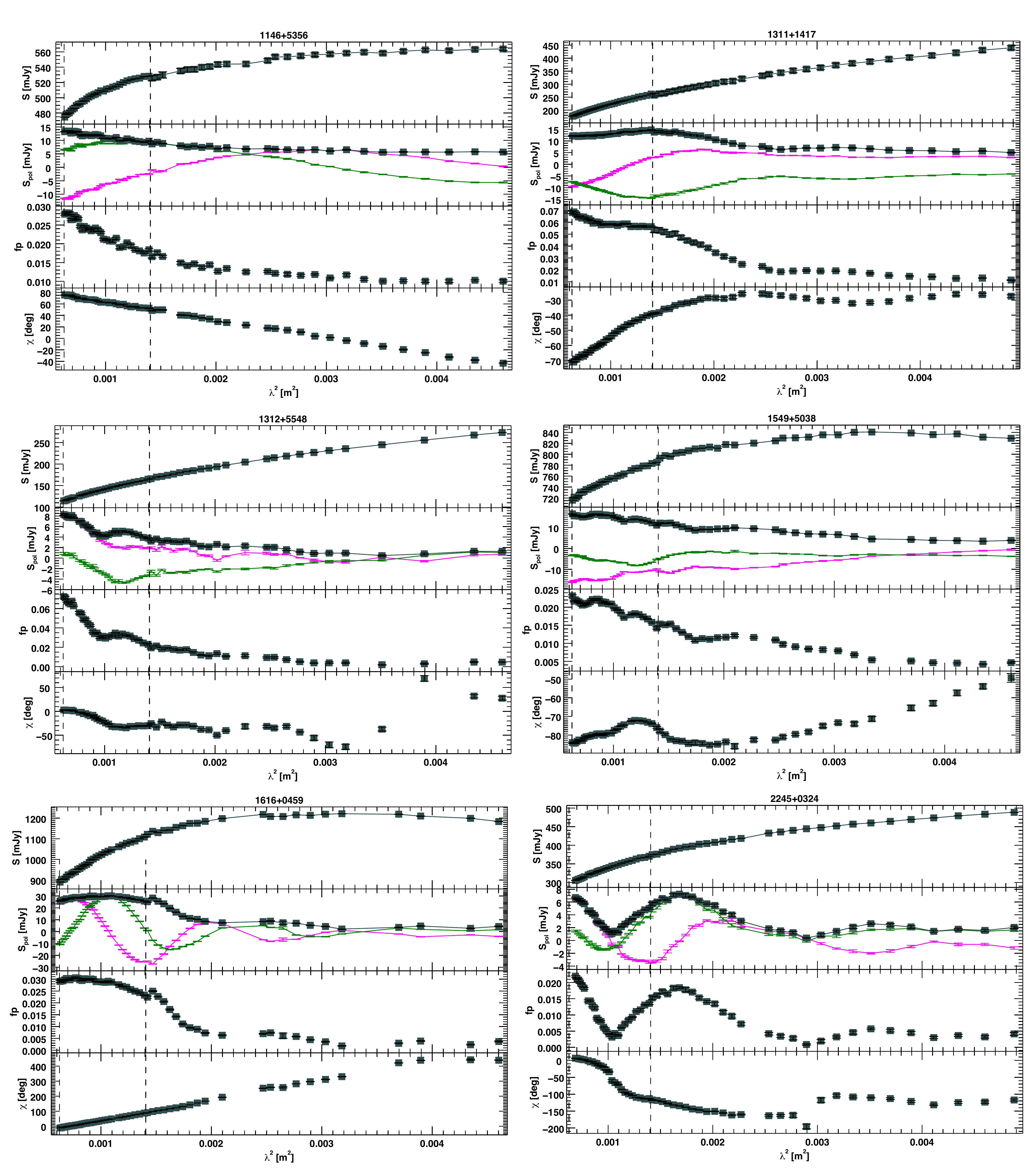}
	\label{pol2}
\end{figure*}

\begin{figure*}[!h]
	\centering
	\caption{Polarization information of the sources at L, C and X bands.}

\includegraphics[width=0.9\textwidth]{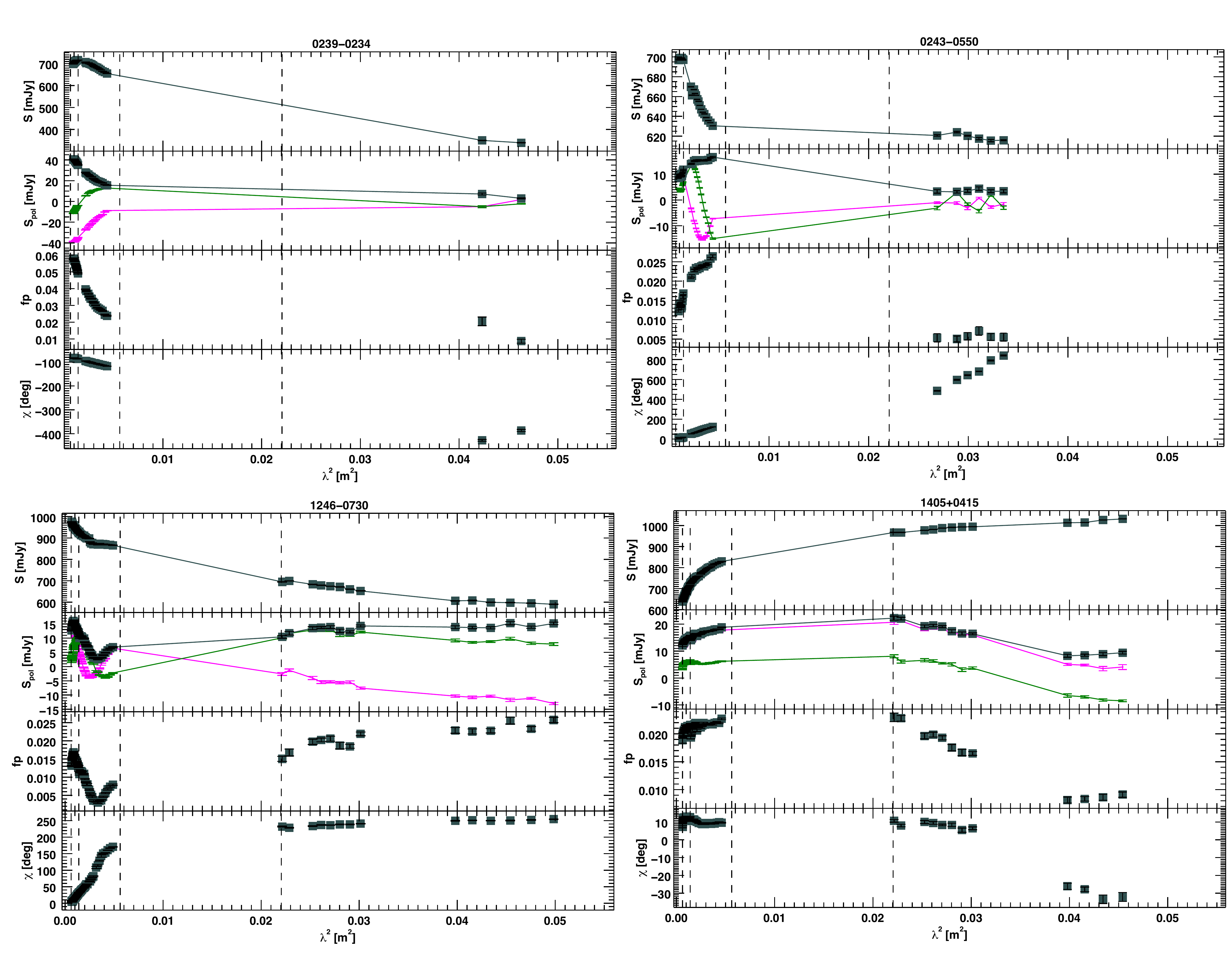}
	\label{pol3}
\end{figure*}

In Fig.s\ \ref{pol1} and \ref{pol2} we show the observed polarization quantities (the Stokes Q and U together with the polarized flux density \textit{S$_{pol}$}, the fractional polarization \textit{fp} and the polarization angle $\chi$) together with the total intensity data points for each source at C and X bands. In Fig.\ \ref{pol3} we show the same information but for those sources for which we have also polarization detection at L band (0239-0234, 0243-0550, 1246-0730 and 1405+0415). Previous single dish data revealed first signs of the complexity of the medium with a deviation from the linear fit in the determination of the RM value \citepalias{Pasetto2016}. Now, with these new broadband data,  the polarized signal clearly shows a complicated behaviour, with the fractional polarization and the polarization angle changing in a non-trivial manner. Our JVLA observations confirmed the previous results from the Effelsberg campaign \citepalias{Pasetto2016}, i.e.\, for most of the sources, the behaviour of the polarization angle deviates significantly from a simple linear trend with $\lambda^2$. Thus, we cannot assign a single RM for these sources in the 4--12 GHz range (for the 10 sources having C and X bands polarization data points) and in the 1--12 GHz range (for the 4 sources which have also polarized data points at L band). The polarisation behaviour for all the targets requires the presence of multiple synchrotron emission and Faraday rotation media. 




\noindent
This statement is reinforced after the qu-fitting (see section \ref{fitting} for more details) of the polarization data. The results of the qu-fitting are reported in Fig.s\ \ref{0751+2716model}$-$\ref{2245+0324model} and in Fig.s\ \ref{0239-0234CXbandmodel}$-$\ref{1405+0415Lbandmodel} in the appendix. The plots show the behaviour of the fractional Stokes parameters $\textit{q}$ and $\textit{u}$, the fractional polarization $\textit{p}$, the polarization angle $\textit{$\chi$}$, vs $\lambda^2$ and the behavior of $\textit{q}$ vs $\textit{u}$. The parameters that result from the depolarization modelling and their statistics are reported, for each of the target, in Tab.\ \ref{Parammodelling} and Tab.\ \ref{Parammodelling4sources}. 
The depolarization modelling reveals the presence of several Faraday components (more than two) to explain the complexity of the polarized signal.
Since all sources have C and X bands polarization data available, we focused on the analysis and discussion of that frequency range, e.i., 4--12 GHz range.
In order to analyze the polarization properties for all sources, i.e., RM, $\Delta RM$ and $\sigma_{RM}$ for each source, we weighted the respective values to the number of the Faraday rotating components , as explained in section \ref{fitting}. The discussion of these parameters are reported in section \ref{discussion}. The RM$_{wtd}$, the subsequent determined RRM, the $\sigma_{RM,wtd}$ and the $\Delta$RM$_{wtd}$ values are reported in Tab.\ \ref{RMrestframe}. The correction of the RRM in the rest frame of the targets is also reported in Tab.\ \ref{RMrestframe}. Overall, these parameters intrinsically reveal, once again, the complexity of the medium surrounding the sources. Indeed the values are considerably large, suggesting a highly magnetized and dense medium in the vicinity of the central engine. The source 1616+0459 shows the highest RRM value, in the rest frame it assumes a value of RM$_{rf}$$\approx$ 2.3$\cdot$10$^4$ rad/m$^2$.


\begin{table*}[h]
\caption{Resulting parameters of the depolarization modelling for the 10 sources with data at C and X bands}
\small
\begin{center}
\begin{tabular}{lllllllllllllllllll}
\hline\hline

 \hline\hline

Source 		& 	p$_{01}$ 				&	p$_{02}$ 				&	p$_{03}$ 					&	$\chi$$_{01}$ 			&	$\chi$$_{02}$ 			&	$\chi$$_{03}$ 				\\	
			& 		[\%]				& 		[\%]				& 		[\%]					&  [deg]					& [deg]						& [deg]							\\	
\hline

0751+2716 	& 25.0 (7.0	)				& 2.1 (0.8	)				& 9.0 (4.0)						&   --22.9 		(5.7 )		& 3.4	 	( 22.9)			& 68.8  (23.0 )					\\	 
0845+0439 	& 7.1  (0.2	)				& 3.3 (0.2	) 				& ... 							&   37.8 		(0.6 )		& --107.2 	( 1.7 )			& ...							\\	 
0958+3224 	& 0.60 (0.02)				& 0.5 (0.3	)				& 0.5 (0.3)						&   --174.2 	(1.7 )		& 5.7 		( 17.2)			& --63.0 (23.0 )				\\	 
1048+0141 	& 0.99 (0.02)				& 3.7 (0.1	)				& ... 							&   76.0 		(1.1 )		& 67.6 		( 1.7 )			& ...							\\	 
1146+5356 	& 1.36 (0.05)				& 1.9 (0.1	)				& ... 							&   --88.1	 	(1.1 )		& --83.1 	( 1.7 )			& ...							\\	 
1311+1417 	& 6.4  (0.4 )				& 2.6 (0.1 )				& 1.3 (0.3) 					&   --105.4 	(2.3 )		& --45.8 	( 2.3 )			& 101.8  (11.5) 				\\	 
1312+5548 	& 3.3  (0.3	)				& 9.0 (2.0	)				& 2.4 (0.5)						&   --12.0 		(4.6 )		& --104.1 	( 11.5)			& --118.0 (11.5)				\\	 
1549+5038 	& 2.8  (0.4	)				& 0.6 (0.1	)				& 1.3 (0.3)						&   --63.0 		(5.7 )		& --22.9 	( 11.5)			& --138.0 (11.5)				\\	 
1616+0459 	& 3.4  (0.1	)				& 0.6 (0.1	) 				& ... 							&   --107.2 	(1.1 )		& --159.1 	( 5.7 )			& ...							\\	 
2245+0324 	& 1.0  (0.2	)				& 1.0 (0.1	)				& 1.3 (0.1)						&   120.3 		(5.7 )		& 23.5 		( 5.2 )			& 22.9   (2.3 )					\\	 
	
\hline\hline

Source 		& 	RM$_{1}$ 					&	RM$_{2}$ 					&	RM$_{3}$ 					&	$\sigma_{1}$			&	$\sigma_{2}$ 			&	$\sigma_{3}$ 				\\
			& 	[rad/m$^2$]					& 	[rad/m$^2$]					& 	[rad/m$^2$]					& 	[rad/m$^2$]				&  	[rad/m$^2$]				& 	[rad/m$^2$]					\\	
\hline

0751+2716 	& --70 		(30)				& 414 		(84)				& --430 (90)					& 530 (70)	   				& 110 (70)					& 260 (50)						\\ 
0845+0439 	&  790 		(10)				& 1600 		(20)	 			& ... 							& 153 (5)					& 232 (14)	 				& ... 							\\ 
0958+3224 	& 3890 		(20)				& 640 		(180)				& 1160 (190)					& ... 						& ... 						& ... 							\\ 
1048+0141 	& --20 		(10)				& 5100 		(30)				& ...							& 10 (280)					& 600 (10)	 				& ... 							\\ 
1146+5356 	& --530 	(10)				& --560 	(50)	 			& ... 							& 100 (10)					& 630 (30)	 				& ... 							\\ 
1311+1417 	&  800 		(40) 				&  80 		(10)  				& --1150 (200) 					&  430 (10)  				&  130 (10) 				& 550 (60) 						\\ 
1312+5548 	& --210 	(40)				& --1655 	(273)				& 3423 (242)					& 360 (20)					& 970 (80)					& 724 (70)						\\ 
1549+5038 	& 90 		(70)				& 2280 		(150)				& 330 (50)						& 440 (40)					& 470 (50)					& 180 (20)	 					\\ 
1616+0459 	& 2460		(20)				& 500 		(50)	 			& ... 							& 380 (10)					& 310 (30)					& ... 							\\ 
2245+0324 	& --2800	(110)				& --1490 	(40)				& 140 (20)						& 430 (40)					& 230 (20)					& 270 (10)	 					\\ 

\hline\hline

Source 		& 	$\Delta RM_{1}$ 				&	$\Delta RM_{2}$ 				&	$\Delta RM_{3}$ 				& AIC 							&	BIC 			&red$\chi^2$	& $\sigma^{2}$ 					\\	
			& 	[rad/m$^2$]					& 	[rad/m$^2$]					& 	[rad/m$^2$]					& 	 							& 	 				& 			& 	 							\\	
\hline

0751+2716 	& ... 							& ... 							& ... 							& 370 							& 404 				& 	0.99	& 1.3 							\\	
0845+0439 	& ... 							& ... 							& ... 							& 475 							& 500 				& 	0.99	& 3.0 							\\	
0958+3224 	&  370 (50)						& 810 (50)						& 850 (70)	 					& 620 							& 660 				& 	0.95	& 11.0 							\\	
1048+0141 	& ...							& ...							& ...							& 455 							& 480 				& 	0.98	& 3.0							\\
1146+5356 	& ... 							& ... 							& ... 							& 350 							& 370 				& 	0.99	& 1.2 							\\	
1311+1417 	& ... 							& ... 							& ... 							& 296 							& 332 				& 	0.99	& 0.6 							\\	
1312+5548 	& ... 							& ... 							& ... 							& 449 							& 483 				& 	0.97	& 3.5 							\\	
1549+5038 	& ... 							& ... 							& ... 							& 450 							& 490 				& 	0.99	& 3.0 							\\	
1616+0459 	& ... 							& ... 							& ... 							& 546 							& 570 				& 	0.97	& 9.2 							\\	
2245+0324 	& ... 							& ... 							& ... 							& 502 							& 537 				& 	0.94	& 5.4 							\\	

\hline\hline

\end{tabular}
\end{center}
\label{Parammodelling}
	\begin{flushleft}
NOTE: \\
Column one of the first second and third panel is the source name; p$_{01}$, p$_{02}$, p$_{03}$ are the initial fractional polarization of the first, second and third Faraday components; $\chi_{01}$, $\chi_{02}$, $\chi_{03}$ are the initial polarization angle of the Faraday components; RM$_{1}$, RM$_{2}$, RM$_{3}$ are the Rotation Measure of the Faraday components; $\sigma_{RM1}$, $\sigma_{RM2}$, $\sigma_{RM3}$ are the Faraday dispersion values of the Faraday components; $\Delta RM_{1}$, $\Delta RM_{2}$, $\Delta RM_{3}$ are the RM gradient of the Faraday components.
\textit{AIC} is the Akaike Information Criterion, it is a measure of the relative quality of statistical models for a given set of data.\\
\textit{BIC} is the Bayesian Information Criterion, it is a criterion for model selection among a finite set of models. The lowest the BIC, the better the model. \textit{red$\chi^2$} is the reduced chi-squared test. 
\textit{$\sigma^2$} is the estimated variance; it is the squared deviation of a variable from its mean, how far a set of data are displaced from their mean.
	\end{flushleft}

\end{table*}%


\subsection{Depolarization modelling at L band}
\label{DepolLband}
As mentioned, we detected polarization at L band for four sources (0239--0234, 0243--0550, 1246--0730 and 1405+0415) by splitting each available spectral window (each with bandwidth of 64 MHz) into smaller bandwidth (BW): 30 MHz and 15 MHz. Roughly for the first half of the L band (from 1 to $\sim$1.5 GHz) we did not detect polarization for all the four sources (in Fig.\ \ref{pol3} we omit the first half of L band in order to save space in the plots).

For sources 0239--0234 and 0243-0550, we detected 2 and 6 data points, respectively when splitting the spw into BW of 30 MHz and no data points when using the 15 MHz BW (because of the poor sensitivity obtained at these smaller bandwidths). These two sources have low polarized flux density at L band (around 3 mJy both).
For  sources 1246--0730 and 1405+0415, we detected 14 and 12 data points respectively when using the 30 MHz BW and 24 and 22 data points when using the 15 MHz BW.
Because fluxes were consistent and the lower signal-to-noise ratio in the 15 MHz BW images, we decided to analyze the data of the 30 MHz BW only. Fig.\ \ref{pol3} shows the polarization together with the total intensity for the four sources for which also L band data have been detected when splitting the spw into BW of 30 MHz. 

\begin{figure*}[h]
\begin{center}
    \caption{Depolarization models for the four sources at C and X bands with the L band data points over plotted}
        \includegraphics[width=0.4\textwidth]{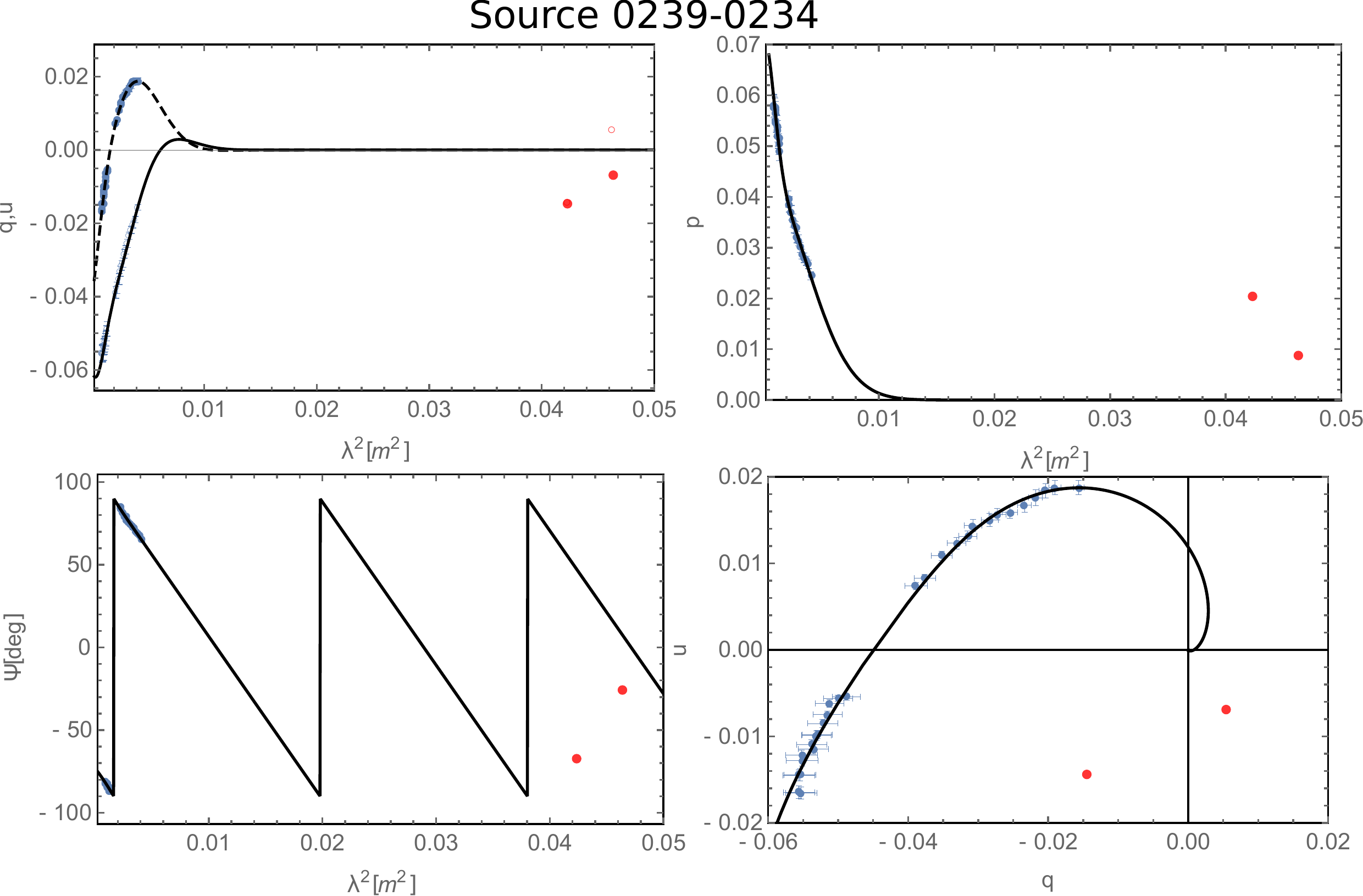}
        \includegraphics[width=0.4\textwidth]{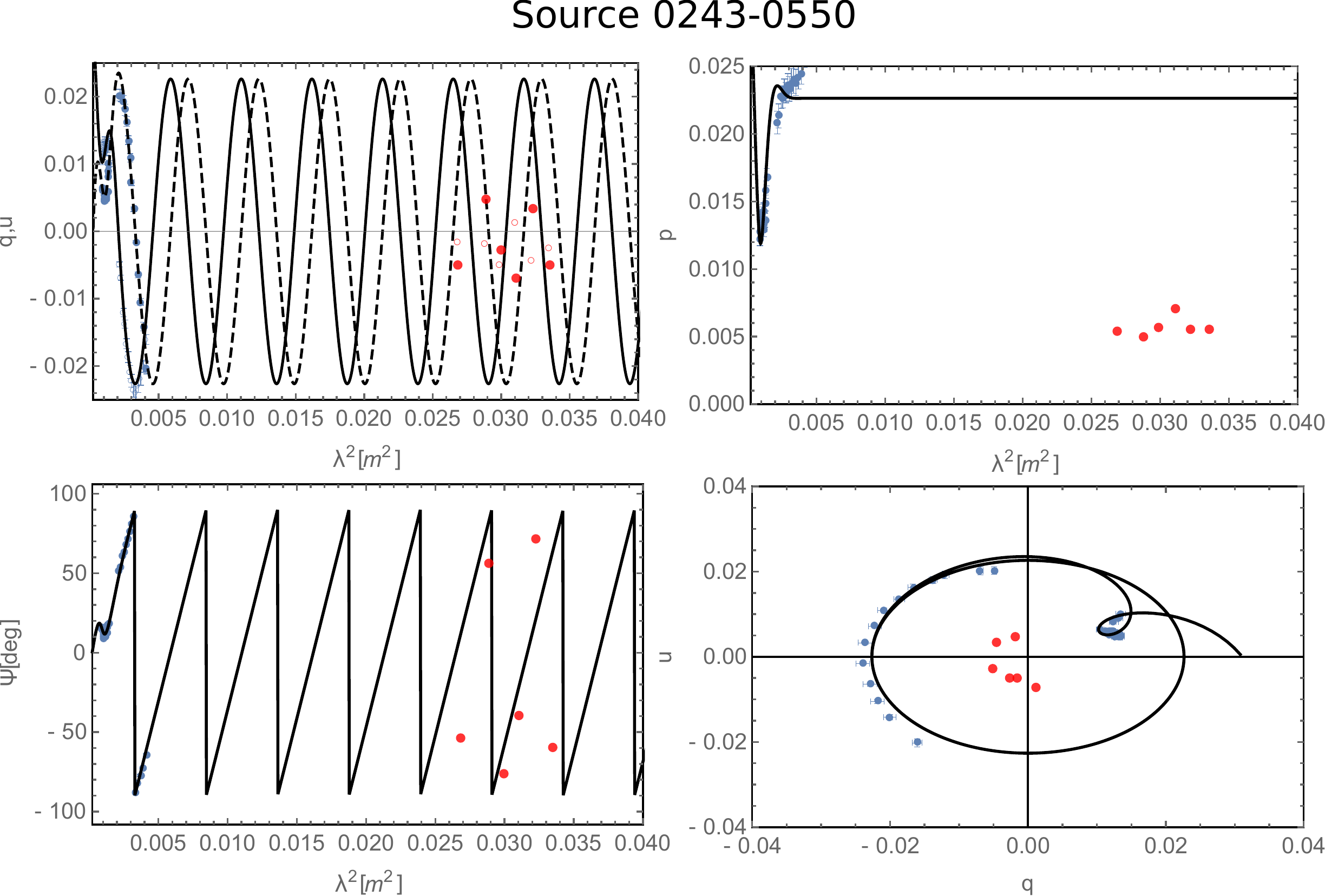}
        \includegraphics[width=0.4\textwidth]{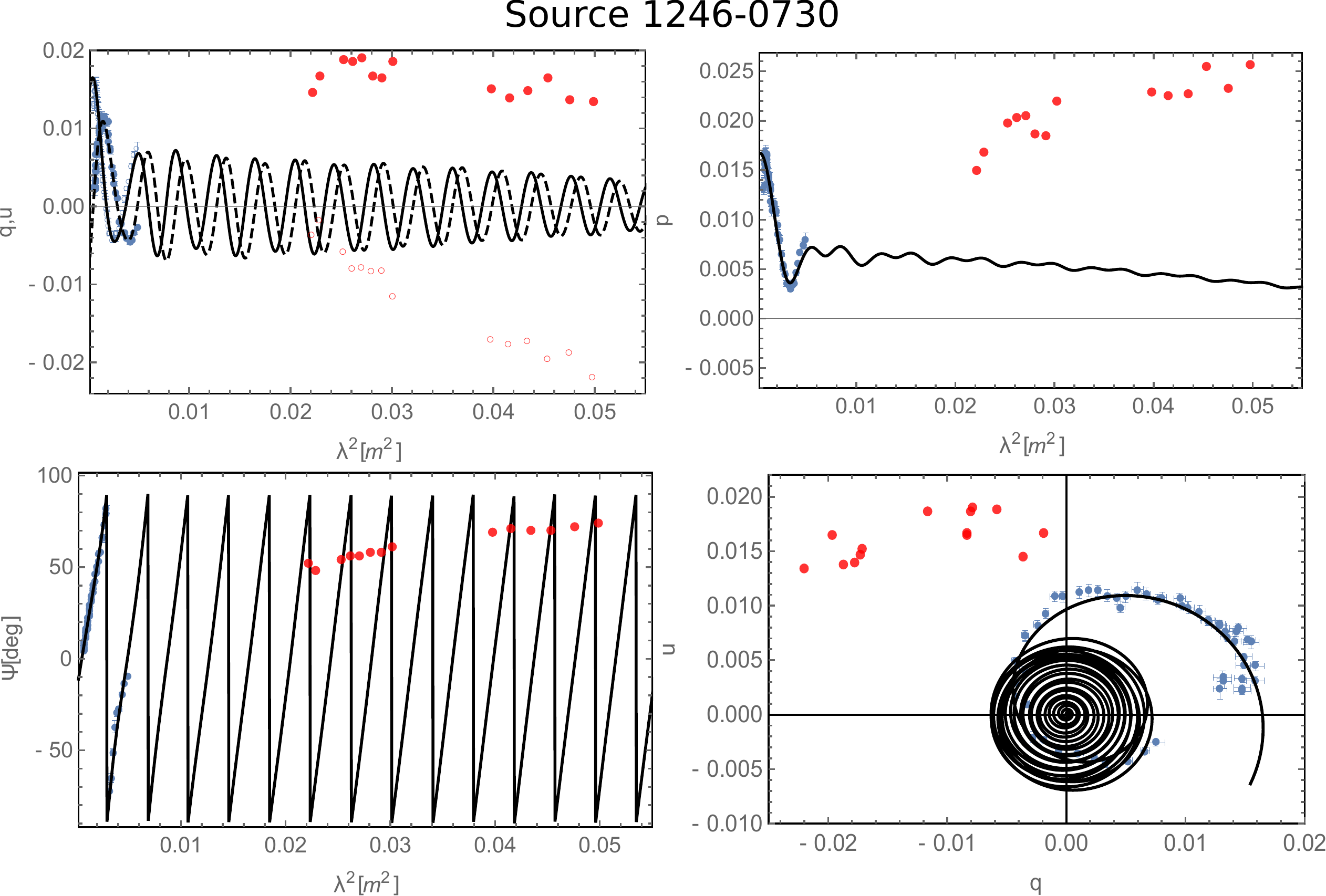}
        \includegraphics[width=0.4\textwidth]{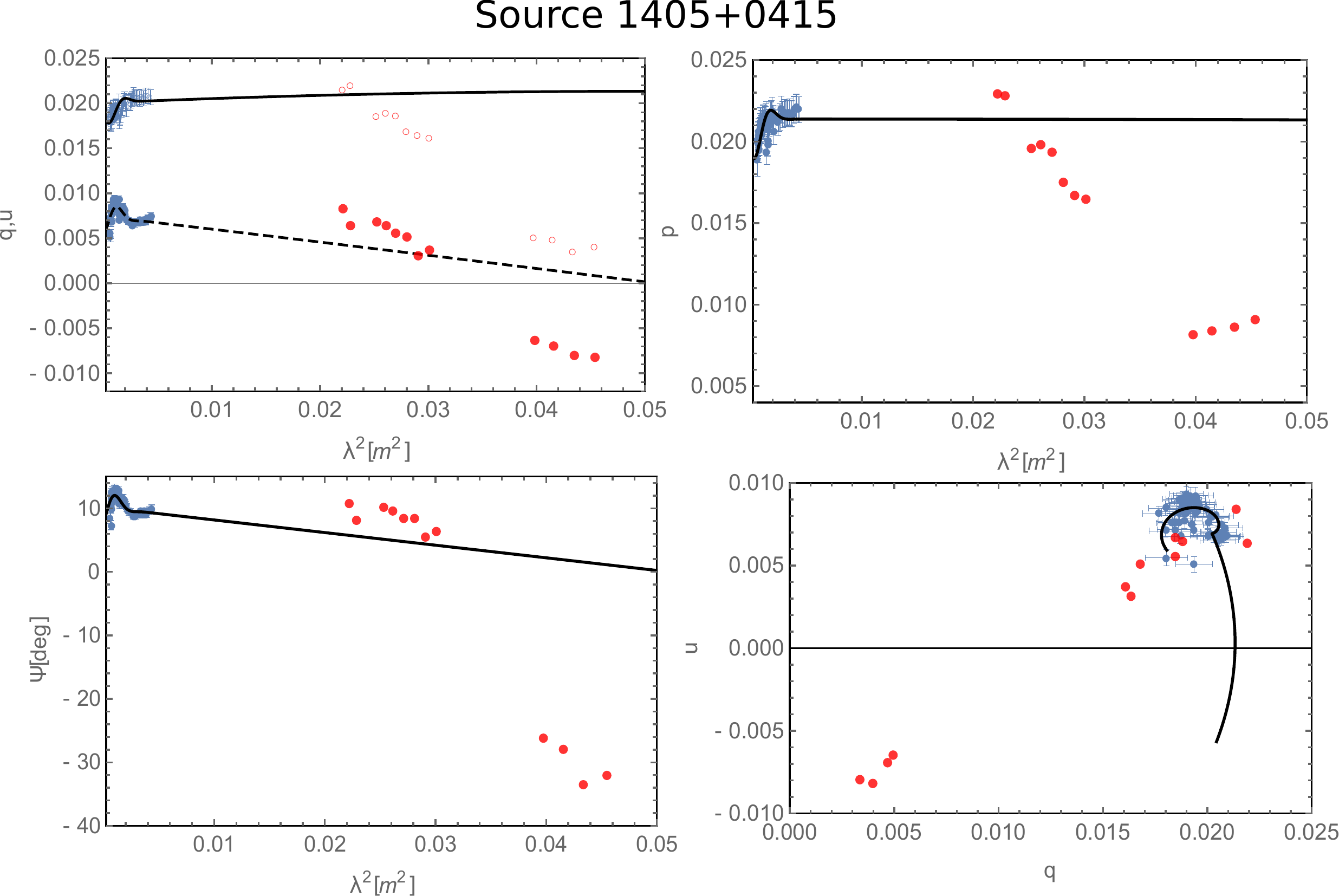}
\label{Depol4targetsTOT}
\end{center}
 \begin{flushleft}
NOTE. - For each source: behaviour of the fractional Stokes parameters (\textit{q} and \textit{u}), the fractional polarization (\textit{p}) and the polarization angle (\textit{$\chi$}) vs.\ wavelength squared (\textit{$\lambda^2$}). Blue data points are C and X bands data, red points are L band data. Black lines are the depolarization models fit considering C and X bands.

\end{flushleft}

\end{figure*}

Depolarization modelling considering the 3 bands (L, C and X bands) have been performed to these targets. However, for none of them we could find any good model that could represent the polarization behaviour in the whole wide band. Indeed, the L band data seem to behave very differently with respect to the C-X bands data. Fig.\ \ref{Depol4targetsTOT} show the C-X bands (blue points) depolarization modelling of these sources with the L band data points (red points) over plotted. The low frequency data do not follow the depolarization described using the high frequency data. Therefore, we performed depolarization modelling considering C-X bands and L band separately. The source 0239-0234 was not considered for L band depolarization modelling, because it only has two data points at low frequency.
Fig. \ref{0239-0234CXbandmodel}-\ref{1405+0415Lbandmodel} in the appendix show the depolarization modelling for each of the targets considering C and X bands together and L band only (for the source 0239--0234 only C-X bands depolarization modelling is available).
The parameters resulting from the fitting are reported in Tab.\ \ref{Parammodelling4sources}.


\begin{table*}[h]
\caption{Resulting parameters of the depolarization modelling for the four source at CX band and L band.}
\small
\begin{center}
\begin{tabular}{lllll}

 \hline\hline

Source 		& 	p$_{01CX}$ 				&	p$_{02CX}$ 				&	p$_{01L}$ 					&	p$_{02L}$ 					\\		
			& 		[\%]				& 		[\%]				& 		[\%]					& 		[\%]					\\		
\hline

0239--0234	& 2.7  (0.2 )				& 4.2 (0.2	)				& ... 							& ... 							\\		 		 
0243--0550	& 2.3  (0.1	)				& 1.9 (0.4 )				& 0.21 (0.08) 					& 0.43 (0.08) 					\\		  
1246--0730	& 1.41 (0.06)				& 0.7 (0.1	)				& 0.4 (0.2) 					& 2.0 (0.2) 					\\		 
1405+0415 	& 0.3  (0.1	)				& 2.14(0.04)	 			& 1.2 (0.5)  					& 2.0 (0.5) 					\\		 	

 \hline\hline

Source 		&	$\chi$$_{01CX}$ 		&	$\chi$$_{02CX}$ 		&	$\chi$$_{01L}$ 				&	$\chi$$_{02L}$ 				\\
			&  [deg]					& [deg]						& [deg]							& [deg]							\\
\hline

0239--0234	&   --68.8		(1.7 )		& --75.6 	( 1.1 )			& ...							& ...							\\
0243--0550	&   --25.2 		(1.7 )		& --177.6 	( 5.7 )			& --132.5 (166.2)				& 172.0 (74.5)							\\
1246--0730	&   --8.0 		(2.7 )		& --49.9 	( 2.3 )			& 143.2 (46.0)					& 23.0 (11.0)							\\
1405+0415 	&   --61.7 		(17.2)	& 10.3 	( 1.1 )			& 8.0 (23.0)					& 17.2 (17.2)							\\

\hline\hline

Source 		& 	RM$_{1CX}$ 			&	RM$_{2CX}$ 					&	RM$_{1L}$ 					&	RM$_{2L}$ 				\\		
			& 	[rad/m$^2$]			& 	[rad/m$^2$]					& 	[rad/m$^2$]					& 	[rad/m$^2$]				\\		
\hline

0239--0234	& --150 	(40)				& --170 	(10) 						& ... 							& ... 						\\		
0243--0550	& 609 		(11) 			& 1630 		(130) 				& 130 (100) 					& 908 (44) 					\\		
1246--0730	& 610 		(20)			& 840 		(20)	 				& --27 (20)			 		& 20 (4)			 			\\		
1405+0415 	& --850 	(250)				& --3 		(4)					& 24 (10) 						& -20 (7) 						\\		

\hline\hline

Source 		&	$\sigma_{RM1CX}$			&	$\sigma_{RM2CX}$ 			&	$\sigma_{RM1L}$ 				&	$\sigma_{RM2L}$ 				\\
			& 	[rad/m$^2$]				&  	[rad/m$^2$]				& 	[rad/m$^2$]					& 	[rad/m$^2$]					\\	
\hline

0239--0234	& 500 (40)					& 130 (10) 					& ... 							& ... 							\\ 
0243--0550	& ...				 		& 523 (62)	 				& ... 							& ... 							\\ 
1246--0730	& ... 						& ...						& ... 							& ... 							\\ 
1405+0415 	& 440 (100)				& 1 (730)					& ... 							& ... 							\\ 

\hline\hline

Source 		& 	$\Delta RM_{1CX}$ 			&	$\Delta RM_{2CX}$ 			&	$\Delta RM_{1L}$ 			&	$\Delta RM_{2L}$ 		\\			
			& 	[rad/m$^2$]					& 	[rad/m$^2$]					& 	[rad/m$^2$]					& 	[rad/m$^2$]				\\		
\hline

0239--0234	& ... 							& ... 							& ... 							& ... 						\\		
0243--0550	& ...							& ... 							& ... 							& ... 						\\		
1246--0730	& 1340 (90)					& 30 (2120)					& ... 							& ... 						\\		
1405+0415 	& ... 							& ... 							& ... 							& ... 						\\		

\hline\hline

Source 		& AIC$_{CX}$  						&	BIC$_{CX}$ 		&red$\chi^2$$_{CX}$	& $\sigma^{2}$$_{CX}$ 			\\
			& 	 							& 	 				& 					& 	 							\\	
\hline

0239--0234	& 113 							& 132 				& 	0.99			& 0.4 							\\	
0243--0550	& 257 							& 276 				& 	0.98			& 4.2 							\\	
1246--0730	& 463 							& 487 				& 	0.98			& 3.6 							\\	
1405+0415 	& 380 							& 402 				& 	0.99			& 1.4 							\\	

\hline\hline

Source 		  & AIC$_{L}$  						&	BIC$_{L}$ 		&red$\chi^2$$_{L}$	& $\sigma^{2}$$_{L}$ 			\\
			  & 	 							& 	 				& 					& 	 							\\	
\hline 
 
0239--0234	  & ... 							& ... 				& 	...				& ... 							\\	
0243--0550	  & 60 								& 64 				& 	0.85			& 5.6 							\\	
1246--0730	  & 126 							& 135 				& 	0.99			& 4.0 							\\	
1405+0415 	  & 109 							& 117 				& 	0.99			& 4.0 							\\	
\hline\hline

\end{tabular}
\end{center}
\label{Parammodelling4sources}
	\begin{flushleft}
NOTE: \\
p$_{01CX}$ , p$_{02CX}$ , p$_{01L}$ , p$_{02L}$  are the initial fractional polarization of the first and second Faraday components in the C-X bands frequency range and L band frequency range; $\chi_{01CX}$, $\chi_{02CX}$, $\chi_{01L}$ and $\chi_{02L}$ are the initial polarization angle of the Faraday components within the C-X bands frequency range and L band frequency range; RM$_{1CX}$, RM$_{2CX}$, RM$_{1L}$ and RM$_{2L}$ are the Rotation Measure of the Faraday components within the C-X bands frequency range and L band frequency range; $\sigma_{RM1CX}$, $\sigma_{RM2CX}$, $\sigma_{RM1L}$ and $\sigma_{RM2L}$ are the Faraday dispersion values of the Faraday components; $\Delta RM_{1CX}$, $\Delta RM_{2CX}$, $\Delta RM_{1L}$ and $\Delta RM_{2L}$ are the RM gradient of the Faraday components.
\textit{AIC} is the Akaike Information Criterion, it is a measure of the relative quality of statistical models for a given set of data.\\
\textit{BIC} is the Bayesian Information Criterion, it is a criterion for model selection among a finite set of models. The lowest the BIC, the better the model. \textit{red$\chi^2$} is the reduced chi-squared test. 
\textit{$\sigma^2$} is the estimated variance; it is the squared deviation of a variable from its mean, how far a set of data are displaced from their mean. All the statistics parameters have been determine whitin the C-X bands frequency range and the L band frequency range.
	\end{flushleft}

\end{table*}%



\begin{table*}[h!]
\caption{Polarization weighted parameters and RM corrected in the rest frame (RM$_{rf}$) considering the C-X bands frequency range.}
\begin{center}
\begin{tabular}{lccccccc}
\hline\hline
Source		&   z   & RM$_{MW}$ 	& 	RM$_{wtd}$   		&  		RRM 			&      		RM$_{rf}$        &    $\sigma_{RM,wtd}$		&$\Delta$RM$_{wtd}$  	\\
	 \hline								                                 		             			  	                                  
  0239--0234 &  1.1  & -100  (100) 	&	-162  (20 ) 	& -62  	 	(101)		& 140	(450 )	& 280  	(20		)	&     --       \\
  0243--0550 &  1.8  & -100  (100) 	&	1071  (104 )	&  1170  	(144) 		& 4970 	(1130)	& 240  	(43810)		&     --       \\
  0751+2716  &  3.2  & -20   (10 )  &	-130   (50) 	& -110 	 	(50	)		& 1254	(870)	& -300  (63		)	&     --       \\
  0845+0439  &  0.3  & -100  (100) 	&	1047   (24 )	&  1150  	(103)		& 1250 	(174)	& 180   (10		)	&     --       \\
  0958+3224  &  0.5  & -10   (10 ) 	&	2021  (192 )	&  2031     (192)		& 2540  (432)	& --				& 660 (160)		\\
  1048+0141  &  0.7  & -100  (100) 	&	4020   (33 )	&  4120  	(110)		& 6140 	(304)	& 480   (60		)	&     --        \\
  1146+5356  &  2.2  & 10    (10 ) 	&	-550   (30 )	&  -560  	(32	)		& 3260 	(330)	& -410 	(20		)	&     --        \\
  1246--0730 &  1.2  & 10    (40 ) 	&	686   (31 ) 	& 680  	 	(50 )		& 1650	(250)	& --				& -890 (710) 	\\
  1311+1417  &  1.9  & 10    (10 ) 	&	370    (50 ) 	& 360  	 	(50	)		& 1670	(410)	& 230  	(20		)	&     --        \\
  1312+5548  &  1.1  & 20    (10 ) 	&	-500   (220 )	&  -520    	(220)		& 1152 	(960)	& -792  (80		)	&     --        \\
  1405+0415  &  3.2  & 30    (60 ) 	&	-107   (43) 	& -140 	 	(74 )		& 1541	(1310)	& -60 	(640	)	&     --        \\
  1549+5038  &  2.2  & 20    (10 ) 	&	440    (70 ) 	& 420  	 	(70 )		& 2430	(682)	& 252   (32		)	&     --        \\
  1616+0459  &  3.2  & 150   (10 ) 	&	2166  (23 ) 	&  2020     (30 ) 		& 22660 (450)	& -370  (12		)	&     --        \\
  2245+0324  &  1.3  & -40   (100) 	&	-120   (130)	&  -1204 	(162) 		& 3241 	(860)	& 50  	(24		)	&     --        \\

\hline\hline

\hline\hline
\end{tabular}
\end{center}
\smallskip

\label{RMrestframe}
\end{table*}


\section{Comments on the individual sources}
\label{Commentsindividualsources}

Here we comment on the observational and modelling results in the context of the AGN environment. We consider and combine all the wide band observational results: the radio spectrum, the polarization information and the depolarization modelling results.
We also used the MOJAVE catalogue \citep{Lister09}, with milliarcsecond imaging, in order to investigate, if possible, the source morphologies and resolved polarization information. When available, other information from the NED database and the literature are reported within the text.\\
\newline
{\bf{- Source 0239--0234 }}\\
\noindent 
We fit this spectrum with four synchrotron components, with the oldest component peaking at a few MHz, and the other three peaking at $\sim$1 GHz, $\sim$4 GHz and $\sim$10 GHz (see Fig.\ \ref{sed1}). The polarization percentage decreases from 6\% at X band to 3\% at C band, dropping down to $\sim$1\% at L band. The source shows a deviation from a linear $\chi({\lambda^2}$) relation  in the determination of the RM value within the whole frequency range (C, X and L bands, see Fig.\ \ref{pol3}). The broadband depolarization model fits the data very well within the C and X bands frequency range finding two Faraday components with random magnetic fields (see Fig.\ \ref{0239-0234CXbandmodel} and Tab.\ \ref{Parammodelling4sources}), with the first Faraday component having a more turbulent magnetic field ($\sigma_{RM1CX}$ of about 500 rad/m$^2$) than the second.

These results suggest the presence of at least two emission components at high frequency that synchrotron emit and Faraday rotate at the same time in the presence of turbulent magnetic field. Therefore, the two synchrotron components that are dominating the radio spectrum at C and X bands, could be those that Faraday depolarize at that range of frequency.
In the previous single dish study, \citetalias{Pasetto2016} measured a RM lower that 500 rad/$m^2$ for this source; in this new work we decided to include this source in order to test whether the broadband spectropolarimetry could reveal Faraday structures. Indeed, the Stokes qu-fitting reveals a much more complex polarized structure.
\\
\newline
{\bf{- Source 0243--0550 }}\\
The SED of this source was fit with four synchrotron components: one peaking at a low frequency ($<$74 MHz) and the other three peaking at higher frequencies ($\sim$1 GHz, $\sim$3 GHz and $\sim$10 GHz; see Fig.\ \ref{sed1}). Overall, the radio spectrum at high frequency is flat. The fractional polarization increases from X to C bands but it drops at L band (Fig.\ \ref{pol3}). This suggests some repolarization mechanism in the X and C bands frequency range. Physical repolarization can be explained with an increase of the ordering of the magnetic field \citep{Sokoloff98}, by partial coverage of the emission by a rotating and depolarizing layer \citep{Mantovani09} and/or through a helical magnetic field with a tangled component \citep{Homan02}.

We fit the C-X bands global trend of the fractional polarization with two Faraday components with random magnetic fields (see Fig.\ \ref{0243-0550CXbandmodel} and Tab.\ \ref{Parammodelling4sources}). Our modelling suggests one component with RM$_{1CX}$ $\simeq$ 610 rad/m$^2$ and with a regular magnetic field, and a second component with RM$_{2CX}$ $\simeq$ 1630 rad/m$^2$ with a more disordered magnetic field ($\sigma_{RM2CX}$$\simeq$520 rad/m$^2$). This suggests an ordering of the magnetic field in the environment towards lower frequency that seems to be consistent with the observed repolarization (Fig.\ \ref{pol3}). 
The repolarization visible in the fractional polarization vs $\lambda^2$ panel, could also be due to ordering of the magnetic field within the source. However, we cannot be sure, since the emission region is unresolved. We can also notice in the polarization plot (Fig.\ \ref{pol3}) and in the depolarization modelling plot (Fig.\ \ref{0243-0550CXbandmodel}) that the data points (the $\textit{q}$ and $\textit{u}$ values, $\textit{p}$ and $\chi$) at short wavelength, exhibit a sharp turnover in a very small $\lambda^2$ interval. This could be an indication of the presence of another Faraday component closer to the central region than those already fitted by the model.

The model predicts the source to show polarized signal at 1.4 GHz (with a fractional polarization of $\sim$ 2\%) but it is not. Further depolarization occurs at longer wavelengths. Indeed, the polarization data available at L band were modelled with two simple Faraday components having two high RM values (see Fig.\ \ref{0243-0550Lbandmodel} and Tab.\ \ref{Parammodelling4sources}). The low frequency data were not in agreement with the depolarization modelling performed at C-X band, suggesting no connection between the X-C band and L band data points. We had to treat the two intervals separately. This suggests that the Faraday components revealed by the Stokes q-u fitting are mapping different regions of the targets: structures close to the central engine at C ancd X band and regions more distant from the central engine at L band.
\\
\newline
{\bf{- Source 0751+2716 -}}\\
This source was fit with a synchrotron component with a break (see Fig.\ \ref{sed1}). The fractional polarization decreases from 15\% to a few \% from X band to C band (Fig.\ \ref{pol1}). The RM at X band is quite high ($\simeq$500 rad/m$^2$), and at C band there seems to be no rotation of the polarization angle, except for a narrow frequency range, in which a dramatic change of the polarization angle occurs and the fractional polarization reaches a minimum (see Fig.\ \ref{pol1}). 

The depolarization modelling suggests that there are at least three different rotating components with random magnetic field intervening the synchrotron radiation.(see Fig.\ \ref{0751+2716model} and Tab.\ \ref{Parammodelling}).\\
\newline
{\bf{- Source 0845+0439 -}}\\
The radio SED of this source was fit by four synchrotron components peaking at $\sim$ 100 MHz, $\sim$ 1 GHz, $\sim$ 4 GHz and $\sim$ 10 GHz (see Fig.\ \ref{sed1}). However, the spectrum is also consistent with a flat spectrum across the whole frequency range. The fractional polarization follows a sinusoidal-like behavior and its polarization angle clearly does not follow a linear $\chi({\lambda^2})$ relation (see Fig.\ \ref{pol1}).

The depolarization model fit reveals the presence of two synchrotron emitting and Faraday rotating regions (Fig.\ \ref{0845+0439model} and Tab.\ \ref{Parammodelling}). We obtain two moderate RM dispersions, $\sigma_{RM}$ ($\sigma_{RM1}$ $\simeq$150 rad/m$^2$ and $\sigma_{RM2}$ $\simeq$ 230 rad/m$^2$) components, with two high values of RM (RM$_1$ $\simeq$ 790 rad/m$^2$ and RM$_2$ $\simeq$ 1600 rad/m$^2$; see Tab.\ref{Parammodelling}). 
In this case, it is possible that the two synchrotron components that contribute to the radio spectrum at high frequency are also those responsible for the Faraday depolarization.\\
\newline
{\bf{- Source 0958+3224 -}}\\
The spectrum of this source, also known as 3C232, was fit by three synchrotron components: the first peaking at low frequency ($\sim$ 100 MHz) and the other two at $\sim$ 300 MHz and $\sim$ 10 GHz respectively (see Fig.\ \ref{sed1}). The fractional polarization follows a sinusoid-like behavior (see Fig.\ \ref{pol1}). 

The polarization properties are well described by three Faraday components with three Faraday RM gradients (Fig.\ \ref{0958+3224model}) with regular magnetic fields. The values of the RMs are very high (RM$_1$$\simeq$ 3900 rad/m$^2$ RM$_2$$\simeq$ 640 rad/m$^2$ and RM$_3$$\simeq$ 1160 rad/m$^2$, Tab.\ \ref{Parammodelling}) with very high RM gradients respectively ($\Delta RM_{1}$$\simeq$370 rad/m$^2$, $\Delta RM_{2}$$\simeq$810 rad/m$^2$, $\Delta RM_{3}$$\simeq$850 rad/m$^2$) suggesting a very dense magnetized media.
These RM gradients could be similar to those found through high resolution imaging of AGN jets \citep[e.g. ][]{OSullivan2009}

\cite{Veron-Cetty2006} classify this object as Seyfert 1.8 (Sy 1.8). This optical classification would be due because of reddening by an obscuring torus or by low ionization of the medium. In the first case, the jet orientation would be most likely in the plane of the sky. The kpc scale observations, from low to high frequency (data from literature and the L, C and X band JVLA data), would be tracing different dominant emission regions. Low frequency data from the literature and from our new broadband, L band JVLA data, are most likely dominated by extended, more diffuse emission from the radio galaxy, i.e. the radio emission from the lobes. The high frequency JVLA data, still unresolved at our kpc scale, should be dominated by the central core. This could explain the steep radio spectral shape at low frequency and its flattening towards high frequency. The source, then, could have a large viewing angle with respect the observer and show different radio emission components on different scales. 

Altogether this information suggests that the radio emission at high frequency comes from the central region of the galaxy and it contains at least three Faraday screens with regular magnetic field that smoothly depolarize at C and X bands. 
\\
\newline
{\bf{- Source 1048+0141 -}}\\
The radio spectrum of this object was fit by two synchrotron components peaking at $\sim$ 100 MHz and at $\sim$ 1 GHz respectively (see Fig.\ \ref{sed1}). The fractional polarization decreases following a sinc-like trend and reaches a roughly constant value of 1\% across C band. The polarization angle increases at X band and reaches an approximately constant value at C band (see Fig.\ \ref{pol1}).

The polarization properties are well fitted by two Faraday components with turbulent magnetic fields (Fig.\ \ref{1048+0141model}) with a low RM$_1$ ($\simeq$--20 rad/m$^2$), due to the constant value of the Stokes parameters at C band, and a very high RM$_2$ ($\simeq$ 5100 rad/m$^2$). Note in Tab.\ \ref{Parammodelling} that the $\sigma_{RM1}$ has a large error; this means that the first Faraday component with random magnetic field is in principle not necessary. 
At long wavelengths the $\textit{q}$ and $\textit{u}$ values do not cross each other, resulting in a constant fractional polarization and polarization angle with a low value of RM, suggesting a less dense magnetized medium. However, the model predicts constant polarization of the source at 1.4 GHz, but it is not. Therefore, further depolarization occurs at longer wavelengths.
At short wavelength the parameters cross very frequently. Moreover, the dashed and the straight lines of the qu-fitting vs $\lambda^2$ are showing that at higher frequencies the q and u continue crossing each other, with the result of a possible increase of the RM value towards higher frequency.\\
\newline
{\bf{- Source 1146+5356 -}}\\
This source was fit by four synchrotron components: one at $\sim$ 200 MHz and the other three at higher frequencies ($\sim$ 2 GHz, $\sim$ 4 GHz and $\sim$ 10 GHz respectively; see Fig.\ \ref{sed1}). Overall the radio spectrum could also be considered flat, consistent with a small viewing angle source. 
The polarized properties seem to follow a simple behavior with the fractional polarization decreasing exponentially with wavelength and the polarization angle following a linear $\chi(\lambda^2)$ relation (see Fig.\ \ref{pol1})s.

The depolarization model fit requires two Faraday components characterized by turbulent magnetic fields (Fig.\ \ref{1146+5356model} and Tab.\ \ref{Parammodelling}). Both the RM values are around --500 rad/m$^2$, but with two different values of the RM dispersion ($\sigma_{RM1}$ of 100 rad/m$^2$ and $\sigma_{RM2}$ of 630 rad/m$^2$; see Tab.\ref{Parammodelling}). This suggests that the magnetic field is more ordered within the first Faraday screen compared to the second Faraday screen that is more dominated by random magnetic fields.\\
\newline
{\bf{- Source 1246-0730 -}}\\
This source was fit by a synchrotron component at low frequency (peaking at frequencies smaller that 100 MHz) and three components peaking at $\sim$ 500 MHz, $\sim$ 3 GHz and $\sim$ 10 GHz respectively (see Fig.\ \ref{sed1}). The spectrum looks flat at higher frequency. The fractional polarization decreases until roughly 6 GHz and then it increases, or repolarizes (Fig.\ \ref{pol3}). The polarization angle follows roughly a linear $\chi(\lambda^2)$ relation among C and X bands but it stays constant at L band.

The polarization properties in the C-X bands range are well fit by two Faraday components with two gradients of RM (Fig.\ \ref{1246-0730CXbandmodel} and Tab.\ \ref{Parammodelling4sources}) one of which has a very high value with $\Delta RM_{1CX}$$\simeq$--1300 rad/m$^2$. This medium could be a large layer within which the radiation is subject to a smooth and large change of the polarization angle rising to a very high value of $\Delta RM$.

The source is monitored in the MOJAVE program \citep{Lister09} and shows a very high apparent velocity with $\beta$$_{app}$= 22c consistent with an abject with a small viewing angle. The MOJAVE polarization image shows that the polarized flux density is located within the central region and the value of the RM measured in that map is in agreement with the previous single dish measurement with RM $\approx$ 700$\pm$150 rad/m$^2$ \citep{Hovatta2012} \citepalias{Pasetto2016}. However, this new modelling reveals the presence at C and X band of at least two Faraday components that depolarize and that can be associated to the two synchrotron components that are equally contributing to the radio spectrum.

At L band the source has been modelled with two simple Faraday components the small depolarization of which came only from the presence of a simple external magneto-ionic material (Fig.\ \ref{1246-0730Lbandmodel}). The two RM values detected at low frequency are not large ($|$RM$_{1L,2L}$$|$$\sim$ 20 rad/m$^2$) as those detected at C and X bands. This is in agreement with the fact that the source shows large fractional polarization ($\sim$ 20 \%), almost constant polarization angle and no cross of the Stokes parameters Q and U at low frequency.

As for the source 0243--0550, it was not possible to model the whole polarized data points, instead we had to treat the low frequency range (L band) and the higher frequency range (C and X band) separately. This suggests again no connection between the Faraday structures revealed by the Stokes q-u fitting in the two frequency ranges. The Faraday components mapped within the C-X bands range are different and closer to the central engine than those revealed by the Stokes q-u fitting at L band.

In this case, considering the radio spectrum and depolarization model fit information at C and X bands and the MOJAVE information, we can argue it is most likely that the depolarization is due to at least two Faraday screens originated in the synchrotron source itself and that are producing a large gradient of RM across the beam. Since the gradient of RM is, in this case, symptomatic of regular magnetic field, it is possible that we are detecting depolarization due to helical magnetic field within the radio jet pointing toward the observer. The low frequency Stokes q-u fitting is mapping two extra layers far from the central engine that are contributing little to the depolarization at that frequency range. \\
\newline
{\bf{- Source 1311+1417 -}}\\
This source was fit by two synchrotron components peaking at $\sim$ 400 MHz and at $\sim$ 2 GHz the second of which shows a break peaking around 8 GHz (see Fig.\ \ref{sed2}). The spectrum is characteristic of a Gigahertz Peaked Spectrum (GPS) source. This kind of source is believed to be an AGN in an early phase of evolution, very compact ($\sim$ 10-1000 pc) and high radio luminosity \citep[L$_{radio}$$\sim$ 10$^{45}$ erg s$^{-1}$, ][]{ODea91}. 
The fractional polarization decreases from  $\sim$ 7$\%$ to $\sim$ 2\% at roughly 7 GHz. At lower frequencies the fractional polarization decreases much more slowly, from $\sim$ 2$\%$ to $\sim$ 1\%. In a similar way the polarization angle increases rapidly until 7 GHz where it reaches an approximately constant value (see Fig.\ \ref{pol2}). 

The fit of the polarized properties find three Faraday components with highly turbulent magnetic fields (Fig.\ \ref{1311+1417model} and Tab.\ \ref{Parammodelling}). The high RM values, one of which is $\approx$ --1100 rad/m$^2$, suggests a very dense magnetized medium. Looking at the $\sigma_{RM}$ values of the qu-fitting, they are all very high compared to their respectively RM values, indicating that the media is not only highly magnetized but also highly turbulent.
As for the source 1048+0141, the q and u values cross at higher frequency with the result of a possible increase of the RM.

\cite{Gugliucci2005} reject this source as a possible GPS target, therefore, we can exclude the young nature of this radio galaxy. We suggest that the source, a complex bended core-jet \citep[from VLBA image at 8 GHz,][]{Gugliucci2005} is surrounded by at least three dense and turbulent Faraday rotating clouds that are pierced by the synchrotron emission from the source.\\
\newline
{\bf{- Source 1312+5548 -}}\\
This source was fit by two synchrotron components: the first peaking at very low frequency ($<$ 100 MHz) and the second at $\sim$ 1 GHz. The latter shows a break at $\sim$ 8 GHz (see Fig.\ \ref{sed2}). The fractional polarization decreases from 7 $\%$ to $\sim$ 0\% exponentially. The polarization angle has large variation within the C and X bands (see Fig.\ \ref{pol2}). 

The depolarization modelling fit finds three Faraday components with very high RM values (RM$_{1}$$\simeq$--200 rad/m$^2$, RM$_{2}$$\simeq$--1600 rad/m$^2$ and RM$_{3}$$\simeq$3400 rad/m$^2$) and highly turbulent fields ($\sigma_{RM1}$$\simeq$--360 rad/m$^2$, $\sigma_{RM2}$$\simeq$--970 rad/m$^2$ and $\sigma_{RM3}$$\simeq$--720 rad/m$^2$)(see Fig.\ \ref{1312+5548model} and Tab.\ \ref{Parammodelling}).
 
A high resolution VLBI image at 5 GHz \citep{Helmboldt2007} reveals a complex source morphology: a two sided radio source with a strongly bent jet. It appears to be a misaligned radio source. This complex morphology is likely related to the complex polarization behavior.\\
\newline
{\bf{- Source 1405+0415 -}}\\

This QSO was fit by a component peaking at very low frequency ($<$ 100 MHz) and one synchrotron component peaking at 1 GHz (see Fig.\ \ref{sed2}). The fractional polarization is approximately constant across the C and X bands with a value around 2\% but it decreases to less that 1\% at L band. The polarization angle from 4 GHz to 7 GHZ remains constant at around 9$\deg$, then it increases up to 12$\deg$ at 10 GHz. At high frequency it decreases, producing a high RM detected with qu-fitting.  

This source could be a case in which an n$\pi$ ambiguity affects the data producing a misleadingly high RM detection or the source could be variable. From single dish observations we determined a very high RM value (thousands of rad/m$^2$) by adding n$\pi$ ambiguity to the polarization angle \citepalias{Pasetto2016}. From the JVLA observations it turns out that this source seems to have mainly low RM value within the whole observational frequency bands (L, C and X bands), although the depolarization modelling identifies a Faraday component with a high RM value. 
Two Faraday components with turbulent magnetic fields fit the data at C and X bands (see Fig.\ \ref{1405+0415CXbandmodel}). Although the second Faraday component seems to be not necessary (the RM$_{2CX}$ and the $\sigma_{RM2CX}$ are very low with high errors), the fit is statistically better when the second components is present (see Tab.\ \ref{Parammodelling4sources}). This fit returns a high value of RM$_{1CX}$ ($\sim$--850 rad/m$^2$) quite turbulent (with a high $\sigma_{RM1CX}$  of $\sim$ 440 rad/m$^2$) and a small values of the RM$_{2CX}$ ($\simeq$--3 rad/m$^2$).
The model also predicts that the source should have a constant fractional polarization at low frequency but it is not. L band polarized data are well fitted by two simple Faraday components the small depolarization of which come from the presence of simple external magneto-ionic medium which only rotate the polarized angle (Fig.\ \ref{1405+0415Lbandmodel} and Tab.\ \ref{Parammodelling4sources}). The two RM values resulted from the Stokes q-u fitting have a small value ($|$RM$_{1L,2L}$$|$$\sim$ 20 rad/m$^2$) consistent with the fact that no cross of the Stokes parameters \textit{q} and \textit{u} occurs within the low frequency band. As for the sources 0243--0550 and 1246-0730, it was not possible to study the depolarization considering the whole observational frequency range. We had to treat the high and the low frequency ranges separately. Once again, this implies that the two frequency ranges (C-X band and L band) are tracing different structures and located far away from the central engine.

It is worth noting that this source is monitored in the MOJAVE program and it shows variability in its total flux density \citep{Lister09}. The polarized flux density could be affected by variability, a characteristic of emission dominated by the central region of a source.\\
\newline
{\bf{- Source 1549+5038 -}}\\
This source was fit by four synchrotron components (peaking at $\sim$ 300 MHz,  $\sim$ 2 GHz, $\sim$ 5 GHz and $\sim$ 11 GHz; see Fig.\ \ref{sed2}). Overall the radio spectrum looks flat. The fractional polarization decreases from 2.5 $\%$ to 0.5\%. The polarization angle and the fractional polarization displays complex behavior across the whole observed band (see Fig.\ \ref{pol2}).  

The source shows complex behavior of the q and u values that are difficult to fit well. However, three Faraday components with turbulent magnetic fields give a reasonable fit (Fig.\ \ref{1549+5038model} and Tab.\ \ref{Parammodelling}). The depolarization modelling returns quite high values of RM and high $\sigma_{RM}$ values, suggesting a dense and turbulent medium. This medium could be associated to some of the synchrotron components that characterize the radio spectrum at high frequency and/or some clumpy regions intercepting the synchrotron radiation.\\
\newline
{\bf{- Source 1616+0459 -}}\\
This source could be fitted with a component peaking at very low frequency ($<$ 100 MHz) and two components at higher frequencies ($\sim$ 2 GHz and $\sim$ 6 GHz respectively; see Fig.\ \ref{sed2}). This source forms part of a group of galaxies at redshift 3.2 \citep{Djorgovski87}. The fractional polarization decreases with wavelength from 3\% to 0.5\%. The polarization angle increases with wavelength following a nearly linear $\chi(\lambda^2)$ relation (see Fig.\ \ref{pol2}). 

The depolarization behavior has been fitted with two Faraday components with turbulent magnetic field (Fig.\ \ref{1616+0459model} and Tab.\ \ref{Parammodelling}). The two Faraday screens have high RM values (around 2100 rad/m$^2$ the first and 500 rad/m$^2$ the second RM screen) and quite high values of the dispersion of the RM $\sigma_{RM}$ ($\approx$ --400 rad/m$^2$ and $\approx$ --300 rad/m$^2$ for the first and second RM dispersion respectively). 

Higher angular resolution VLBI images at 5 and 8 GHz reveal a very compact source with a linear scale of the order of $\sim$ 8 pc/mas \citep{OSullivan2011}
Note that, when corrected for the redshift, the RM values of the first Faraday component is $\approx$4$\times$10$^4$ rad/m$^2$, the highest RM value in our sample. Since the source is part of a group of galaxies, with intergalactic medium surrounding the system, the two Faraday components that depolarize the radiation could be associated to: 1) two different layers in the foreground (depolarization due to a gradient of RM in a foreground screen or external Faraday dispersion/beam depolarization) 2) two internal layers emitting and rotating at the same time (because of the presence of two synchrotron components in its radio spectrum at high frequency).\\
\newline
{\bf{- Source 2245+0324 -}}\\
This source was fit by three synchrotron components (peaking at $\sim$ 1 GHz, $\sim$ 4 GHz and $\sim$ 8 GHz; see Fig.\ \ref{sed2}). However, this is also consistent with a convex shape spectrum indicating a possible GPS nature of the source. The fractional polarization seems to follow a sinc-like trend. At C band the polarization angle remains almost constant, with very low RM value (see Fig.\ \ref{pol2}).

Three Faraday components with turbulent magnetic fields fit the data (Fig.\ \ref{2245+0324model} and Tab.\ \ref{Parammodelling}). Two of the resulting RM values are very high (thousands of rad/m$^2$) with high RM dispersions ($\sigma_{RM}$), suggesting very dense and turbulent magnetized regions.

These three Faraday components can be considered associated with the three synchrotron components of the radio spectrum.
\section{Discussion}
\label{discussion}

We observed in full polarization a sample of AGN at L, C and X bands with the JVLA. The sources were selected from \citetalias{Pasetto2016}, as those sources with no detected polarized flux density or blanked at 1.4 GHz (from the NVSS survey) and with large RM values (with RM$>$ 500 rad/m$^2$). The lack of polarized flux density at longer wavelength could be due to strong depolarization due to high RM. In fact, previous single dish observations revealed a sample of sources with large RM values and strong depolarization \citepalias{Pasetto2016}. However, it was not possible to study in detail the complexity of the polarization data because of the insufficient data coverage. Therefore, we selected the most interesting sources, 14 sources with previous single dish RM values larger than 500 rad/m$^2$, to be observed using the broadband capability available at the JVLA.

The sources of our sample show very complex behaviour of the Stokes parameters. For four sources (0239--0234, 0245--0550, 1246-0730 and 1405+0415) we detected sufficient polarized signal also at L band when splitting the L band spw (64 MHz BW) into smaller BW (e.g, 30 MHz). We fit the total intensity using combinations of synchrotron components, and the polarization data using qu-fitting \citep[as suggested by][]{Farnsworth11, OSullivan2012} with simple depolarization equations. The majority of the radio spectra were fit by several synchrotron components (all but the source 0751+2716 and the source 1311+1417). The fractional Stokes \textit{q} and \textit{u} parameters, the polarized intensity, the fractional polarization and the polarization angle, were fit by combining several Faraday components and considering the high frequency range (C-X band range) and the low frequency range (L band), when available, separately suggesting that the two frequency intervals are tracing different magneto-ionic plasma located differently from the AGN core: medium close to the central engine at C and X bands and structures far away from the central engine at L band.
None of the sources in our sample were fit by one Faraday component, instead all the sources needed at least two Faraday components to describe the complex behaviour in polarization. Moreover, although all the sources are unresolved at the higher JVLA resolution (0.6 arcsec), it seems, from a visual inspection, that there is a correspondence between the number of synchrotron components fitted in the total intensity radio spectrum and the number of Faraday components fitted in the polarization domain in the C-X band frequency range ($\sim$8 sources show a clear and a marginal correspondence while the remaining 6 other sources do not, see Tab.\ \ref{FaradaySynch}). Higher angular resolution observations, performed using VLBI technique, will help us to understand whether this is actually true (to be presented in a forthcoming paper).
We also noticed that the depolarization occurring in the 4--12 GHz range for the majority of the sources of our sample (12 sources) is explained by the equation \ref{fitmath} with only contribution from the $\sigma_{RM}$ required. In this case, the depolarization is mainly related to the presence of turbulent magnetic fields, thus with possible scenarios of IFD or EFD/Bd. The remaining two sources: 0958+3224 and 1246--0730 need the contribution of $\Delta RM$ alone. Therefore, the depolarization is explained by the presence of a regular magnetic field. In this case the possible scenarios are the DFR and the gradient of RM for the internal case ($\Delta RM_{int}$). Interestingly, none of the sources required a combination of $\sigma_{RM}$ and $\Delta RM$ to describe the depolarization bahaviour, e.g. $\Delta RM_{ext}$. 
Some support for the physical interpretation of these type of models already exists. For example, \citetalias{OSullivan2017} found a preference for the intrinsic polarization angle derived from the model-fitting to be aligned with the jet orientation, as often observed in FRI radio galaxies \citep[e.g.][]{SaSa88}.
However, in the study presented in this paper, higher angular resolution observations are required to determine how the true underlying polarization and Faraday rotation distributions compare to the model fit results. We plan to report in the future new VLBI observations for a sub sample of these sources to investigate this issue.

\begin{table*}[h]
\caption{Faraday components vs Synchrotron components correspondence.}
\begin{center}
\begin{tabular}{lcccc}
\hline\hline
Source  		& Faraday	& Polarization 	& Synchrotron & Correspondence\\
Name		& components	& parameter & components& [y-n-a]\\
\hline
0239--0234	&	2	&	$\sigma_{RM}$	&	2&y\\
0243--0550	&	2	&	$\sigma_{RM}$	&	2&y\\
0751+2716 	&	3	&	$\sigma_{RM}$	&	1&n\\
0845+0439 	&	2	&	$\sigma_{RM}$	&	2&y\\
0958+3224 	&	3	&	$\Delta$RM		&2-3&a\\
1048+0141 	&	2	&	$\sigma_{RM}$	&	1&n\\
1146+5356 	&	2	&	$\sigma_{RM}$	&	3&n\\
1246--0730	&	2	&	$\Delta$RM		&	2-3&a\\
1311+1417 	&	3	&	$\sigma_{RM}$	&	1-2&n\\
1312+5548 	&	3	&	$\sigma_{RM}$	&	1&n\\
1405+0415 	&	2	&	$\sigma_{RM}$	&	1&n\\
1549+5038 	&	3	&	$\sigma_{RM}$	&	3&y\\
1616+0459 	&	2	&	$\sigma_{RM}$	&	2&y\\
2245+0324 	&	3	&	$\sigma_{RM}$	&	3&y\\
\hline\hline

\end{tabular}
\end{center}
\label{FaradaySynch}
	\begin{flushleft}
NOTE: \\
In the table -- \textit{y} means there is correspondence between synchrotron components and Faraday components between C and X bands; \textit{a} means there is almost a correspondence between synchrotron components and Faraday components between C and X bands: \textit{n} means there is not correspondence between synchrotron components and Faraday components between C and X bands.\\
A number of 8 sources shows correspondence, clear correspondence (y) and a marginal correspondence (a), between the number of Faraday components used to model the depolarization effects and the synchrotron components used to fit their radio spectrum. The remaining 6 sources do not show correspondence (n).
	\end{flushleft}

\end{table*}%

\begin{figure}[h]
\begin{center}
    \caption{Cumulative plot of the polarization-weighted parameters: RRM (red), $\sigma_{RM,wtd}$ (green) and the $\Delta RM_{wtd}$(blue)}
        \includegraphics[width=0.49\textwidth]{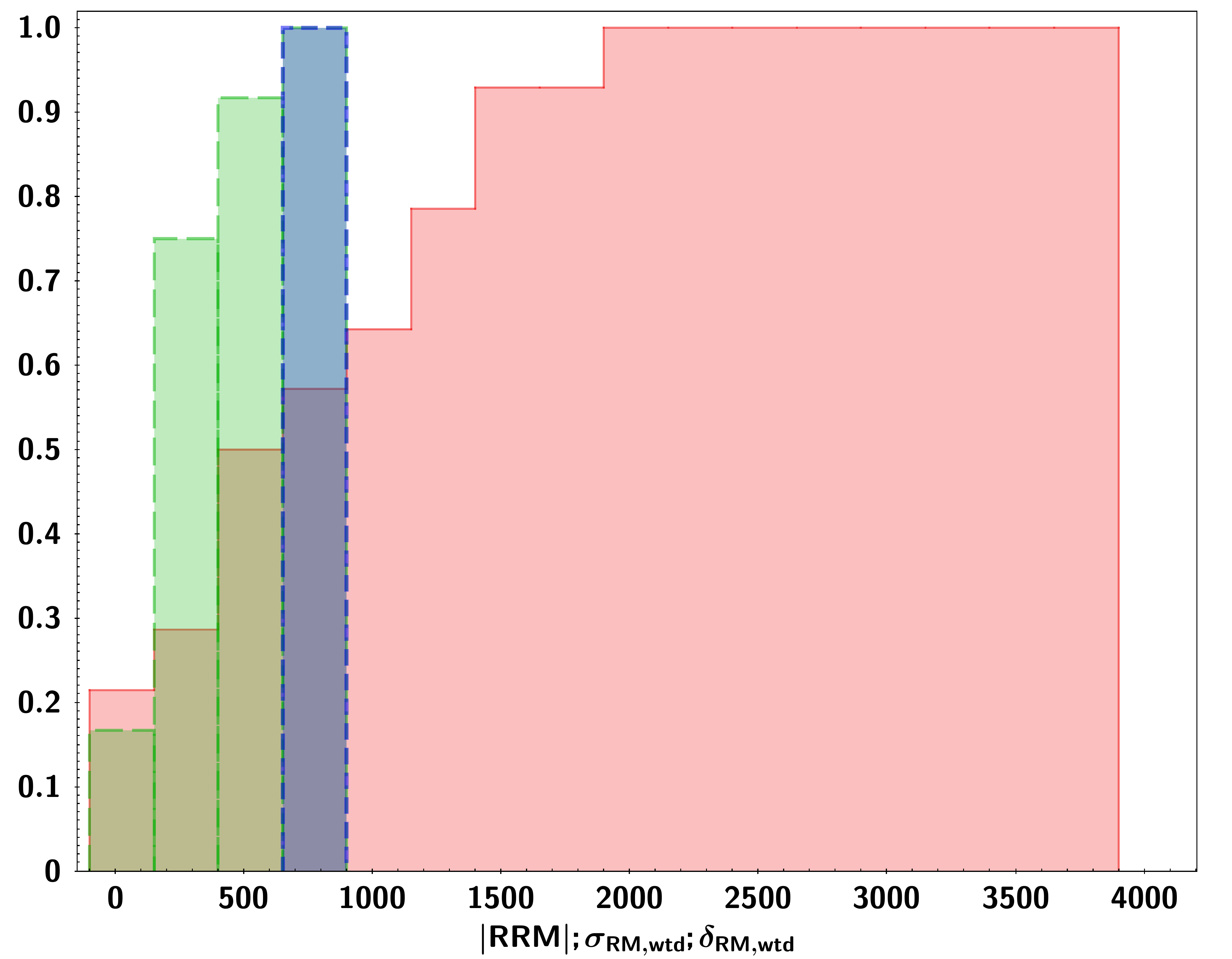}
 \label{cumupativepolparam}
\end{center}
\end{figure}

The initial selection criteria (mainly the lack of polarization at 1.4 GHz in the NVSS survey) results in a sample of sources with very large Faraday rotation parameters (RRM, $\sigma_{RM}$ and $\Delta RM$) within the C and X bands frequency range. The sources can be considered Faraday thick. In only one case (the source 0243-0550), we have found one Faraday component with a $\sigma_{RM2CX}$ value that is close to zero; this could be associated with the repolarization that the source is subject to from X to C band (see Fig.\ \ref{0243-0550CXbandmodel} and Tab.\ \ref{Parammodelling4sources}). 
The median value of the RRM in our sample is 617$\pm$88 rad/m$^{2}$, the median values of $\sigma_{RM,wtd}$ and $\Delta RM_{wtd}$ are 263$\pm$28 rad/m$^{2}$ and 772$\pm$430 rad/m$^{2}$ respectively. Fig.\ \ref{cumupativepolparam} shows the cumulative plot of the polarization-weighted parameters. The red color represents the cumulative distribution of the RRM (for all the sources), the green color represents the cumulative distribution of the $\sigma_{RM,wtd}$ (for 12 sources) and the blue color represents the cumulative distribution of the $\Delta RM_{wtd}$ (for 2 sources).

In order to underline the peculiarity of our sources, we compare our work with that of \citetalias{OSullivan2017}, a broadband polarization (1 to 10 GHz) study of 100 sources selected to be brightly polarized at 1.4 GHz. 
Both studies required only the $\sigma_{RM}$ parameter to describe the depolarization for the majority of sources in the samples, and both required several Faraday components (more than two) to the data \citepalias[only 20\% of the sources in the sample have been fitted with one Faraday component due to relatively low S/N ratio:$<$ 25, ][]{OSullivan2017}.
On the over hand, the median values of the polarized parameters in \citetalias{OSullivan2017} are much lower than ours. In fact, the RRM in our sample is $\sim$ 60 times larger that the RRM in \citetalias{OSullivan2017} and the $\sigma_{RM,wtd}$ and $\Delta RM_{wtd}$ are $\sim$ 18 times and $\sim$ 13 times larger that the values in \citetalias{OSullivan2017}.

It is worth noting that in our sample of unpolarized sources at 1.4 GHz we identified two sources (1246--0730 and 1405+0415) that are blazars, monitored in the MOJAVE program \citep{Lister09}. Blazars are usually highly polarized at low frequency \citep{Marscher1980}. This result implies that blazars polarization properties could also be affected by high RM and strong depolarization due to the presence of a complex medium of interfering Faraday screens.

\begin{figure}[h]
\begin{center}
    \caption{Sketch of the depolarization that occurs in an AGN.}
        \includegraphics[width=0.49\textwidth]{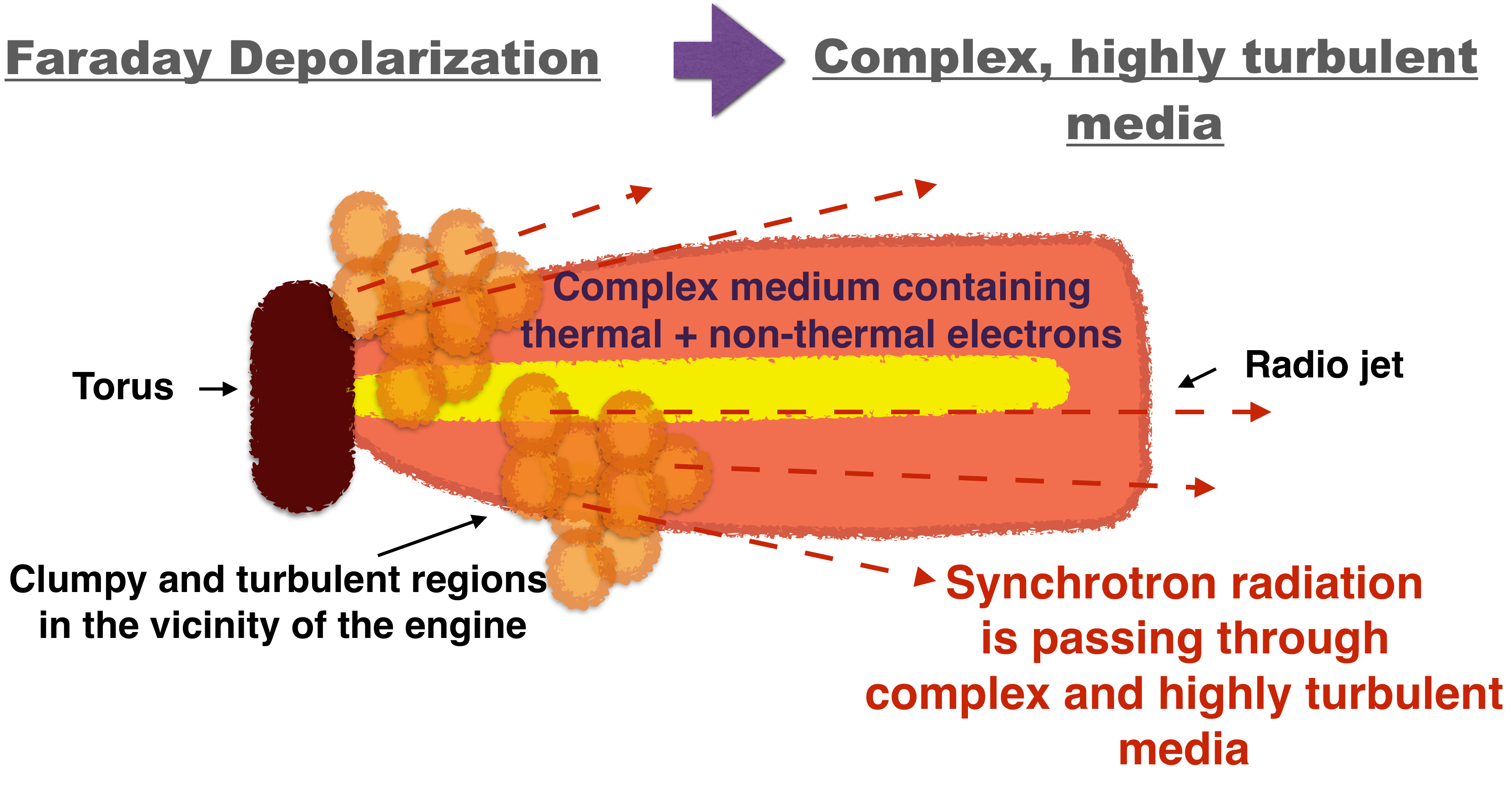}
 \label{plotdepolarization}
\end{center}
\end{figure}

We can conclude that the selection of unpolarized radio sources at low frequency (1.4 GHz in the NVSS) illuminates sources with strong depolarization due to very high RM values, and thus, with complex magneto-ionic media. 
These AGN all seem to be associated with very turbulent magnetic fields. The magneto-ionic media that depolarizes these sources can be internal or external with respect to the synchrotron emitting component but very close to the central engine of the targets. In fact, any contribution from very large structures, e.g. intergalactic medium, the Galactic foreground, that could be important at low frequency, here it can be neglected because of the relatively high radio frequency used for this analysis (C and X bands) and the emission from small scale structures. In fact, we can estimate an upper limit on the linear size of the sources, considering the highest angular resolution reached with these observations (0.6$"$ at B configuration) and the redshift of the targets. The upper limit on the linear size is $\approx$ 5 kpc, much smaller if compared with the polarized sources at 1.4 GHz selected by \citetalias{OSullivan2017}, which have linear sizes of the order of $\approx$ 100 kpc. Therefore, our observations are sensitive of emission coming from components smaller that $\approx$ 5 kpc. As a consequence, the sources in our study are probing more of the dense magneto-ionic medium of the host galaxy, while most of polarized regions of the sources in \citetalias{OSullivan2017} likely extend outside the host galaxy. The magneto-ionic medium that causes the Faraday rotation and depolarization in this sample, can be considered to be local to the source with high electron densities and strong, turbulent magnetic fields. Therefore, most likely it is situated close to the central engine and not far away where the electron density would be much more lower. This is true for the four sources which show also L band polarization data. The depolarization modelling at low frequency reveals structures with low RM values, most likely located far away from the central engine and less dense. Indeed, L band polarized data are sensitive to structure of $\approx$ 30 kpc wide, therefore, structures that could include the far surroundings of an AGN.
The complex media detected at C and X bands can be visualized with highly turbulent clouds in the proximity of the emitting radio source; therefore the radiation coming from the radio jets, the dominant synchrotron emitting components containing non-thermal electrons, could be embedded or pass through very complex and turbulent clouds. Fig.\ \ref{plotdepolarization} shows an example of what can happen in the vicinity of the radio source.  

 \section{Summary and conclusions}
 \label{JVLAsummary}

We have observed at L, C and X bands (1-12 GHz) with the JVLA a sample of AGN which are unpolarized at 1.4 GHz \citep[in the NVSS survey ][]{Condon98} and have high RMs detected in previous single dish observations \citepalias{Pasetto2016}. We have collected total intensity and polarized intensity data among the whole observational frequency range. We analyzed broadband spectro-polarimetric data across the JVLA C and X bands (for four source we have also detected polarized signal at L band). 
 We summarize the overall results as follows:
 
 \begin{itemize}
 
 \item We fit the total intensity radio spectra with several synchrotron components. Overall the sources display a complex spectrum between L and X bands and no strong variability has been detected between previous single dish radio spectra fitting \citepalias{Pasetto2016}.
 
 \item Polarization data, i.e., the Stokes Q and U, the polarization flux density (S$_{Pol}$), the fractional polarization ($\textit{p}$) and the polarization angle ($\chi$), show complex behaviour in the 4 to 12 GHz range. In particular, the $\textit{p}$ data do not show a simple exponential decay, as expected from the simplest depolarization case, i.e., equation \ref{simple}, and $\chi$ deviates significantly from a linear trend.

 \item Depolarization modelling in the C and X band range has been performed for all the targets. We used several Faraday components, at least two Faraday screens, to represent the complexity of the sources. We cannot assign a single RM to the sources, but several Faraday screens contribute to the complexity of the targets. It seems that there is a correspondence between the synchrotron components used to fit the radio spectra and the Faraday components used for the depolarization modelling. However, to verify whether this statement is correct, a higher angular-resolution polarimetric study, presented in a forthcoming paper, is needed.
 
 \item We detected polarization signal for four sources (0239--0234, 0243--0550, 1246--0730, 1405+0415) when splitting the available L band spws  (64 MHz BW) into smaller BW (i.e., 30 MHz BW). The L band data does not follow the trend of the C and X bands data; we could not include in the depolarization modelling of the high frequency data also the low frequency data points. We treted the two frequency intervals (4--12 GHz and 1--2 GHz) separately. This suggests that the low frequency polarized data are tracing different structures with respect to the high frequency data. In fact, the L band data (with a resolution of $\sim 4.5 \arcsec$) are sensitive to structure with sizes of the order of 30 kpc, larger than those detected by the high frequency data sensible to structure with sizes of the order of 5 kpc.

 \item For the majority of our targets (12 sources) the depolarization is caused by turbulent magnetic fields. Possible equations that contain the contribution of turbulent cells of magnetic field, the Faraday dispersion $\sigma_{RM}$, are the internal Faraday dispersion (equation \ref{eqnIFD}), or the External Faraday Depolarization/Beam depolarization (equation \ref{eqnEFD}). Only two sources have been fitted considering the contribution of a regular magnetic field ($\Delta$RM). None of the sources have been fitted using the contribution of both regular and turbulent magnetic fields.
 
 \item Nevertheless, the lack of polarized flux density at 1.4 GHz, results in a sample of sources with very large Faraday rotation parameters (RRM, $\sigma_{RM}$ and $\Delta$RM). The median values of the RRM, $\sigma_{RM}$ and $\Delta$RM are: 617$\pm$88 rad/m$^2$, 263$\pm$28 rad/m$^2$ and 772$\pm$430 rad/m$^2$ respectively. The highest value of RRM detected is 2020$\pm$30 (for the source 1616+0459) which reaches values of $\sim$2$\times$10$^4$ in the rest frame. 

\item Two sources (1246--0730 and 1405+0415) are blazars, monitored in the MOJAVE program \citep{Lister09}. Most likely, blazar type sources can be characterized by high RM and strong depolarization due to a complex intervening medium.

\item We provide upper limits on the linear sizes of the sources of $\approx$ 5 kpc. The sources are thus probing dense magneto-ionic media with high electron density and strong, turbulent magnetic fields, most likely situated close to the central engine.

\end{itemize}

These broadband JVLA data show, without any doubt, the complexity of radio sources both in total intensity and in polarized intensity. Thanks to the high spectral resolution of these data, it has been possible to follow the dramatic changes of the polarization information of these AGN across a wide range of frequency and to model, with very good accuracy, the complexity of the polarization behaviour. The new qu-fitting technique applied to broadband polarization data can be used to map the medium of these radio sources; specifically, it can be use to spectrally resolve polarized components of unresolved radio sources. 

Radio spectro-polarimetric observations are an excellent tool to unveil magnetised structures in radio AGN. Overall, this study paves the way for future broadband and spectro-polarimetric studies with large-area surveys (e.g. VLASS, ASKAP-POSSUM) that will measure the polarization and Faraday rotation properties of hundreds of thousands of radio-loud AGN. 
This will greatly improve the statistical study of the magnetized properties of radio AGN and their environments.

\begin{acknowledgements}
A.P. acknowledges support from CONACyT 238631
G.B. acknowledges financial support under the INTEGRAL ASI-INAF agreement 2013-025.R01.
This research made use of the NASA/IPAC Extragalactic Database (NED), which is operated by the Jet Propulsion Laboratory, California Institute of Technology, under contract with the National Aeronautics and Space Administration.
This research has made use of data from the MOJAVE database that is maintained by the MOJAVE team \citep{Lister09}\\
This work was supported by UNAM-PAPIIT IA101214 and IA102816.\\
We thank Prof. Robert Antonucci for his careful revising of the paper and useful comments.
\end{acknowledgements}

\bibliographystyle{aa} 
\bibliography{paper} 


\clearpage

\begin{appendix} 
\section{Depolarization modelling plots: sources with C-X bands polarization data.}

\begin{figure*}[h]
\begin{center}
    \caption{Depolarization model for the source 0751+2716: 3 components model}
        \includegraphics[width=0.4\textwidth]{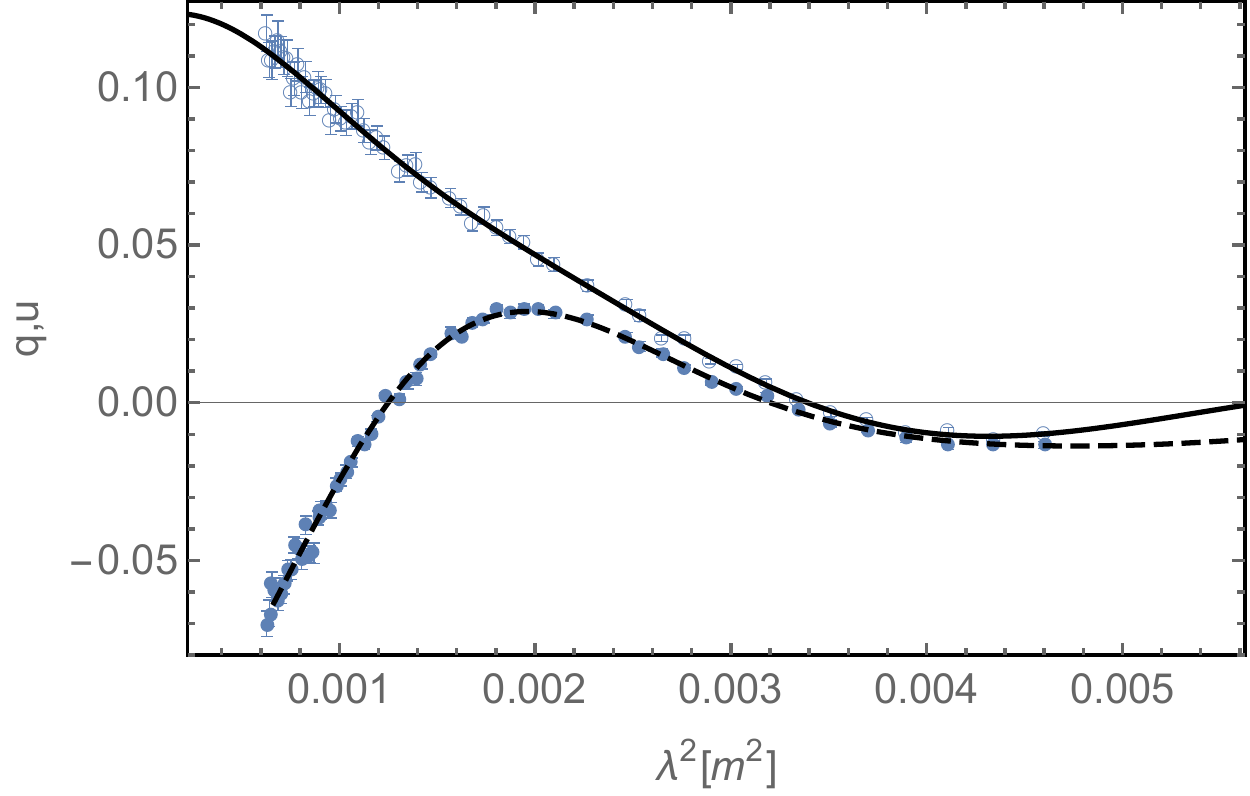}
        \includegraphics[width=0.4\textwidth]{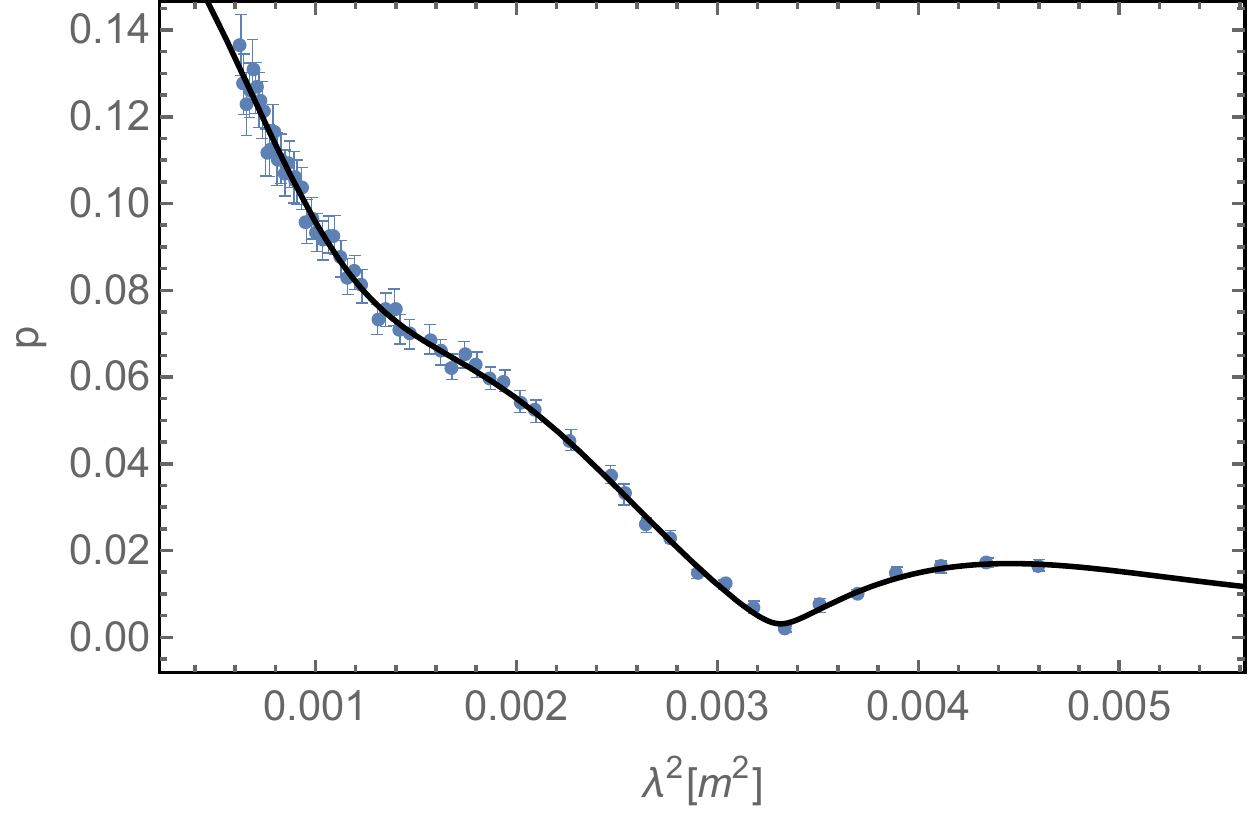}
        \includegraphics[width=0.4\textwidth]{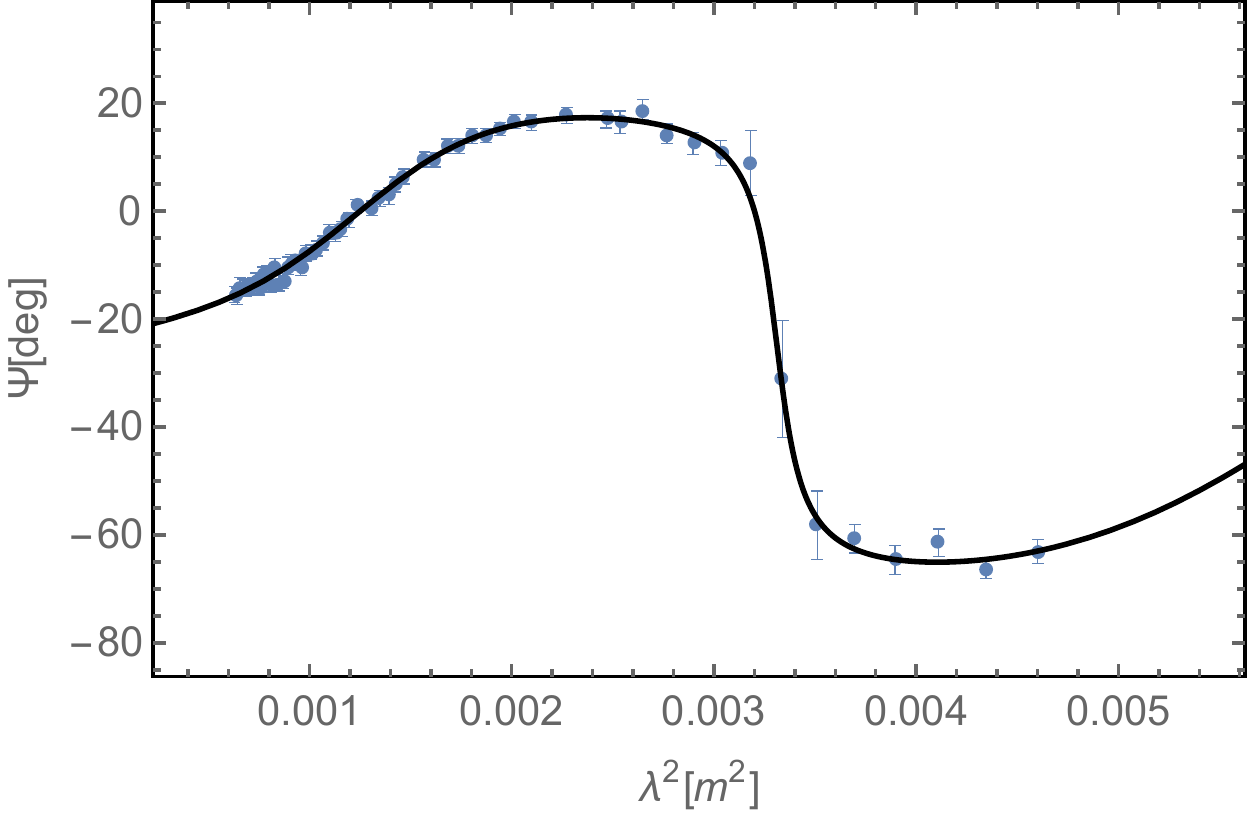}
        \includegraphics[width=0.4\textwidth]{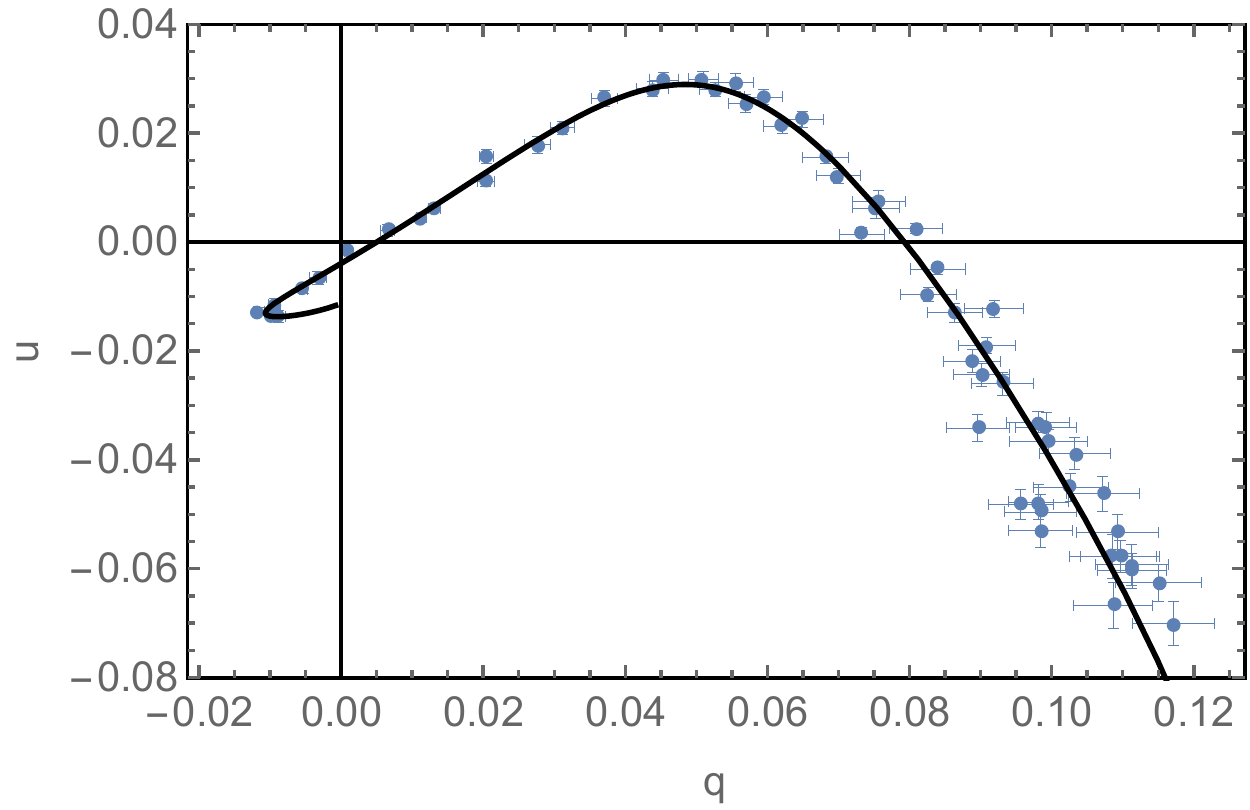}
\label{0751+2716model}
\end{center}
\end{figure*}

\begin{figure*}[h]
\begin{center}
    \caption{Depolarization model for the source 0845+0439: 2 components model}
        \includegraphics[width=0.4\textwidth]{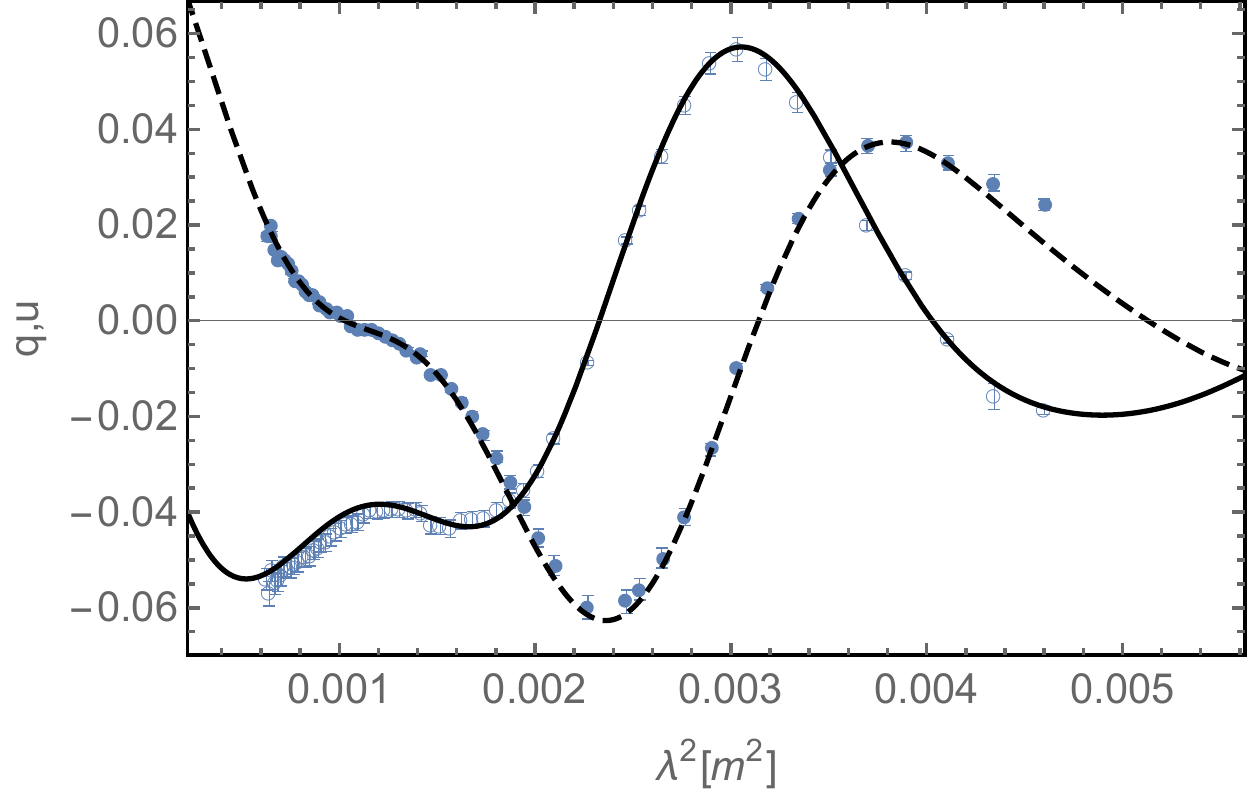}
        \includegraphics[width=0.4\textwidth]{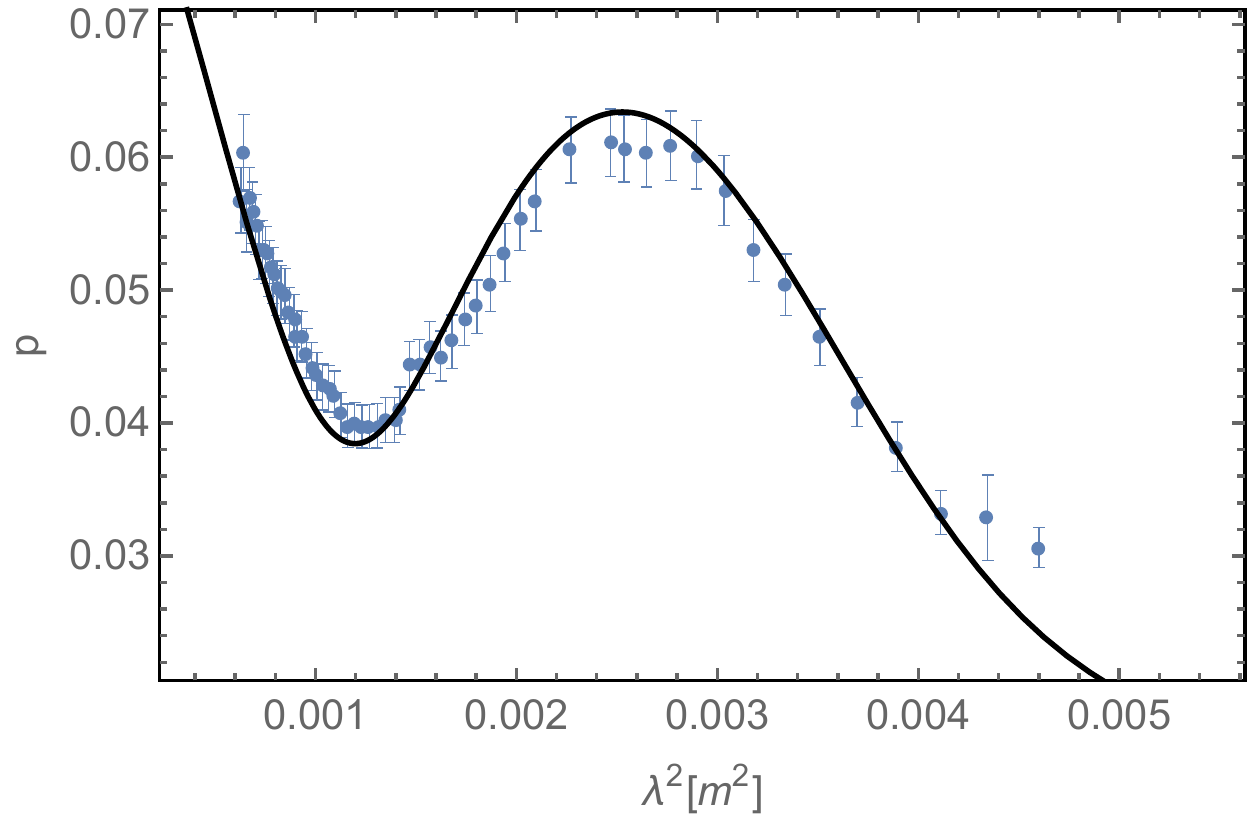}
        \includegraphics[width=0.4\textwidth]{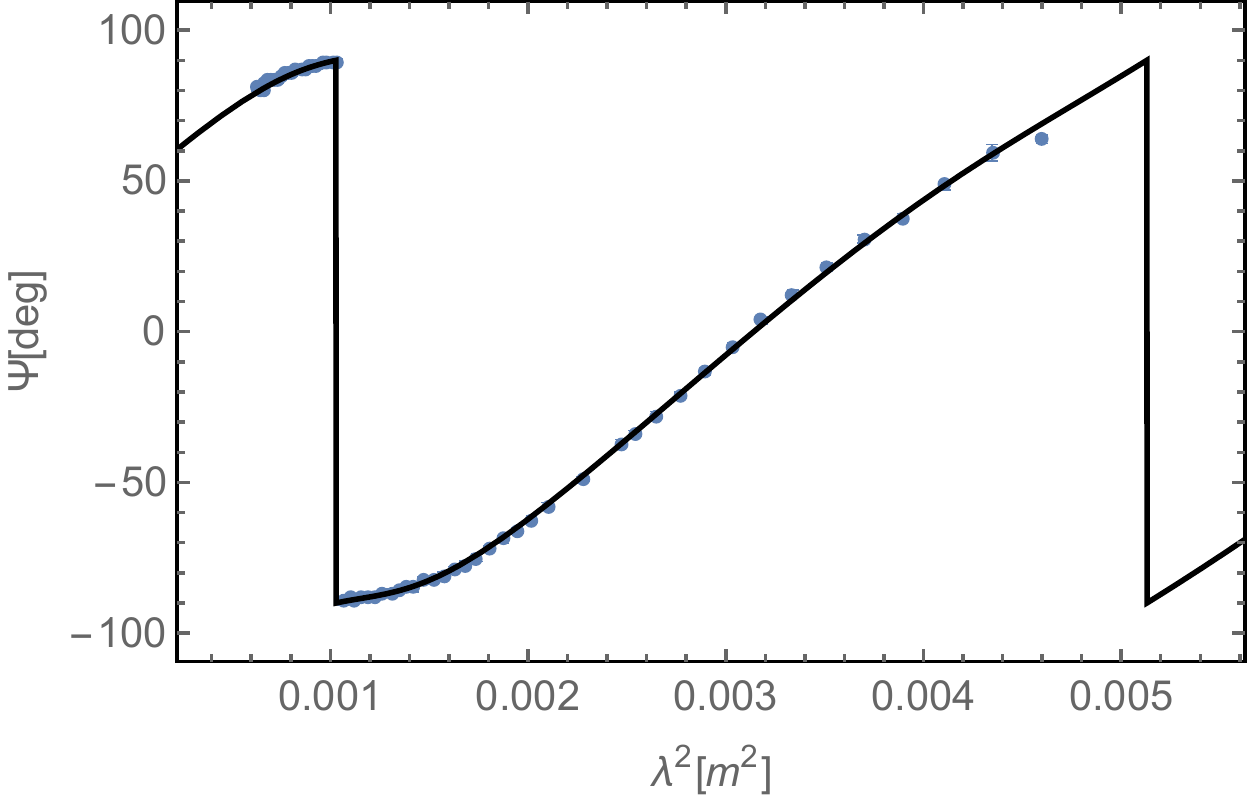}
        \includegraphics[width=0.4\textwidth]{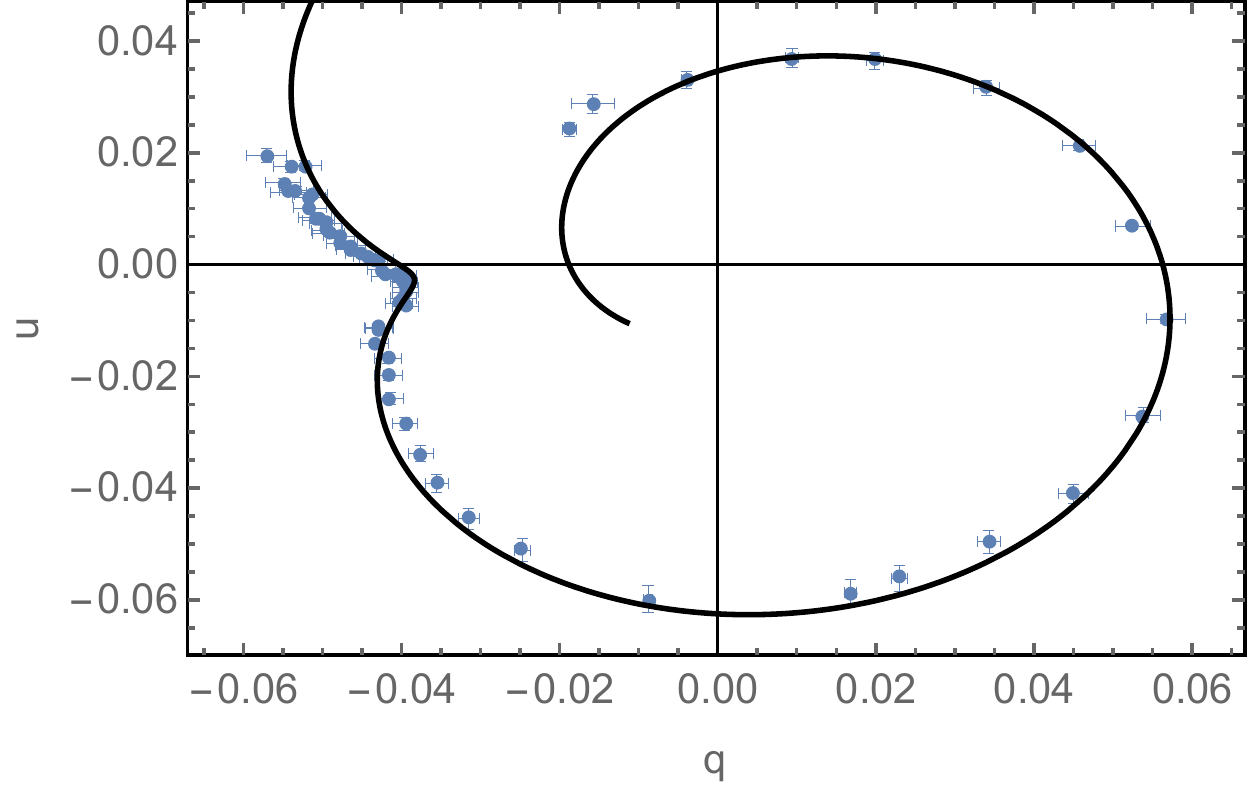}
\label{0845+0439model}
\end{center}
\end{figure*}

\begin{figure*}[h]
\begin{center}
    \caption{Depolarization model for the source 0958+3224: 3 components model}
        \includegraphics[width=0.4\textwidth]{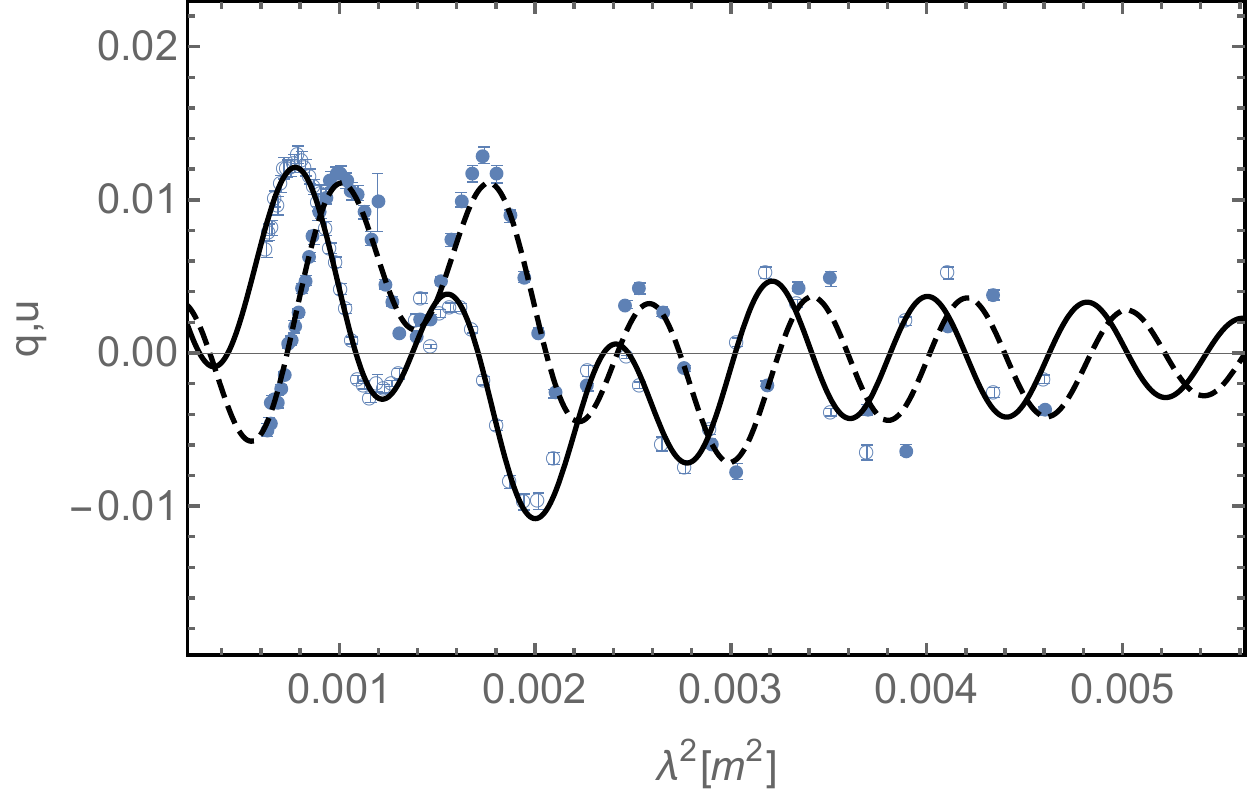}
        \includegraphics[width=0.4\textwidth]{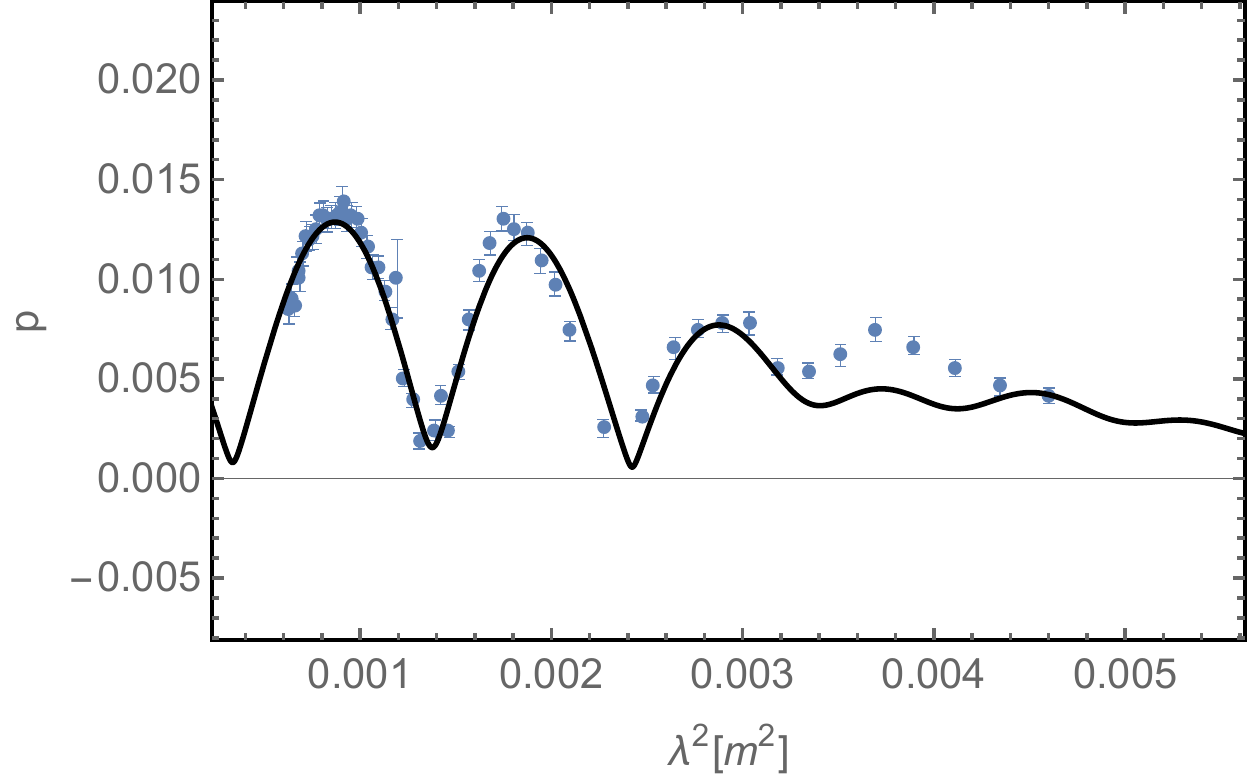}
        \includegraphics[width=0.4\textwidth]{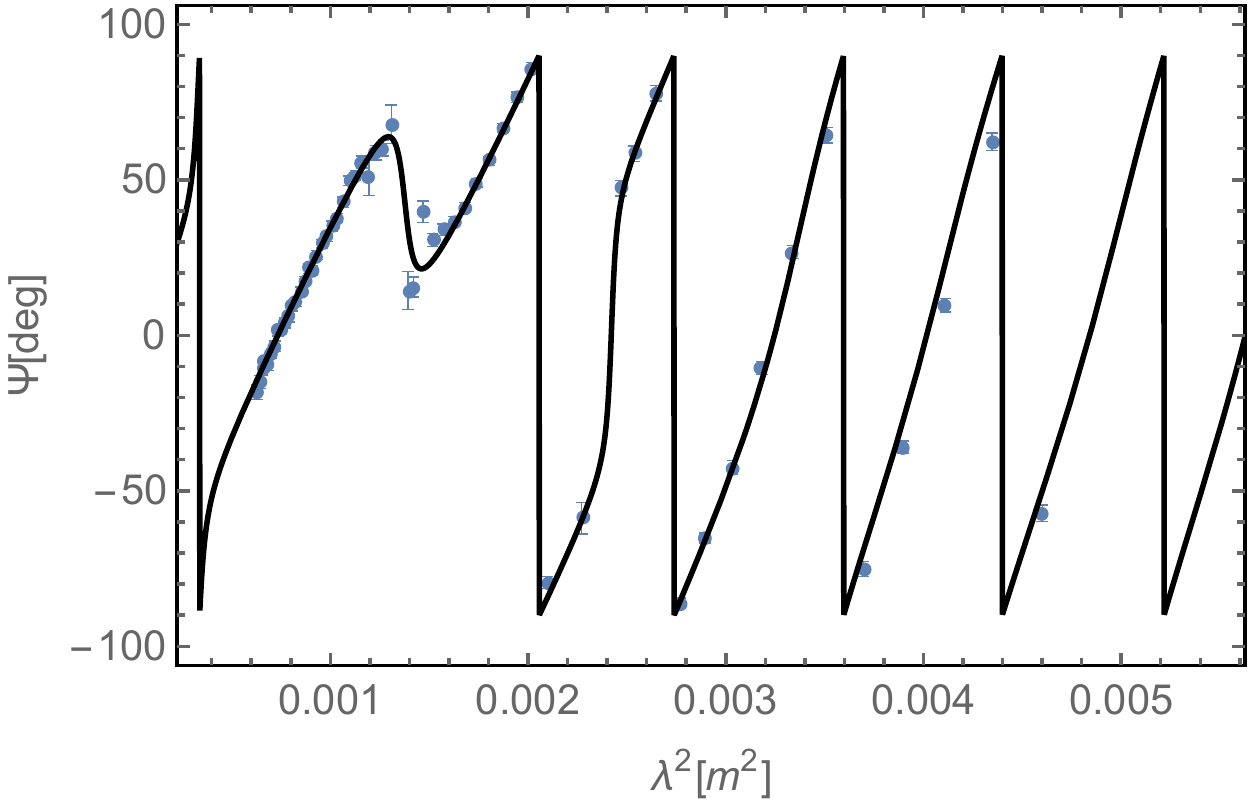}
        \includegraphics[width=0.4\textwidth]{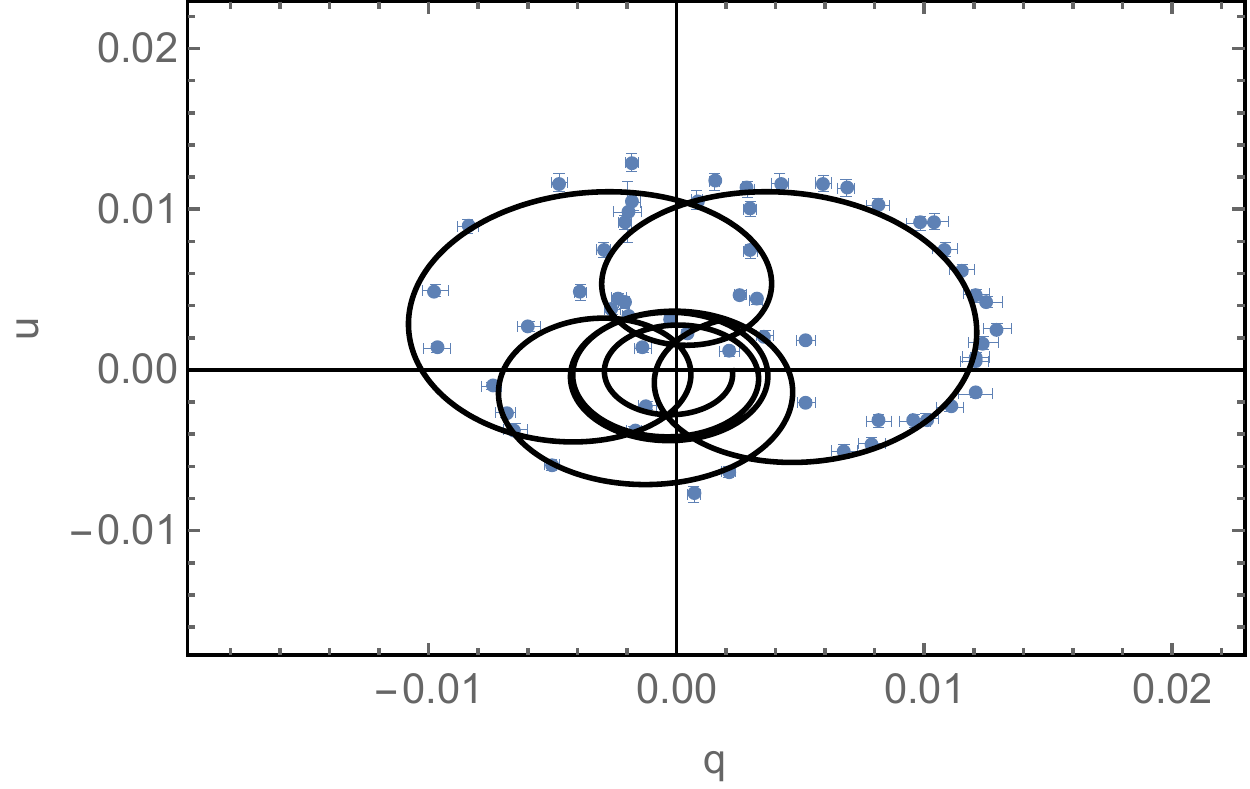}
\label{0958+3224model}
\end{center}
\end{figure*}

\begin{figure*}[h]
\begin{center}
    \caption{Depolarization model for the source 1048+0141: 2 components model}
        \includegraphics[width=0.4\textwidth]{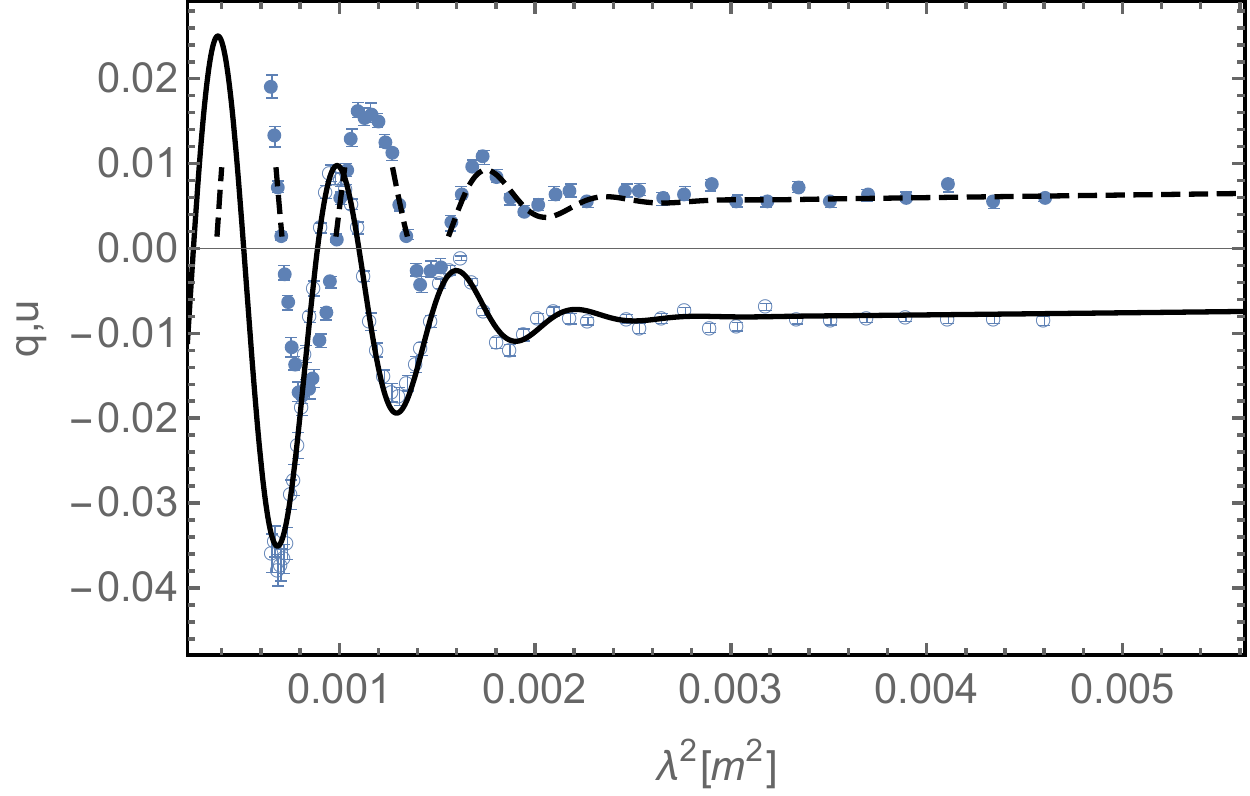}
        \includegraphics[width=0.4\textwidth]{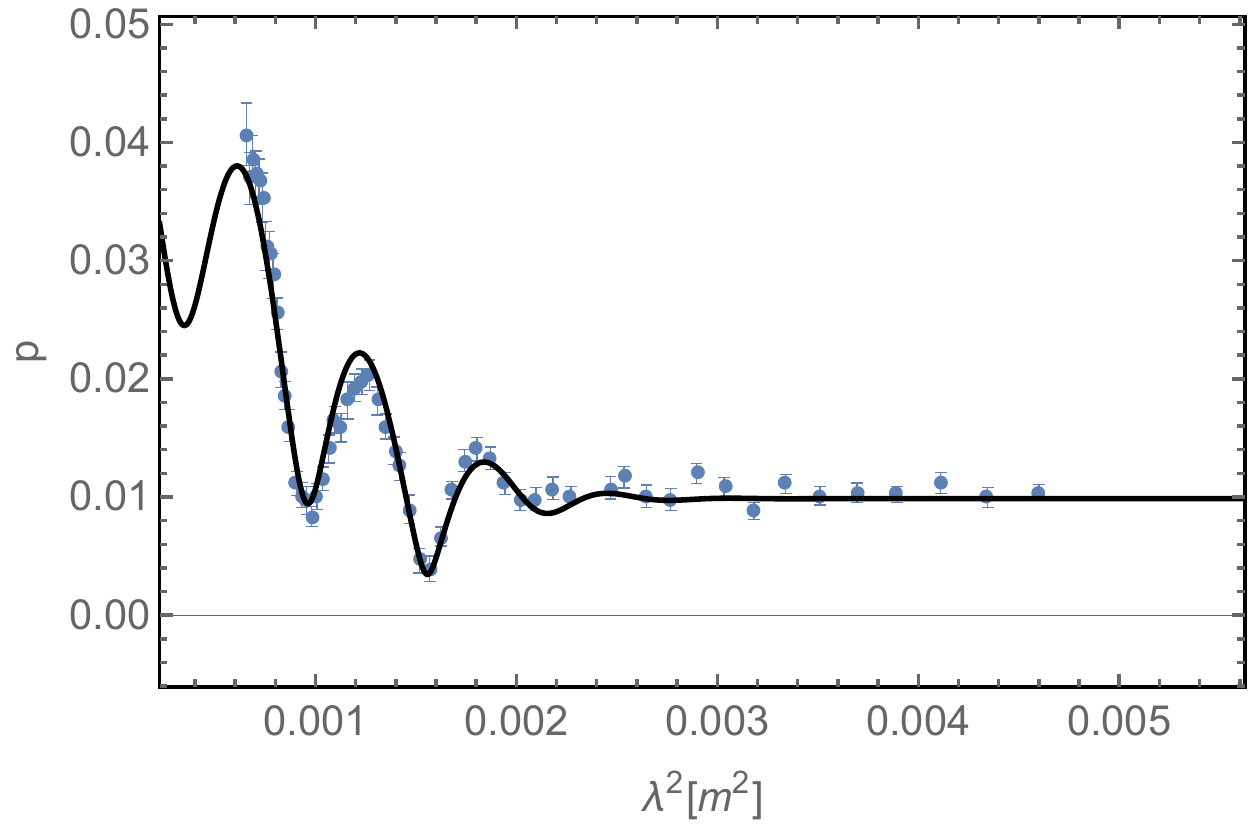}
        \includegraphics[width=0.4\textwidth]{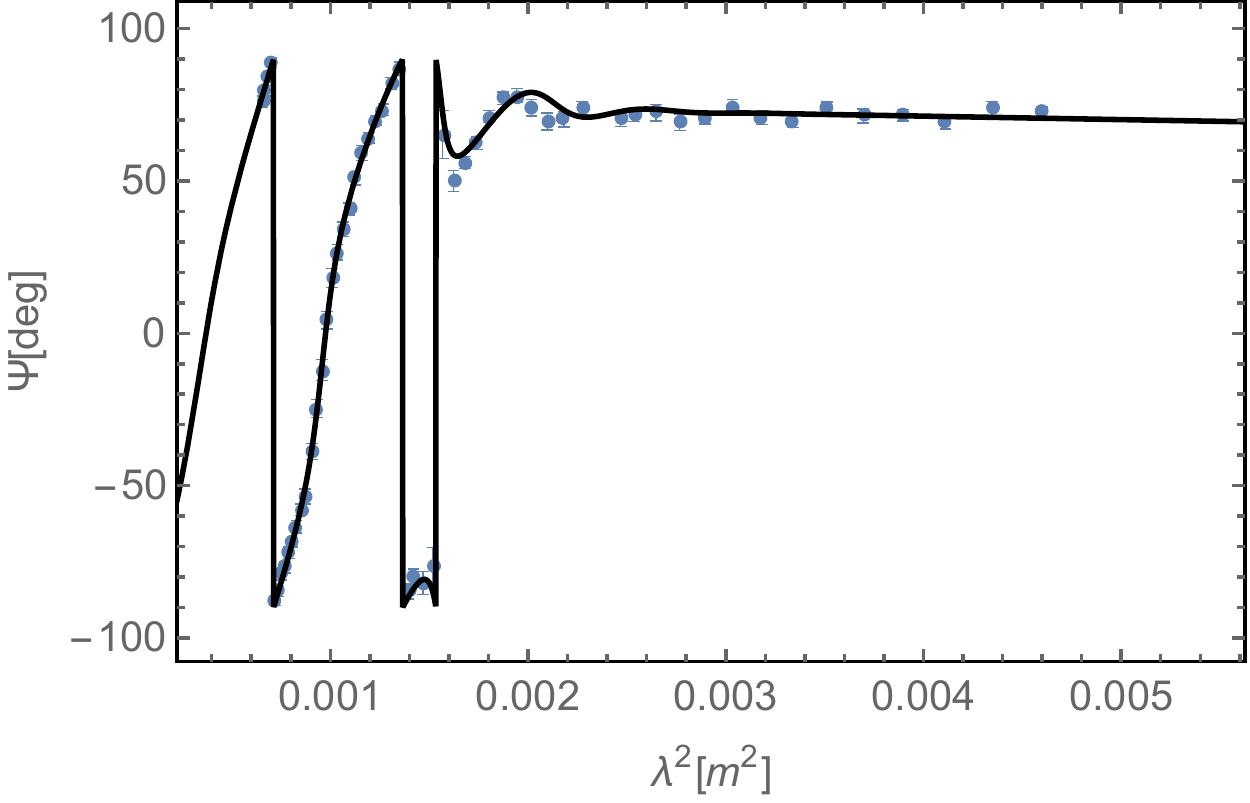}
        \includegraphics[width=0.4\textwidth]{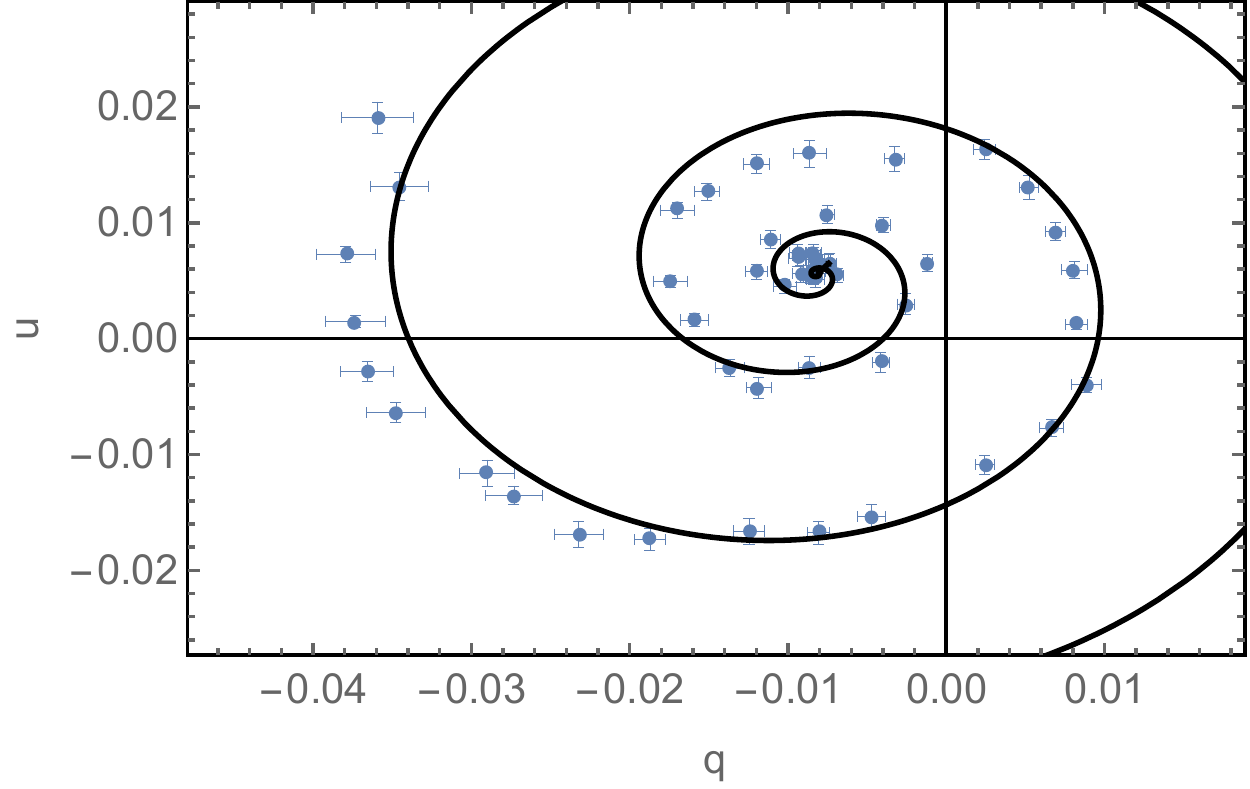}
\label{1048+0141model}

\end{center}
\end{figure*}

\begin{figure*}[h]
\begin{center}
    \caption{Depolarization model for the source 1146+5356: 2 components model}
        \includegraphics[width=0.4\textwidth]{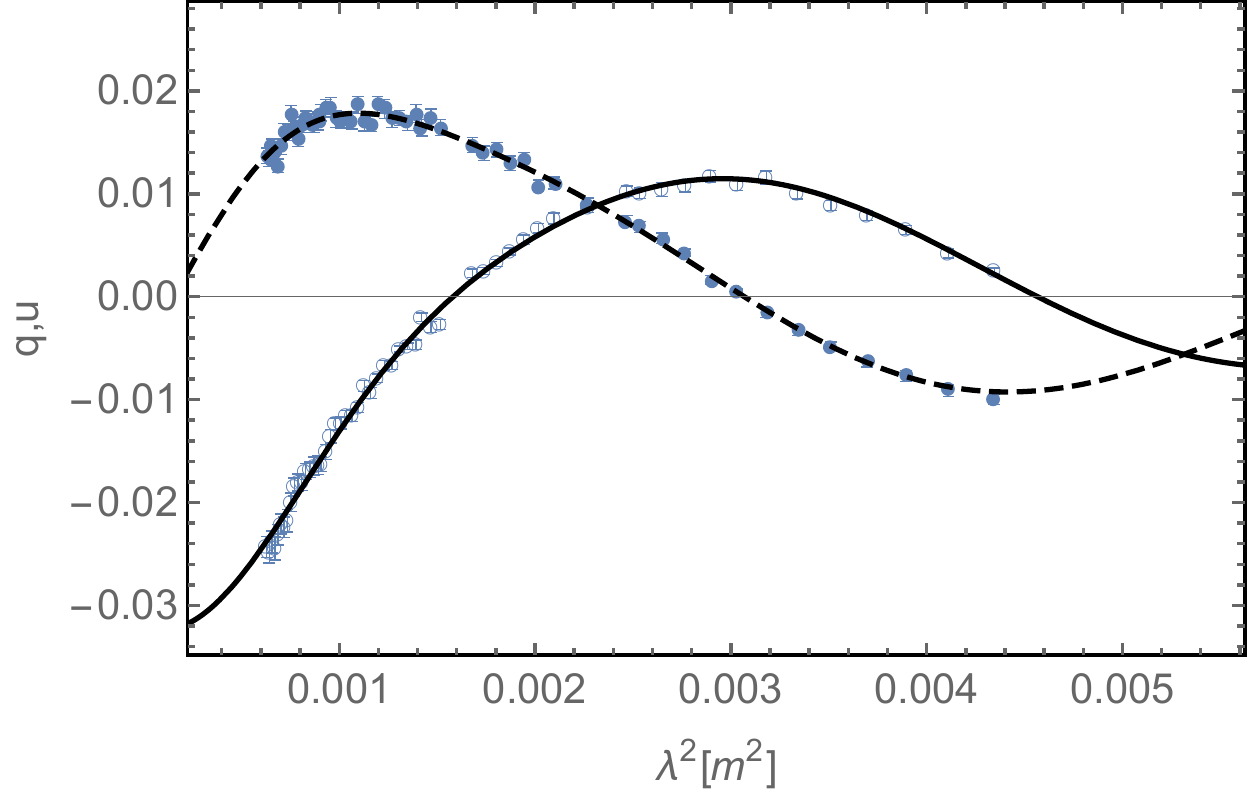}
        \includegraphics[width=0.4\textwidth]{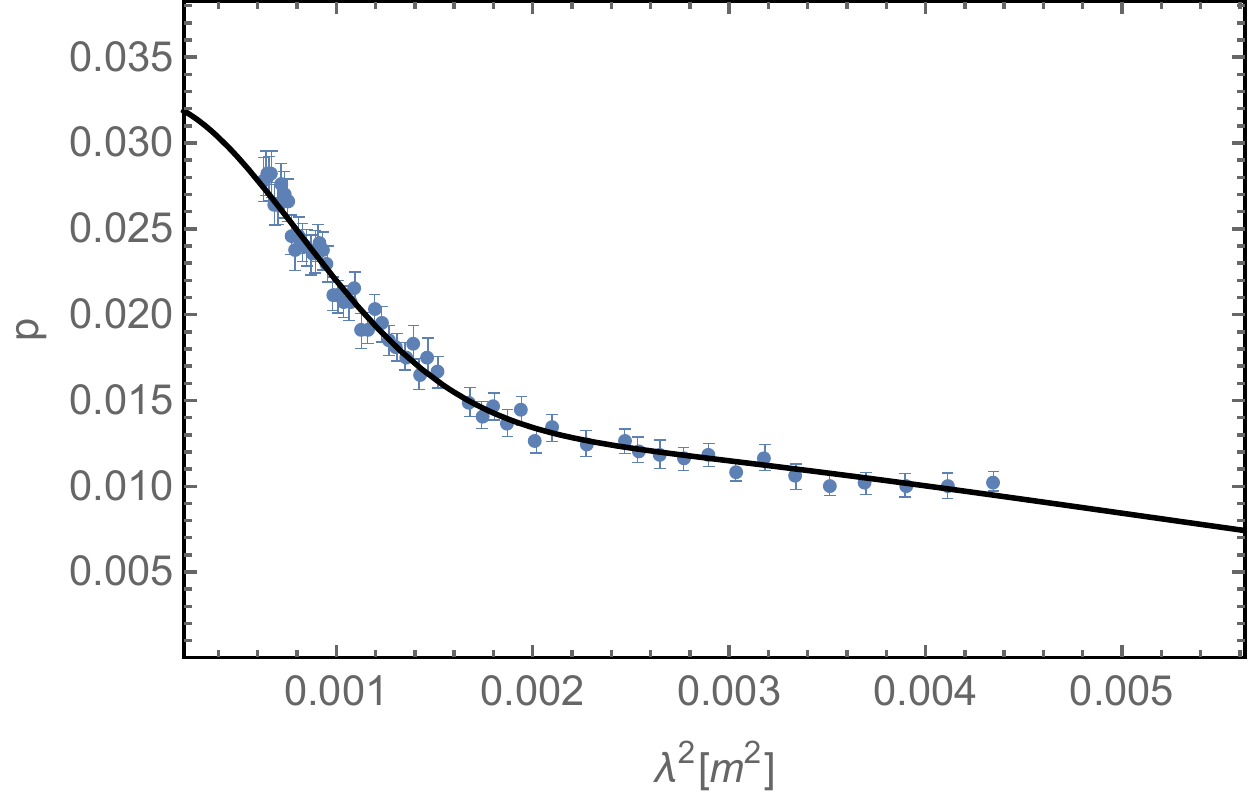}
        \includegraphics[width=0.4\textwidth]{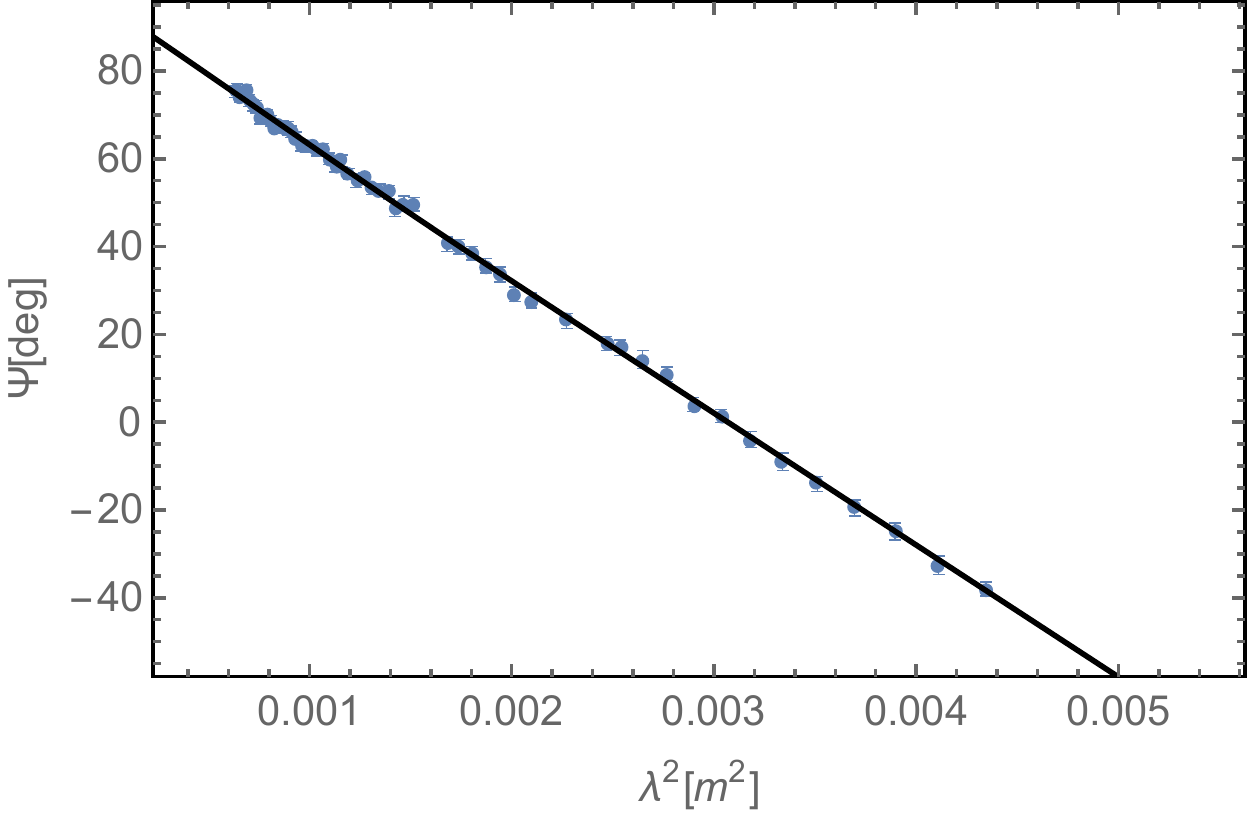}
        \includegraphics[width=0.4\textwidth]{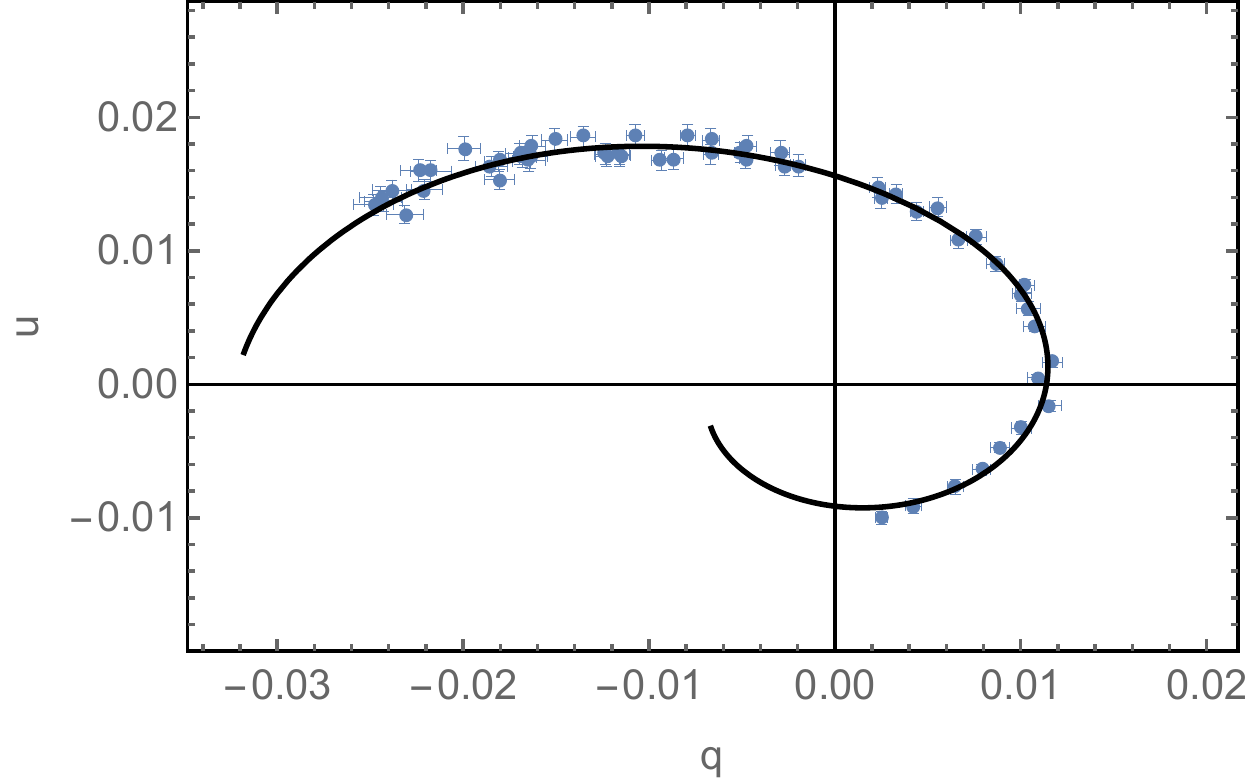}
\label{1146+5356model}
\end{center}
\end{figure*}

\begin{figure*}[h]
\begin{center}
    \caption{Depolarization model for the source 1311+1417: 3 components model}
        \includegraphics[width=0.4\textwidth]{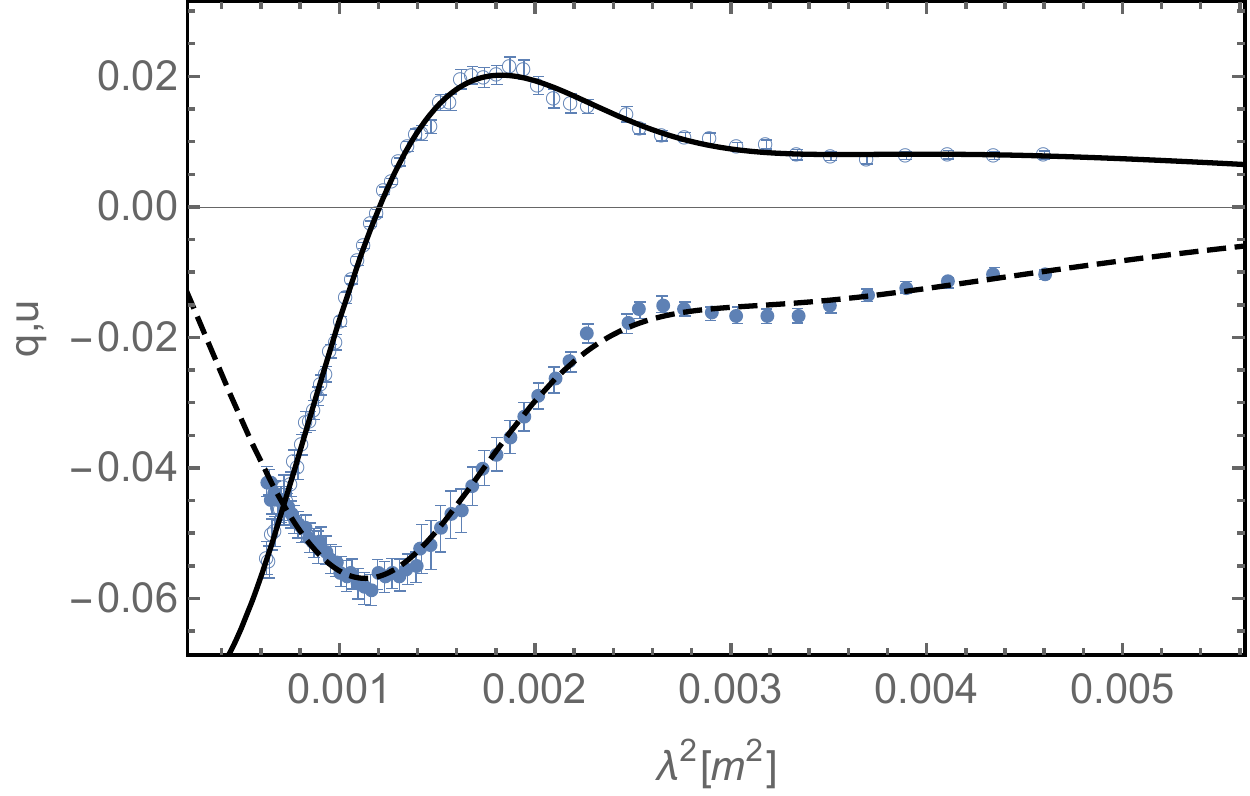}
        \includegraphics[width=0.4\textwidth]{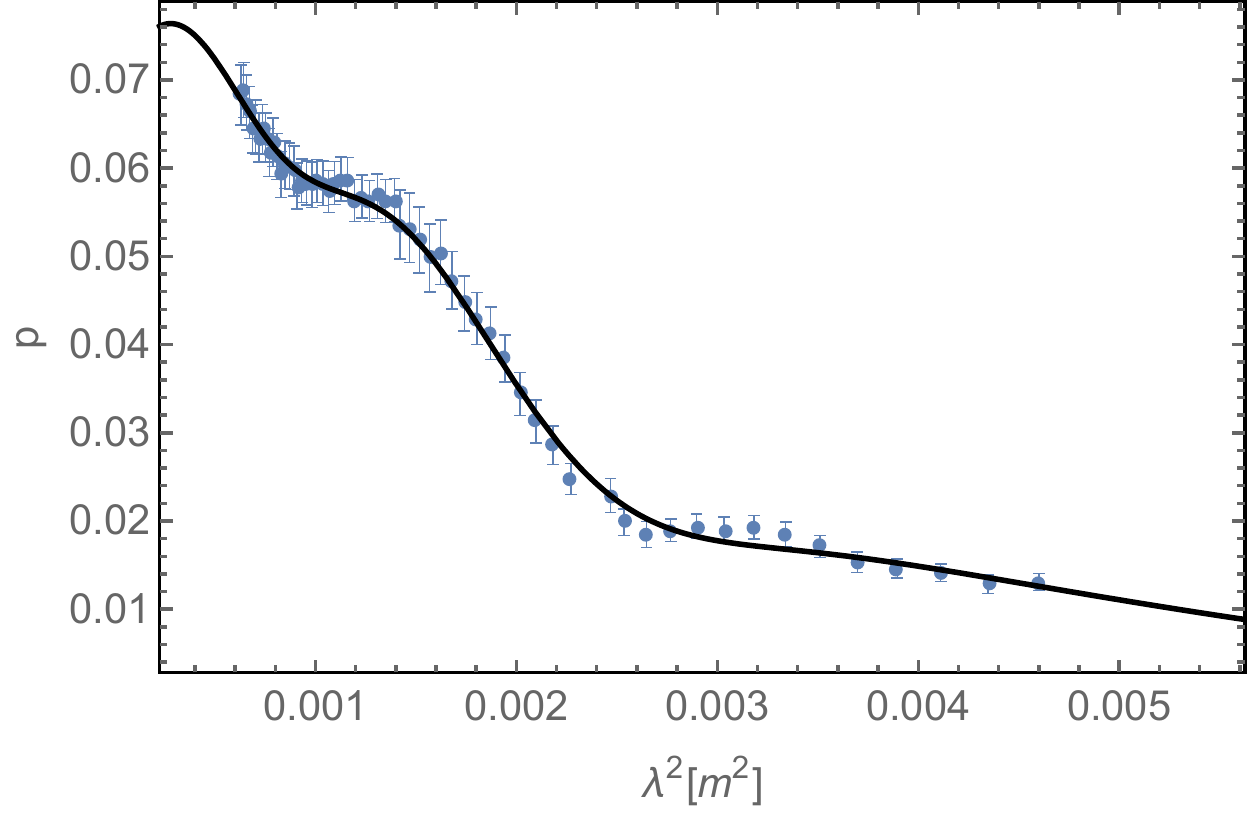}
        \includegraphics[width=0.4\textwidth]{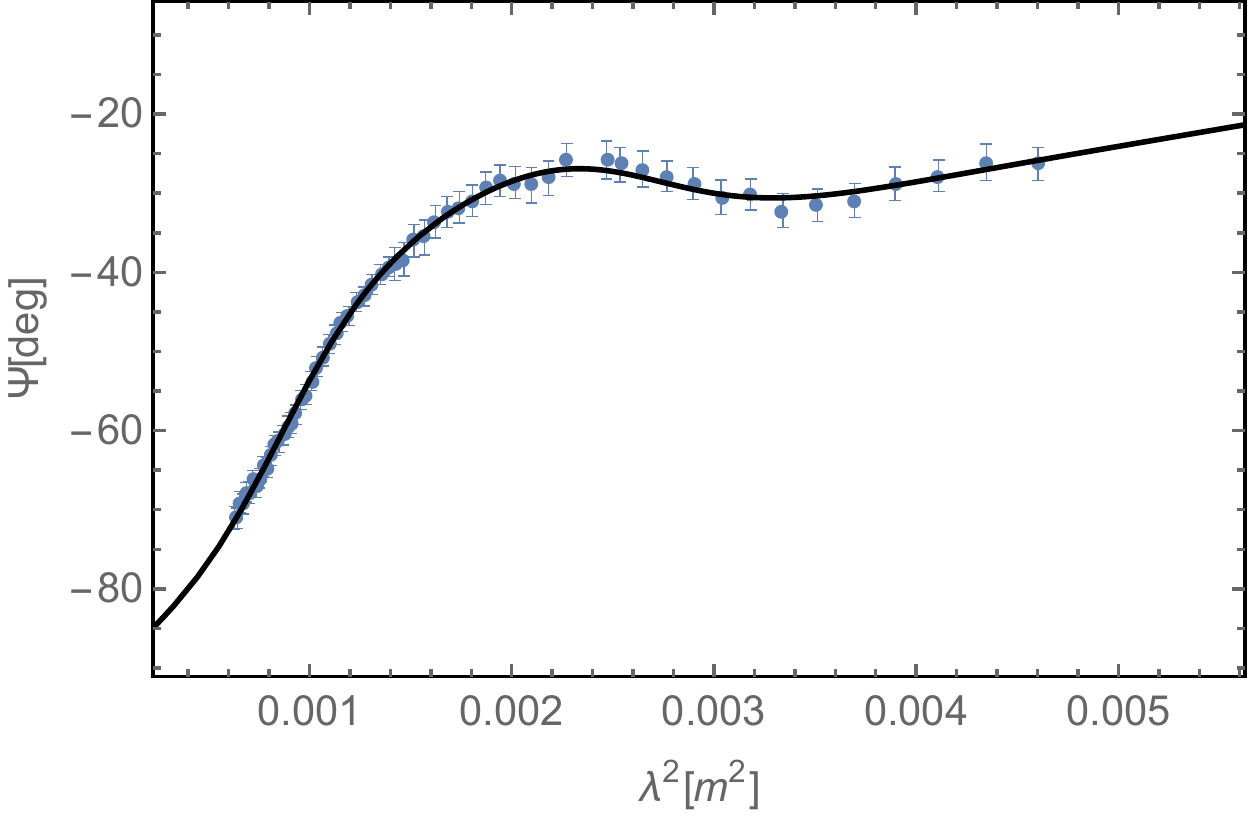}
        \includegraphics[width=0.4\textwidth]{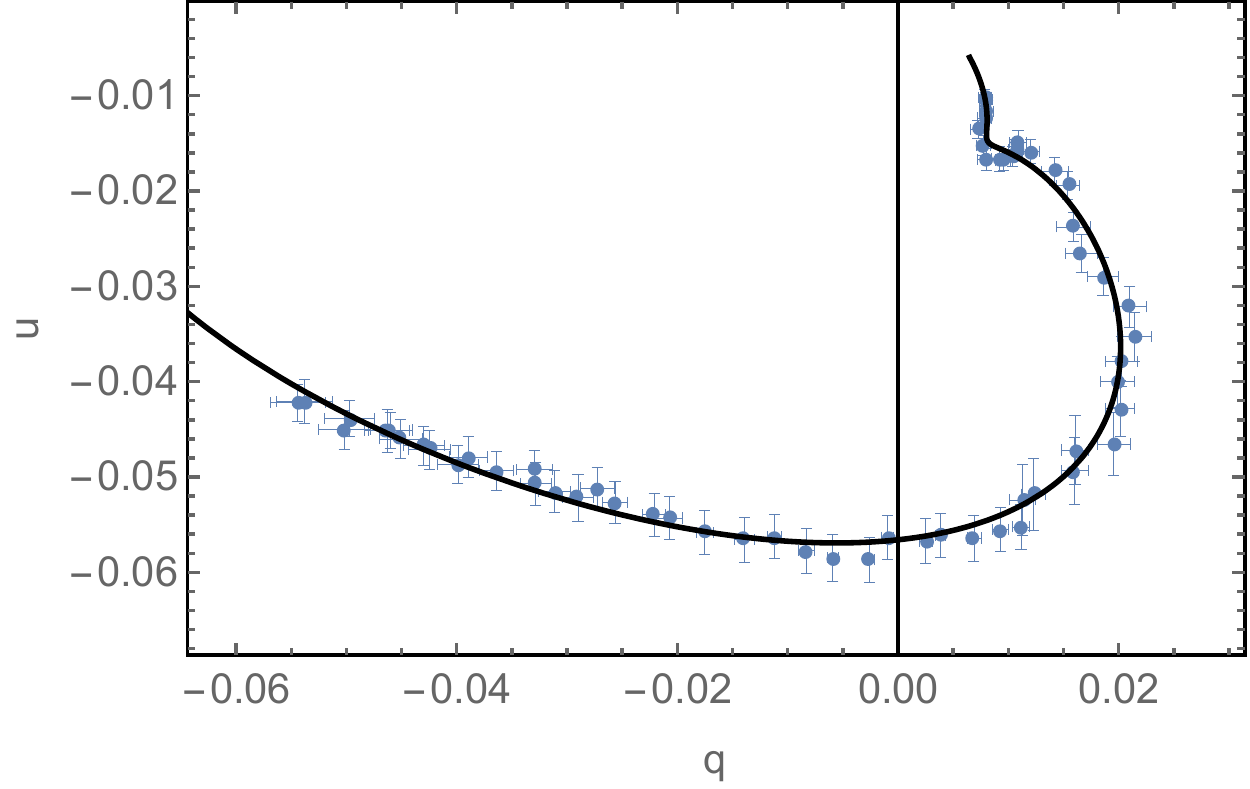}
\label{1311+1417model}
\end{center}
\end{figure*}

\begin{figure*}[h]
\begin{center}
    \caption{Depolarization model for the source 1312+5548: 3 components model}
        \includegraphics[width=0.4\textwidth]{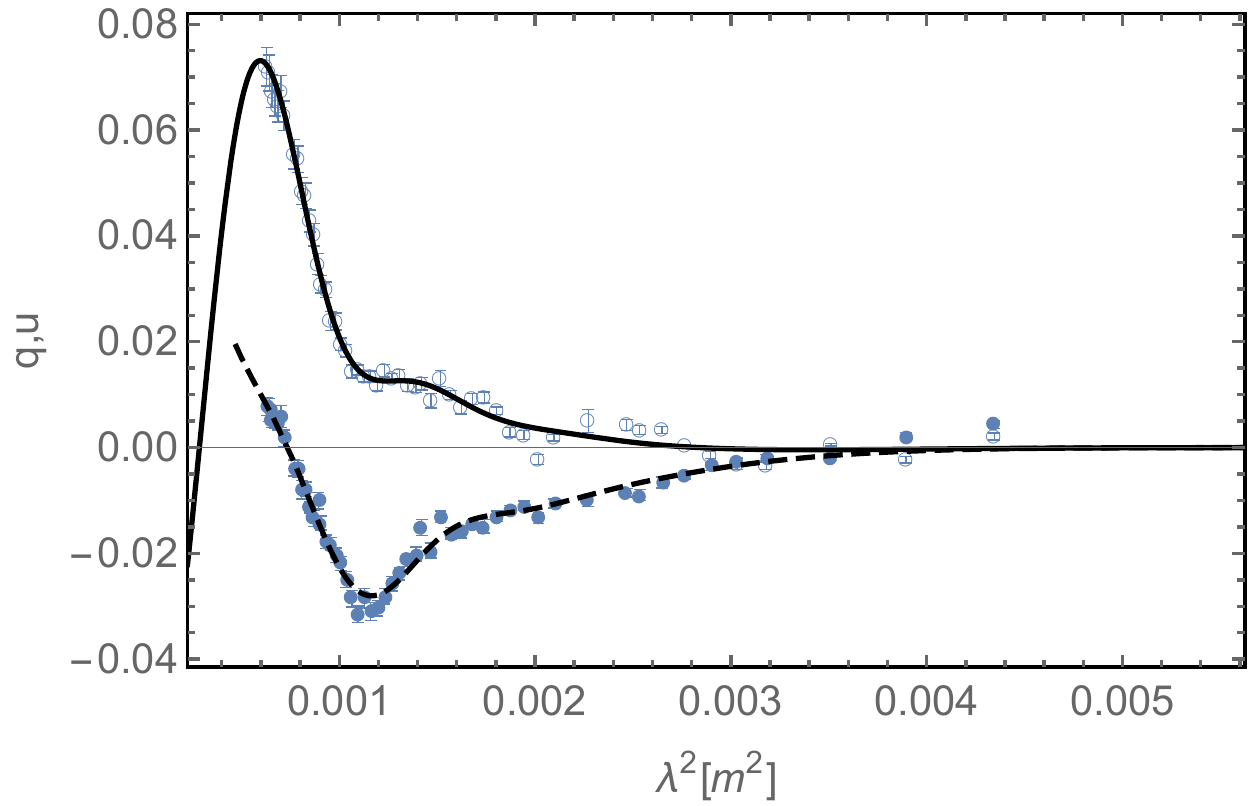}
        \includegraphics[width=0.4\textwidth]{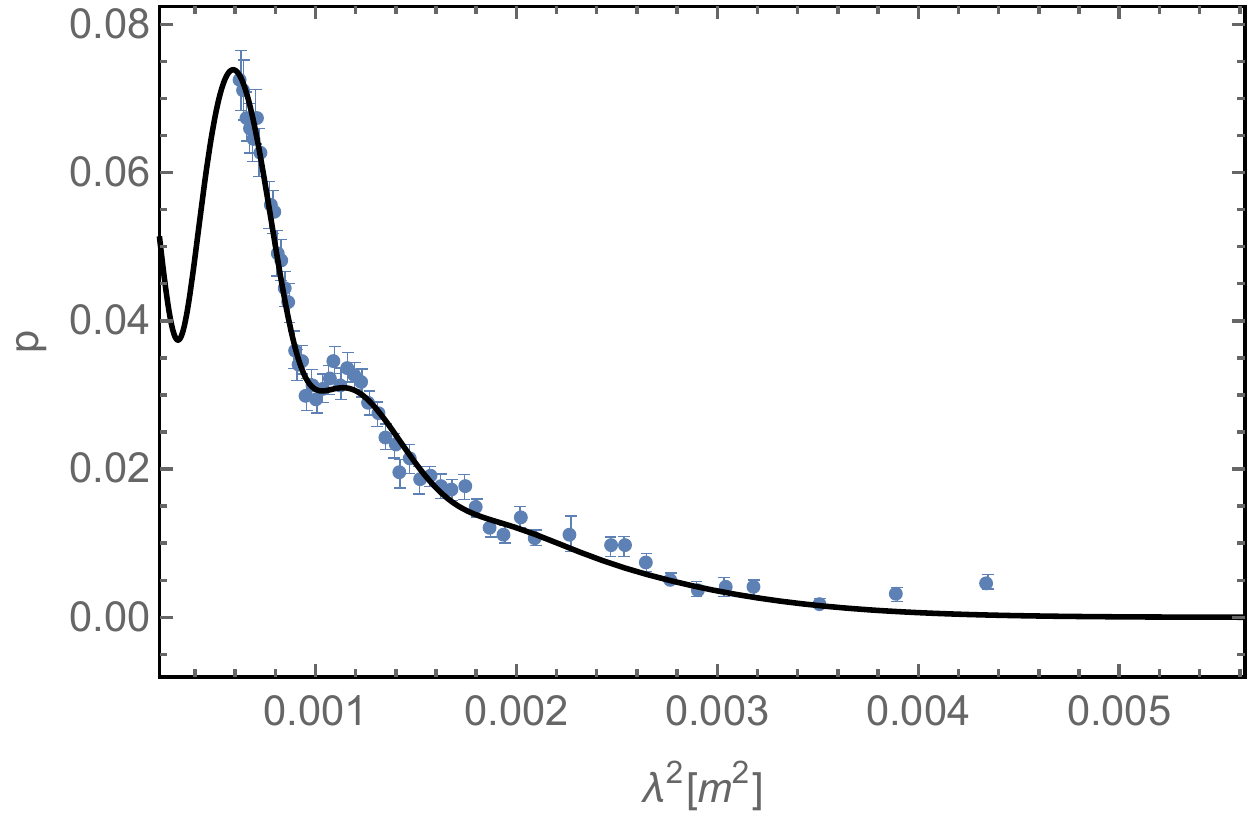}
        \includegraphics[width=0.4\textwidth]{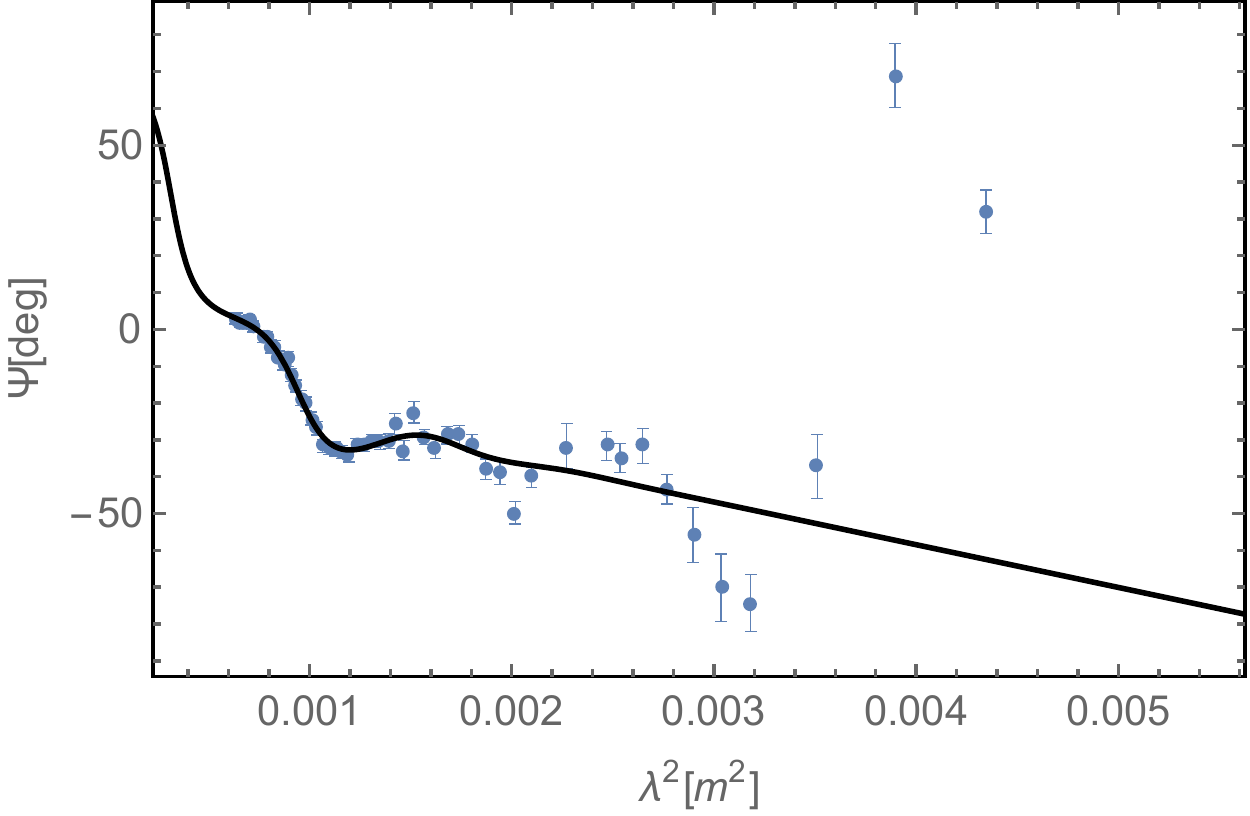}
        \includegraphics[width=0.4\textwidth]{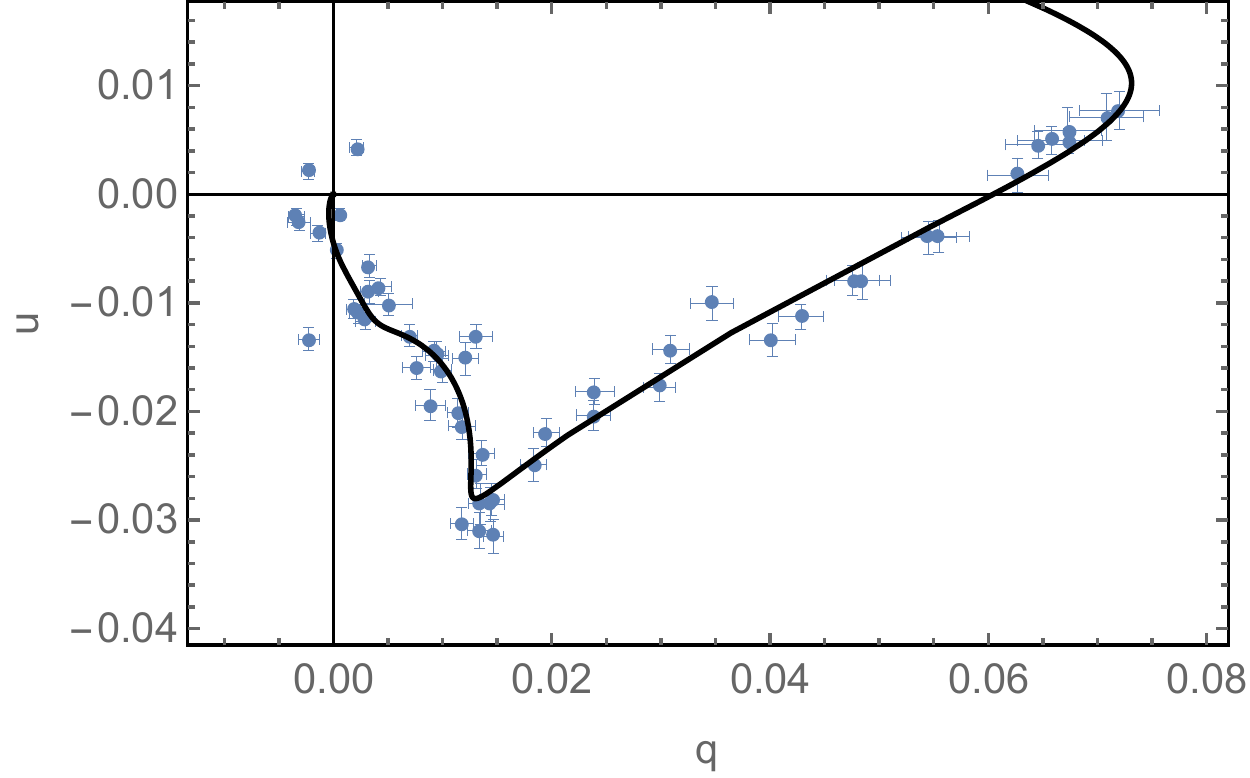}
\label{1312+5548model}
\end{center}
\end{figure*}

\begin{figure*}[h]
\begin{center}
    \caption{Depolarization model for the source 1549+5038: 3 components model}
        \includegraphics[width=0.4\textwidth]{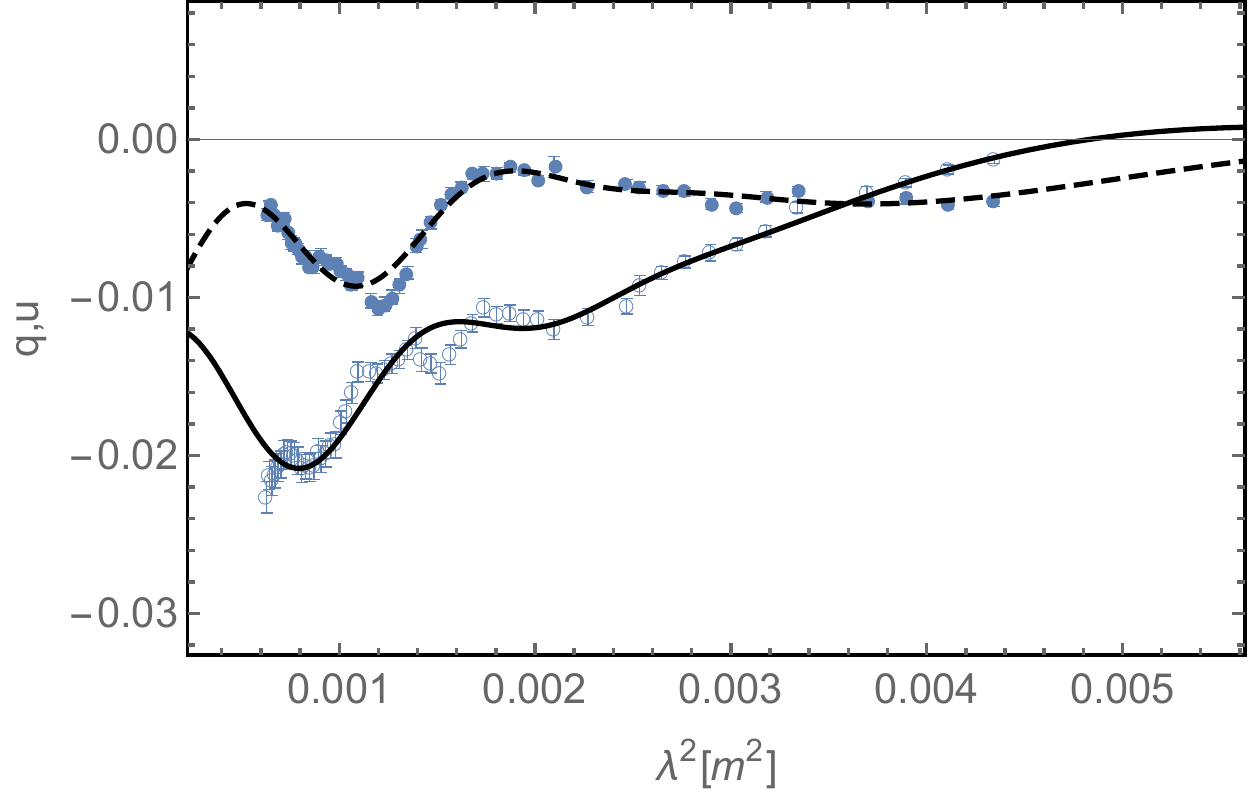}
        \includegraphics[width=0.4\textwidth]{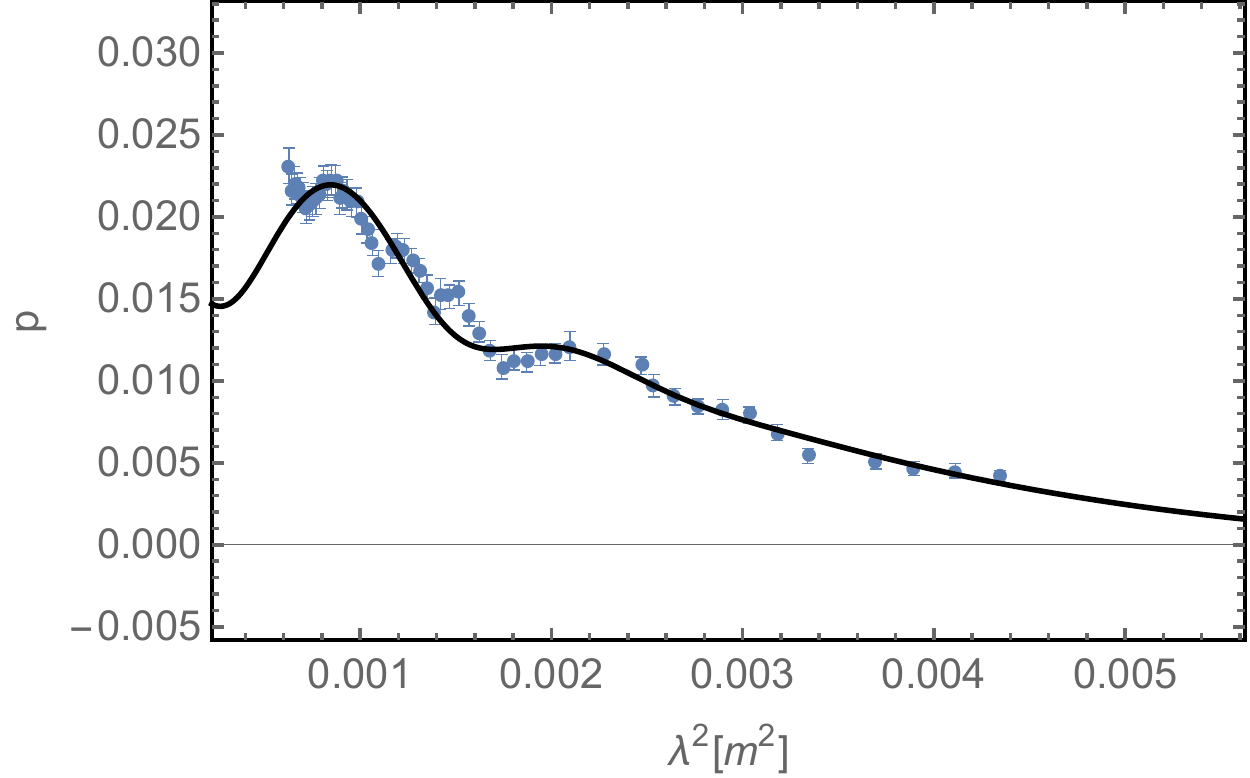}
        \includegraphics[width=0.4\textwidth]{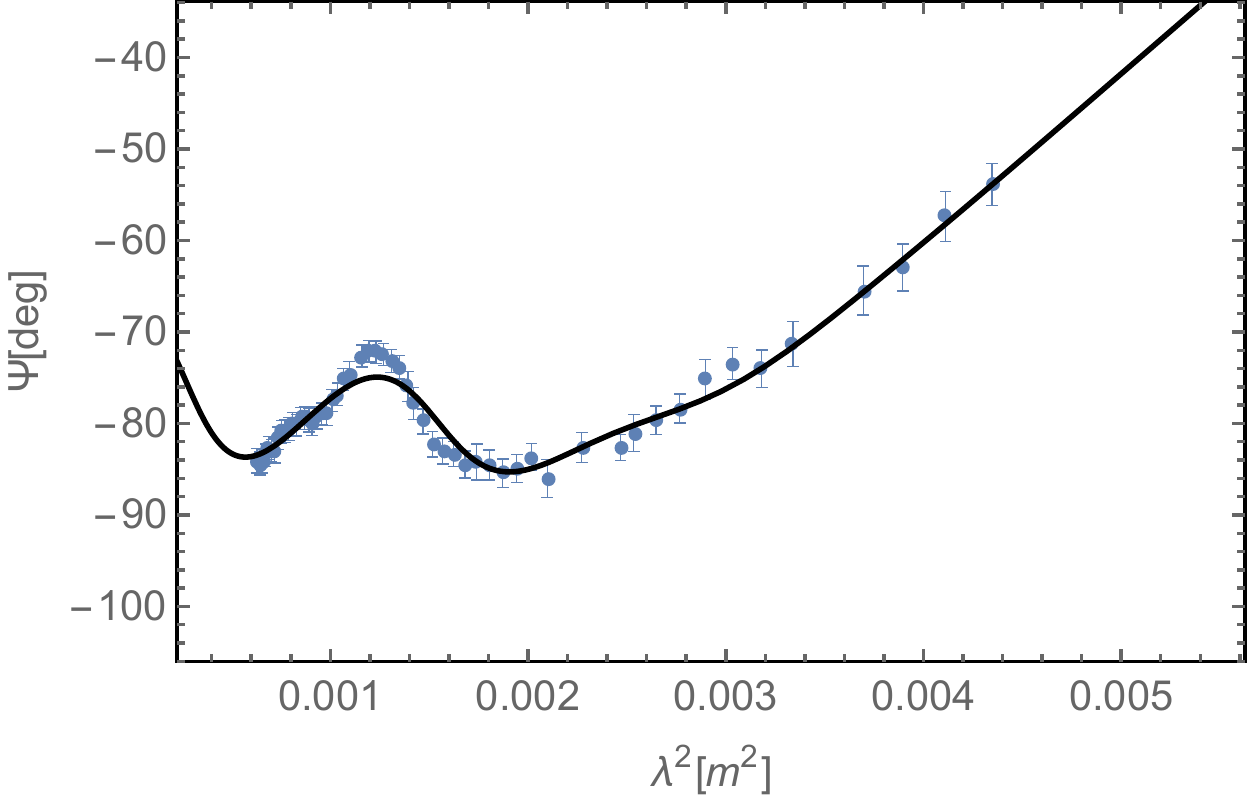}
        \includegraphics[width=0.4\textwidth]{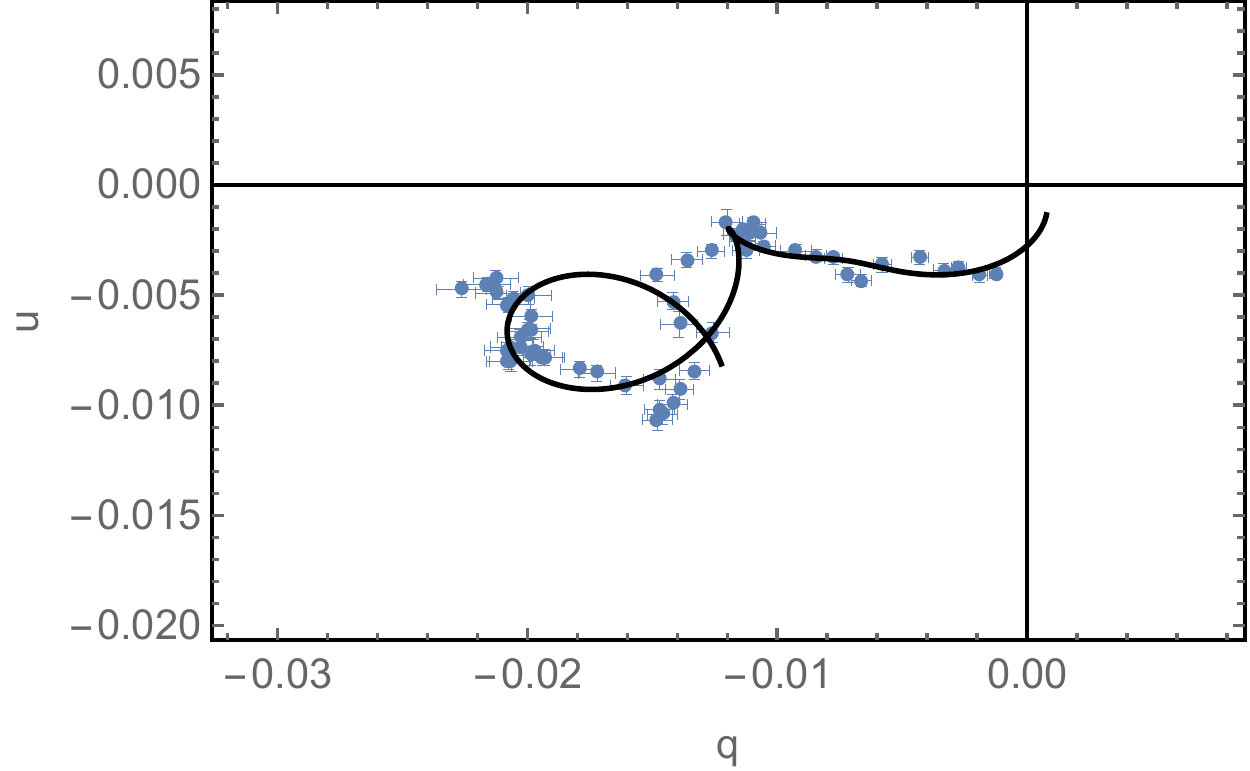}
\label{1549+5038model}
\end{center}
\end{figure*}

\begin{figure*}[h]
\begin{center}
    \caption{Depolarization model for the source 1616+0459: 2 components model}
        \includegraphics[width=0.4\textwidth]{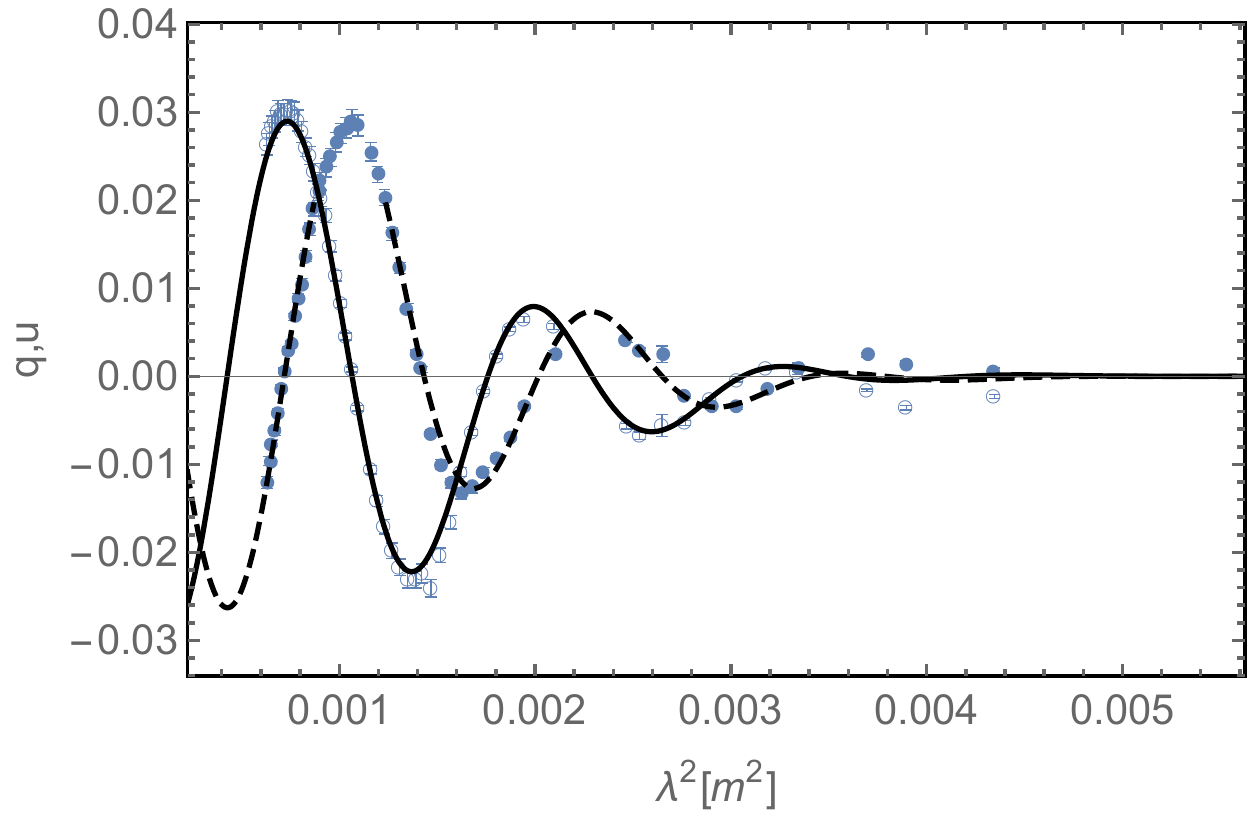}
        \includegraphics[width=0.4\textwidth]{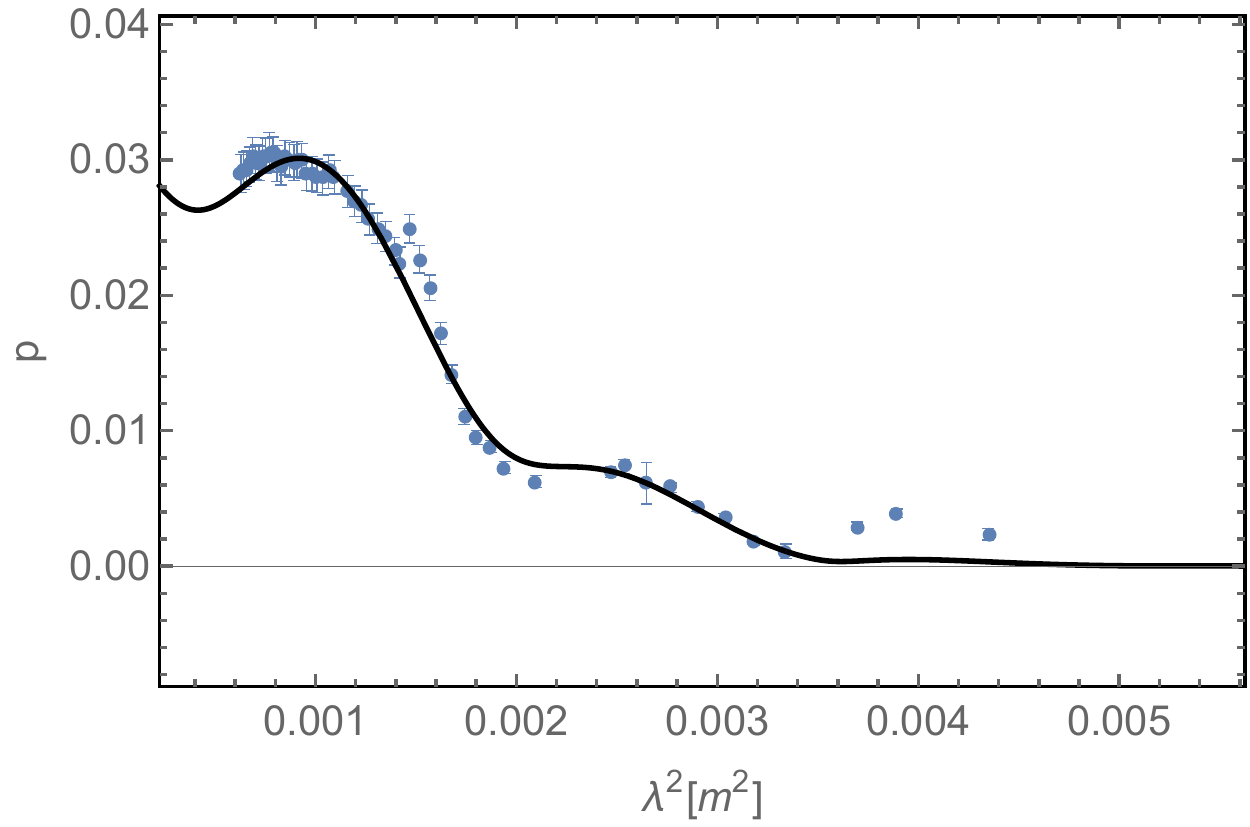}
        \includegraphics[width=0.4\textwidth]{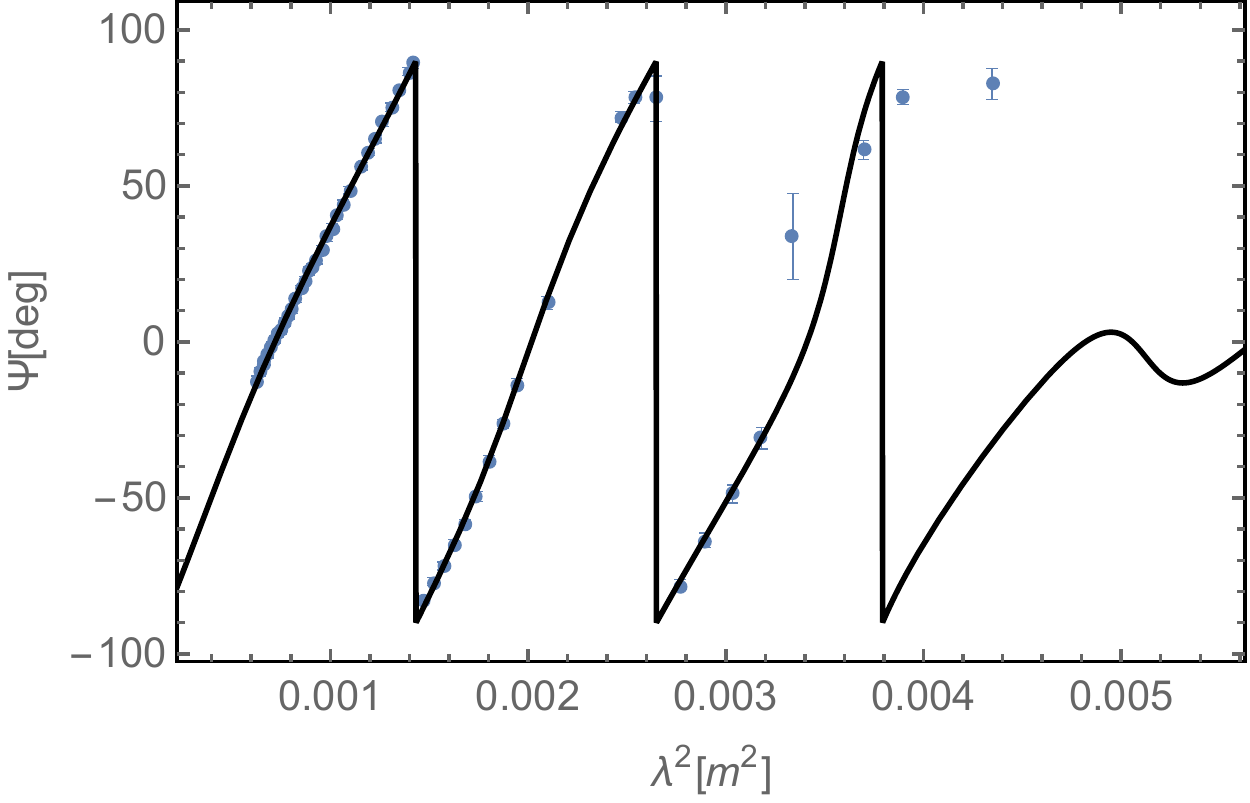}
        \includegraphics[width=0.4\textwidth]{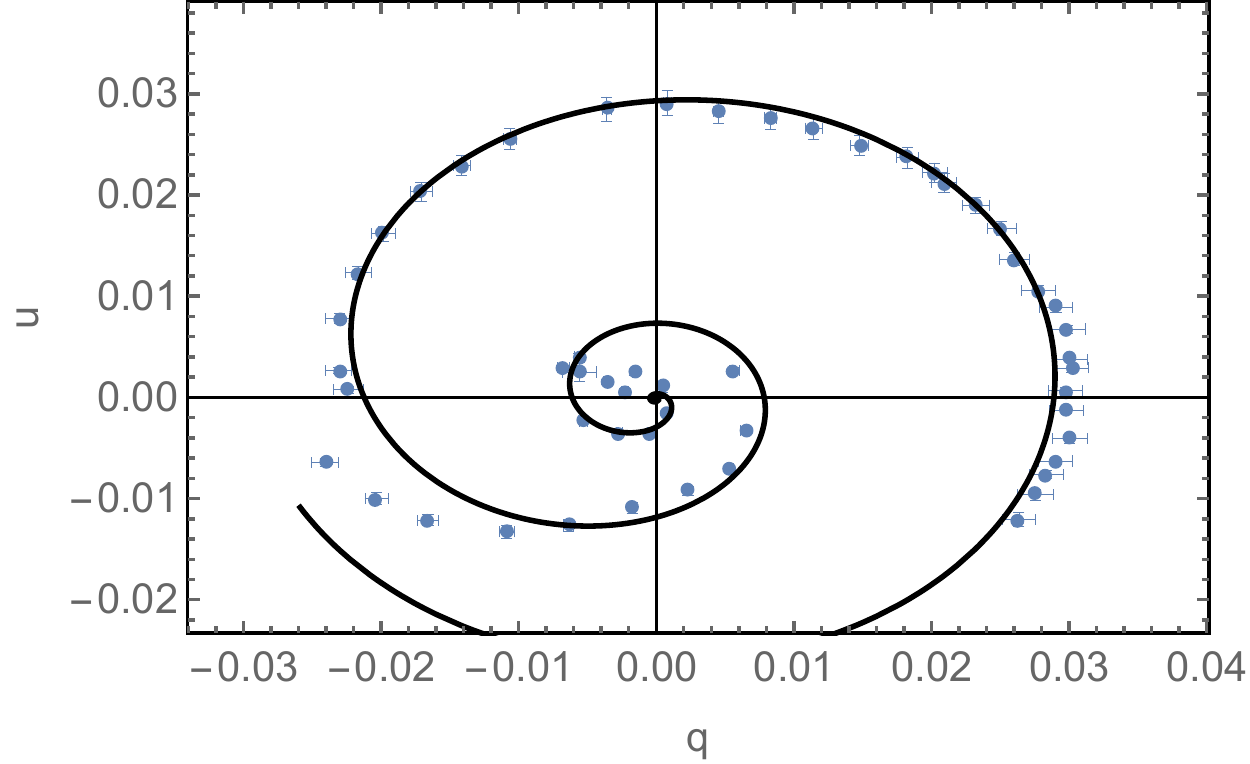}
\label{1616+0459model}
\end{center}
\end{figure*}

\begin{figure*}[h]
\begin{center}
    \caption{Depolarization model for the source 2245+0324: 3 components model}
        \includegraphics[width=0.4\textwidth]{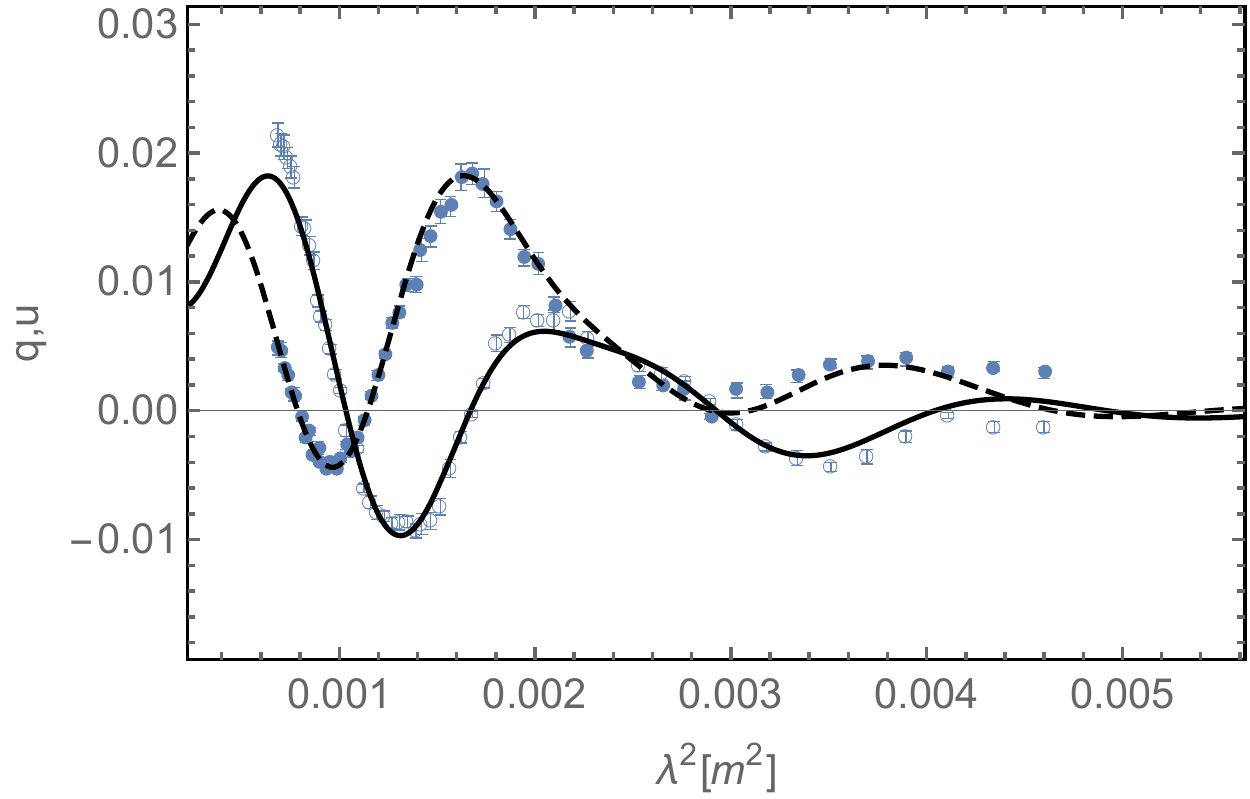}
        \includegraphics[width=0.4\textwidth]{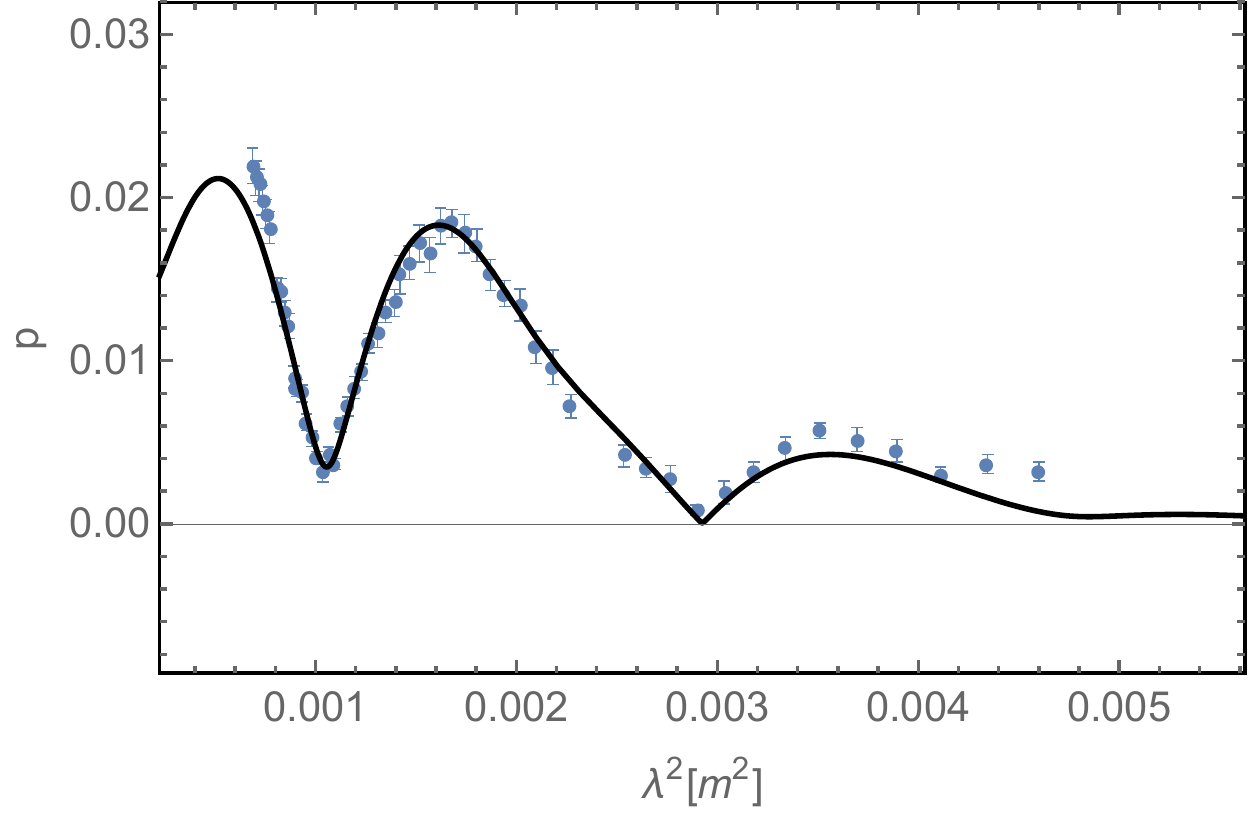}
        \includegraphics[width=0.4\textwidth]{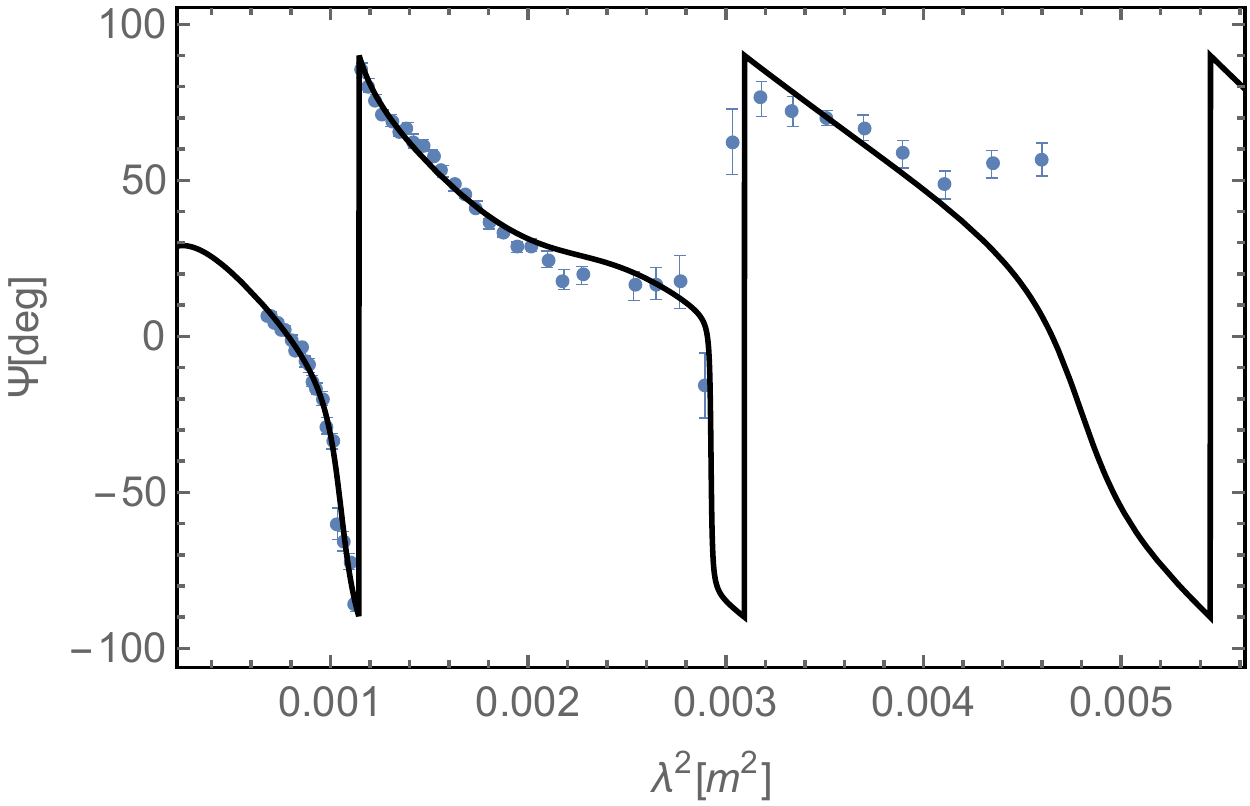}
        \includegraphics[width=0.4\textwidth]{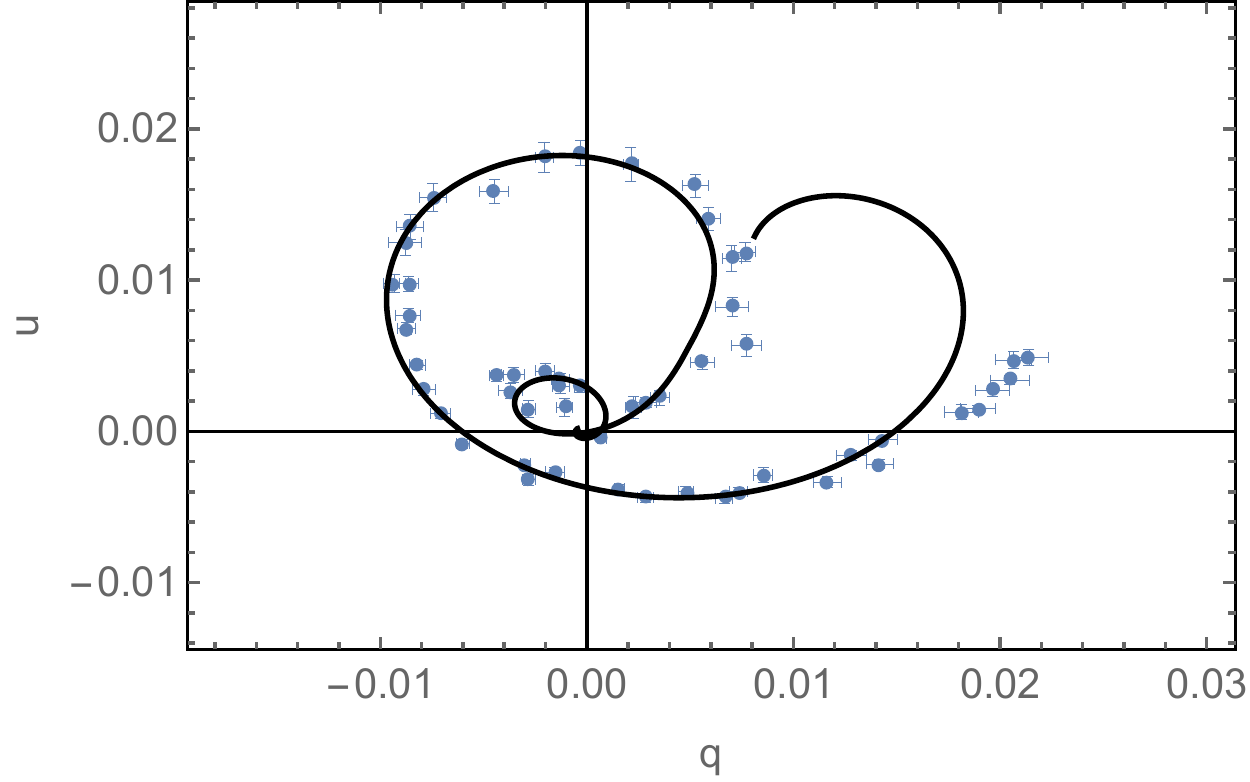}
\label{2245+0324model}
\end{center}
\end{figure*}


\clearpage
\section{Depolarization modelling plots: sources with C-X bands and L band polarization data.}


\begin{figure*}[h]
\begin{center}
    \caption{Depolarization model for the source 0239--0234 at C and X bands: 2 components model}
        \includegraphics[width=0.4\textwidth]{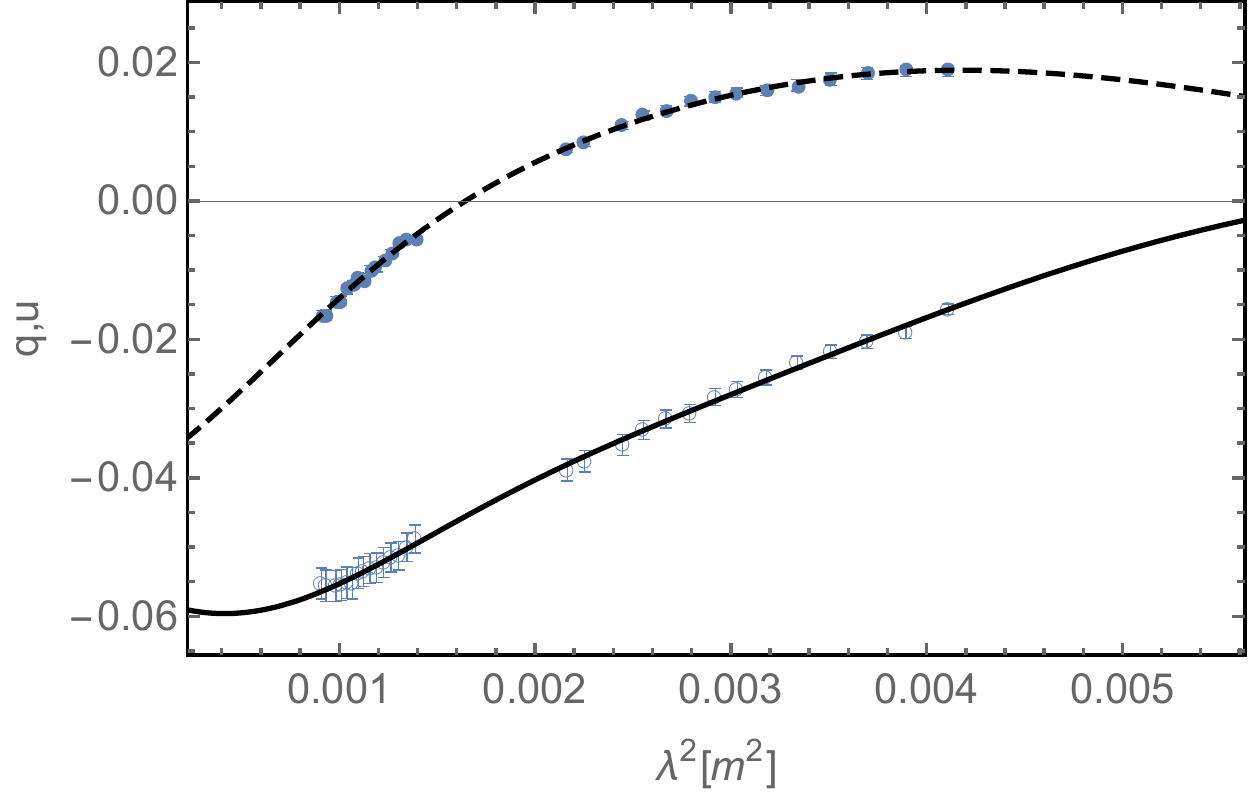}
        \includegraphics[width=0.4\textwidth]{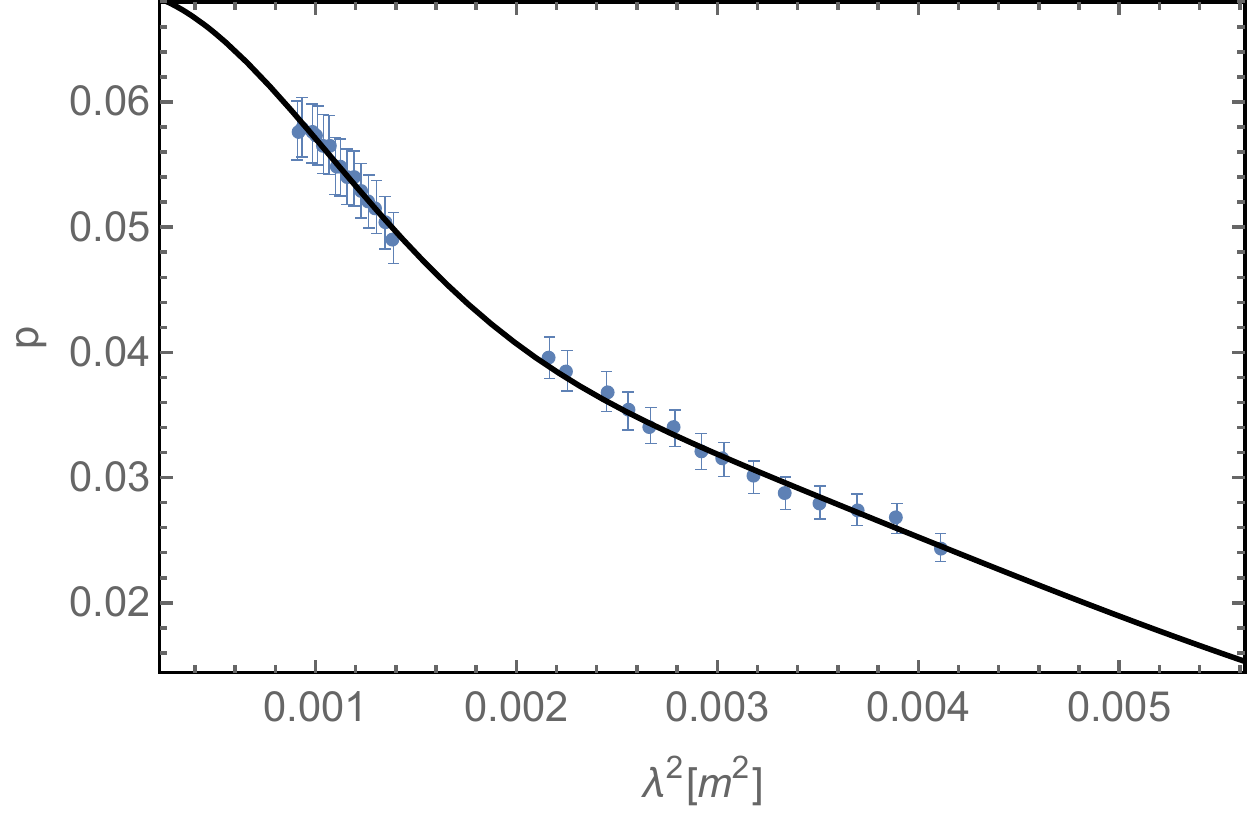}
        \includegraphics[width=0.4\textwidth]{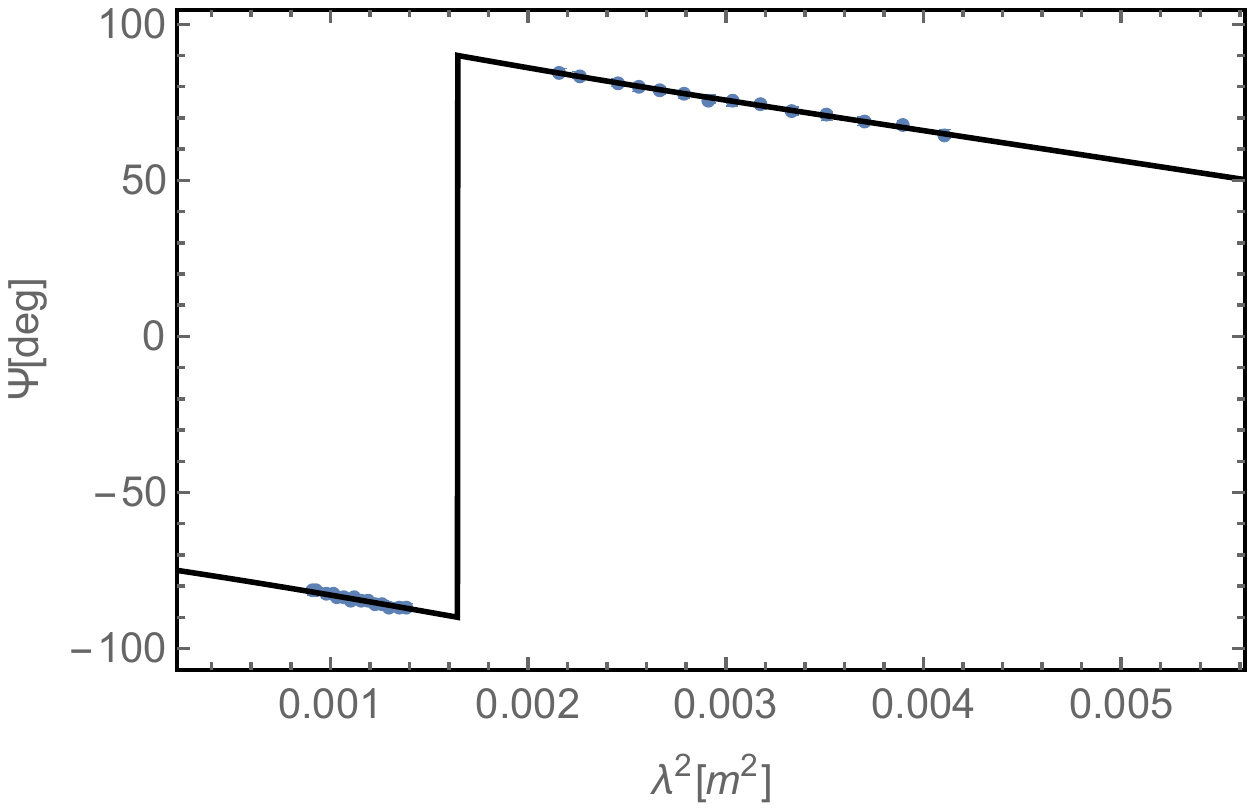}
        \includegraphics[width=0.4\textwidth]{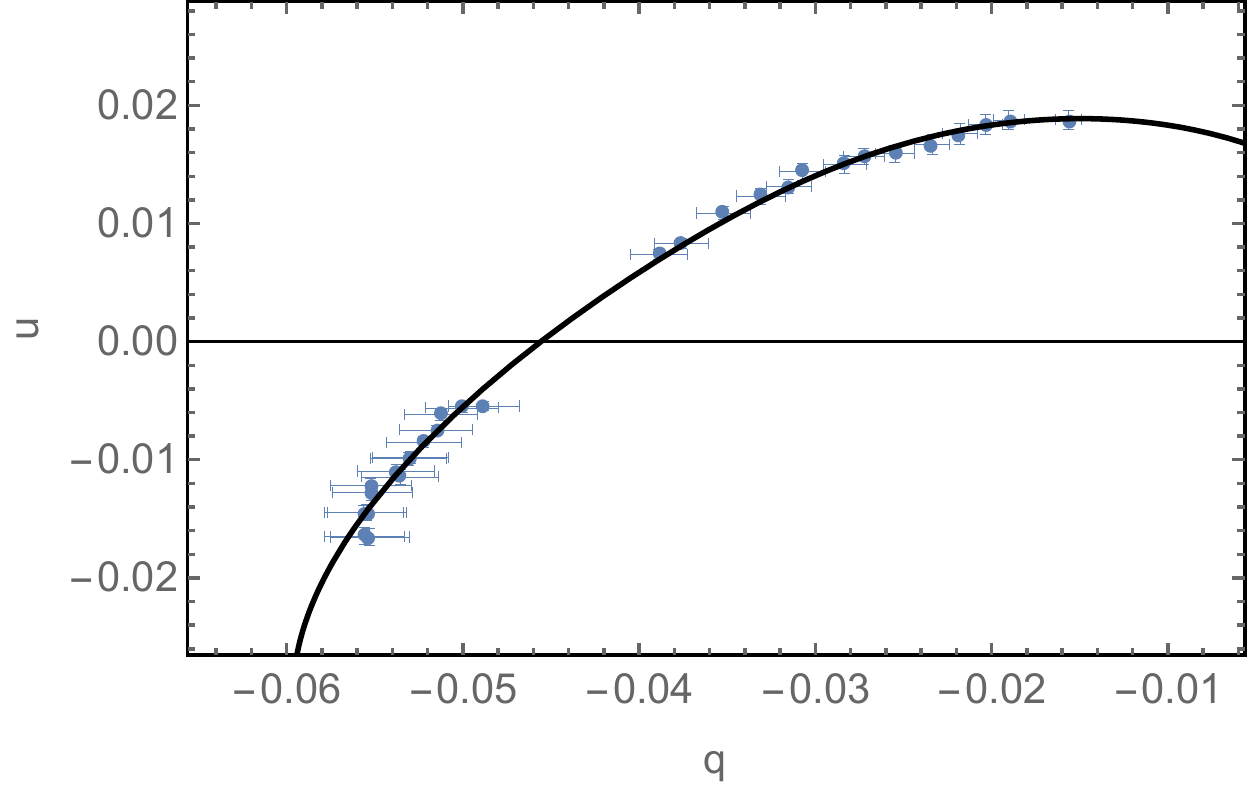}
\label{0239-0234CXbandmodel}
\end{center}
 \begin{flushleft}
NOTE. - L band depolarization modelling is not available because of insufficient data points at this frequency band.
\end{flushleft}

\end{figure*}
\clearpage

\begin{figure*}[h]
\begin{center}
    \caption{Depolarization model for the source 0243-0550 at C and X bands: 2 components model}
        \includegraphics[width=0.4\textwidth]{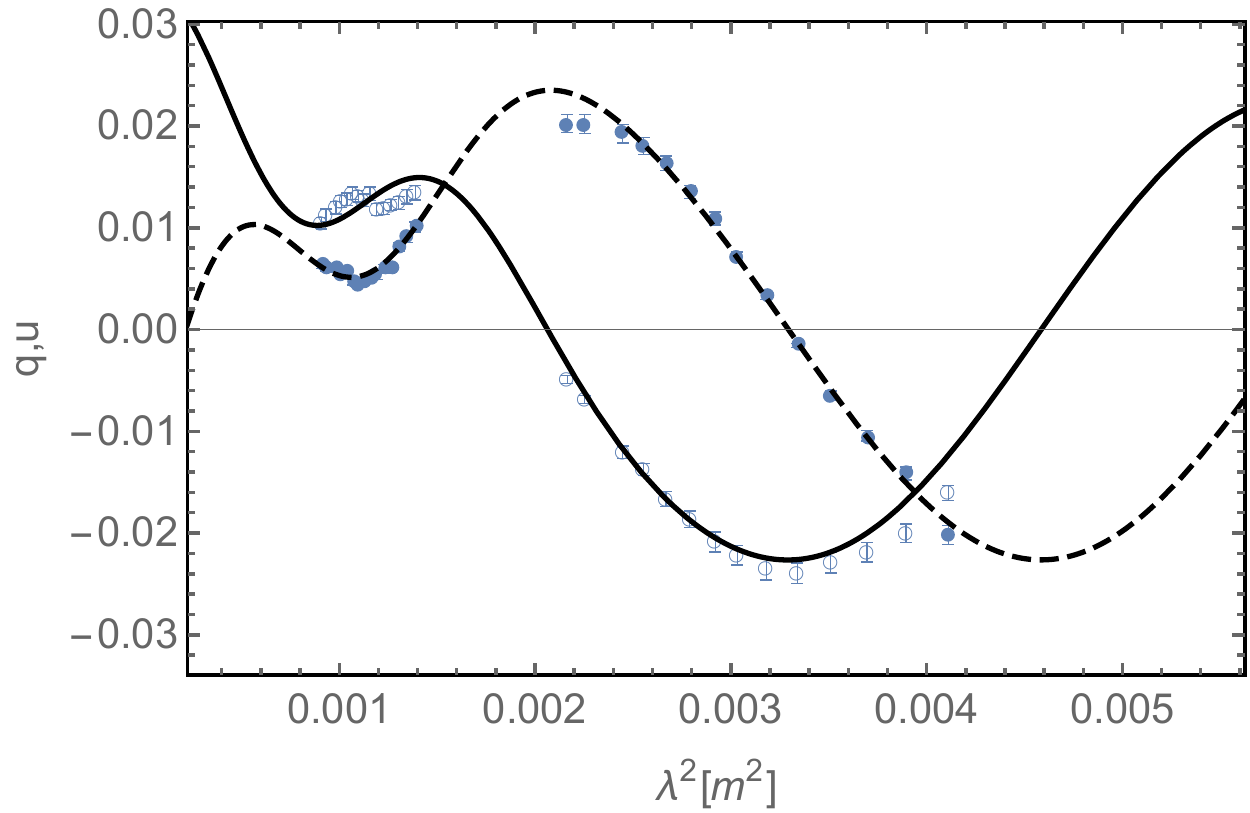}
        \includegraphics[width=0.4\textwidth]{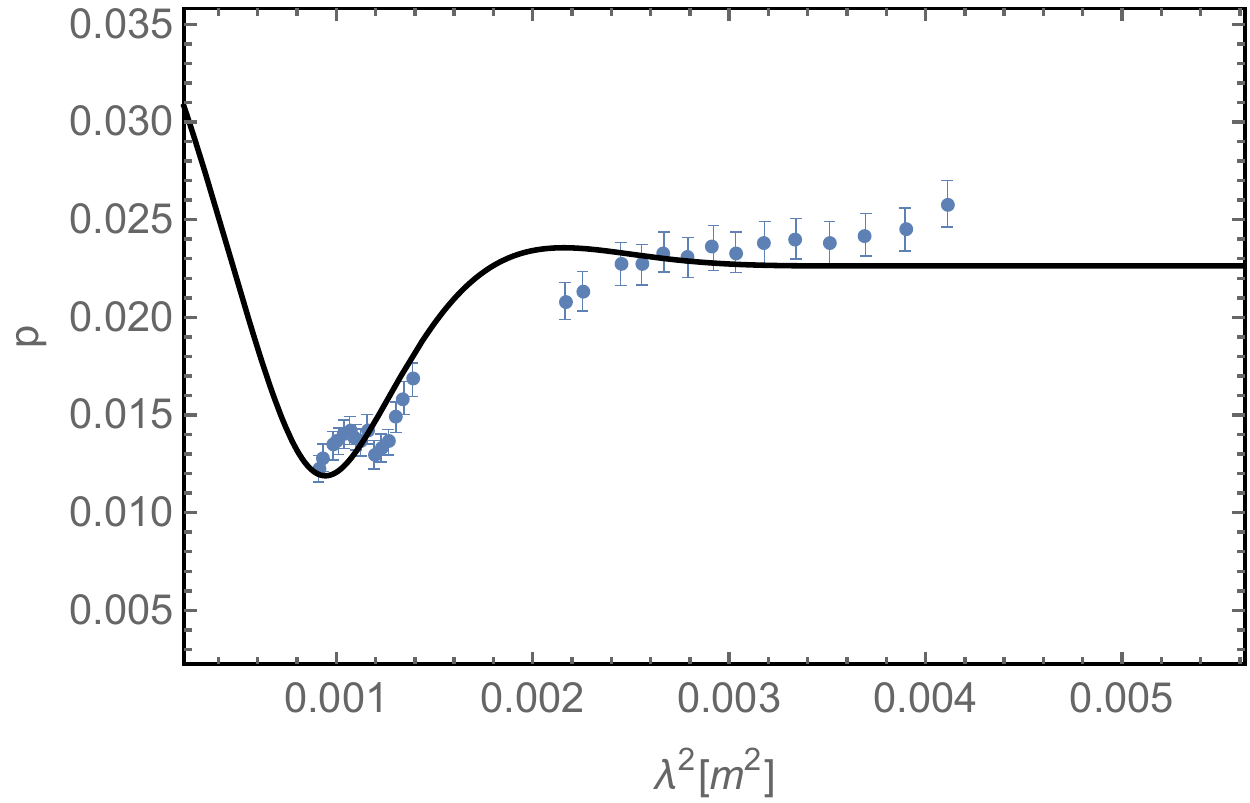}
        \includegraphics[width=0.4\textwidth]{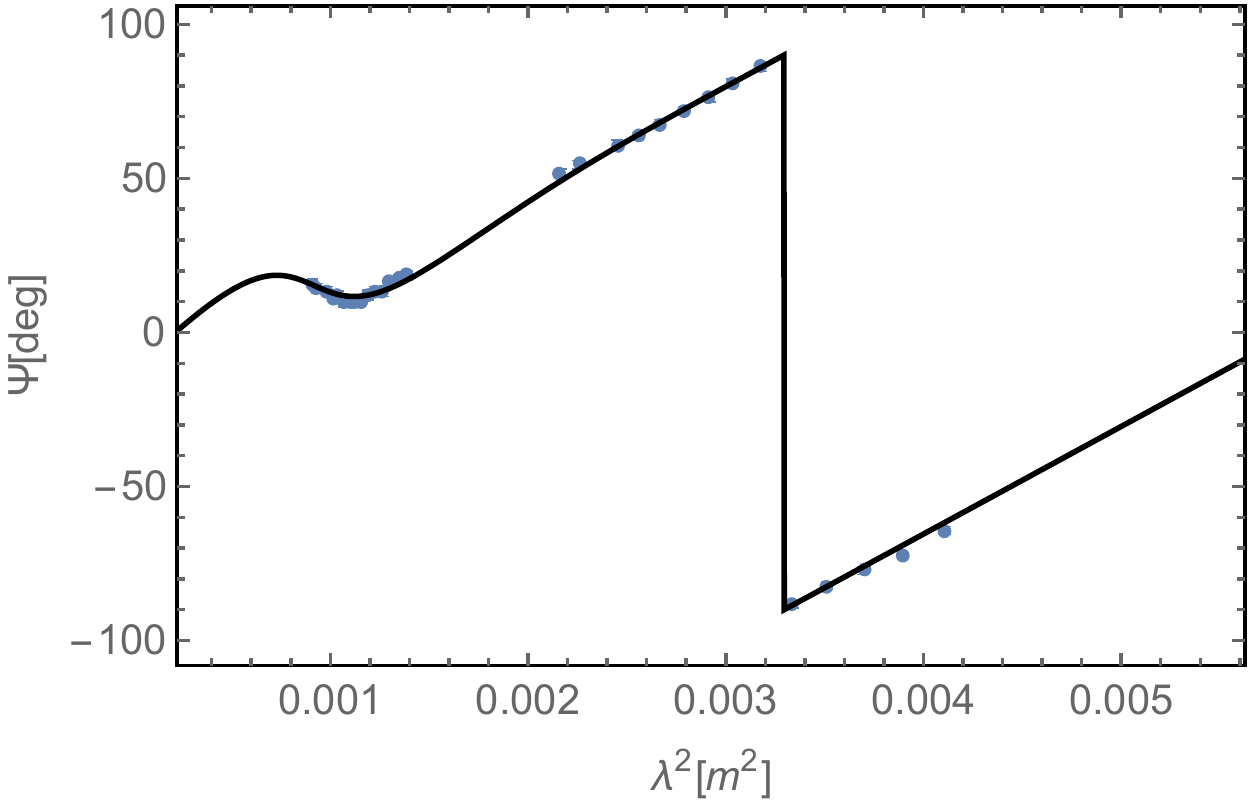}
        \includegraphics[width=0.4\textwidth]{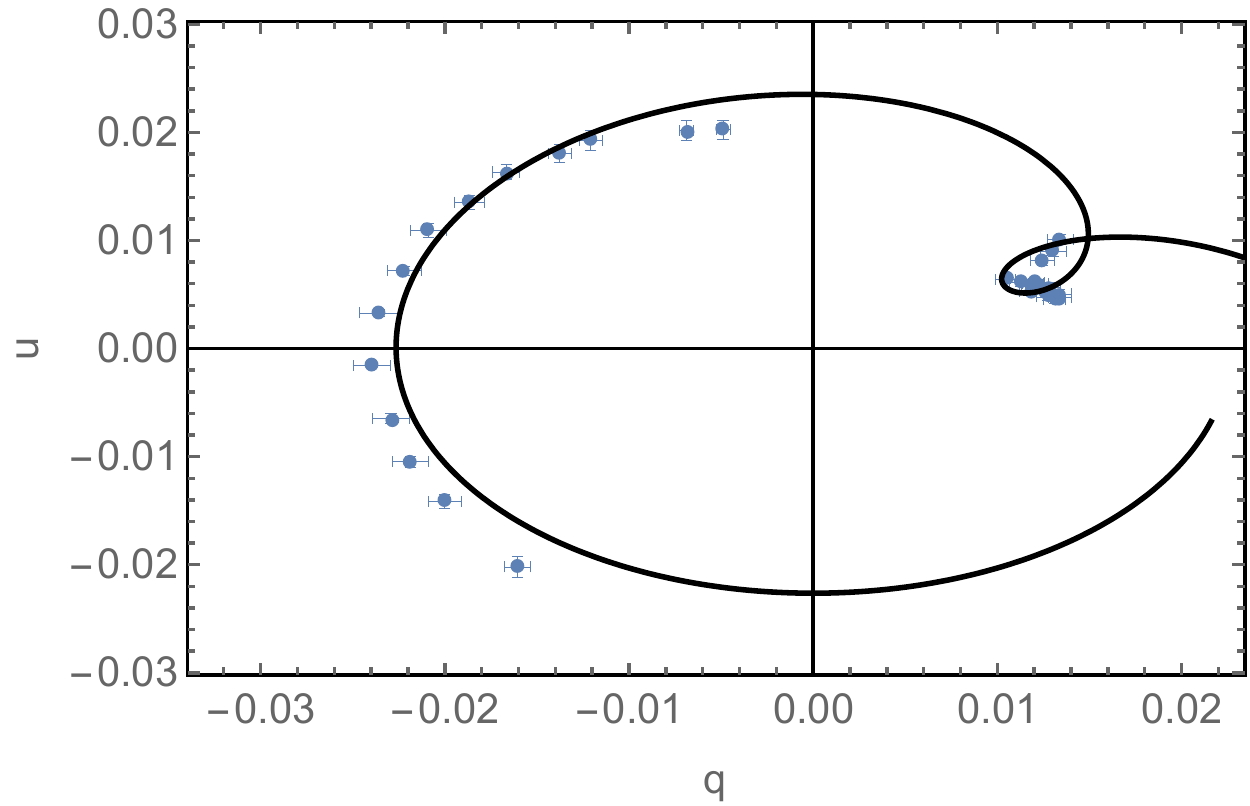}
\label{0243-0550CXbandmodel}
\end{center}

\end{figure*}

\begin{figure*}[h]
\begin{center}
    \caption{Depolarization model for the source 0243-0550 at L band: 2 components model}
        \includegraphics[width=0.4\textwidth]{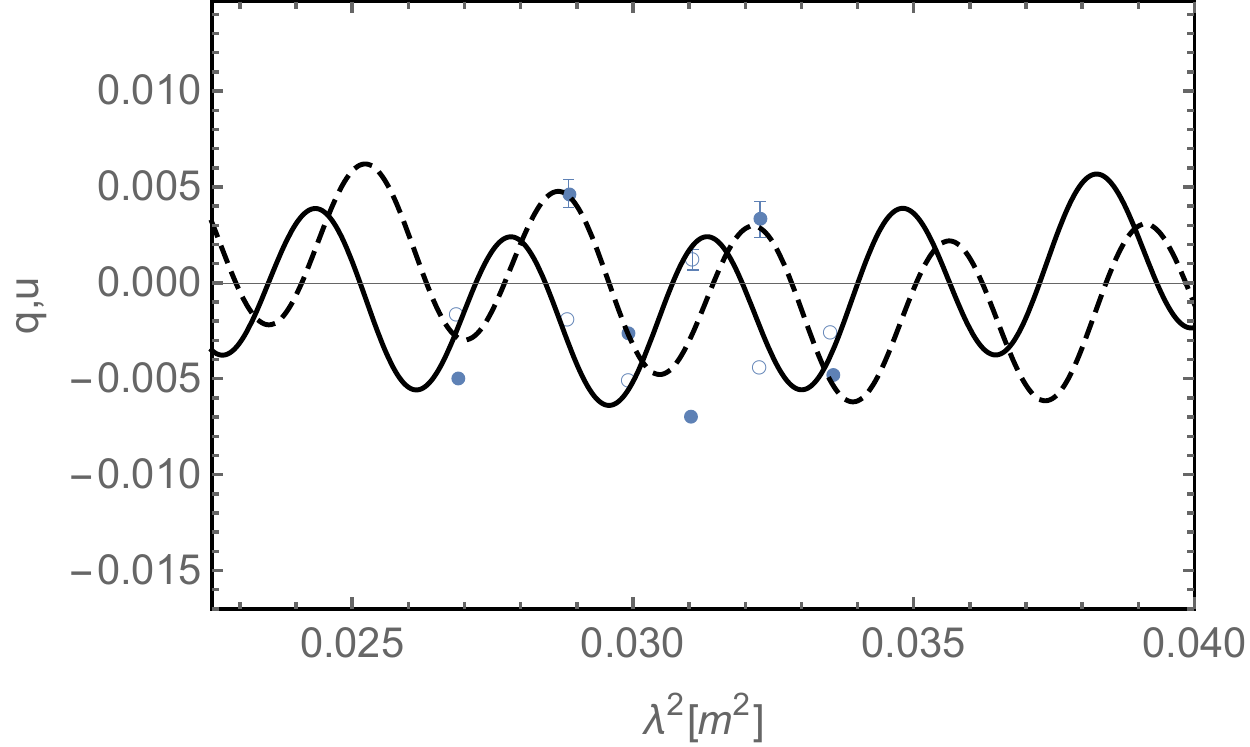}
        \includegraphics[width=0.4\textwidth]{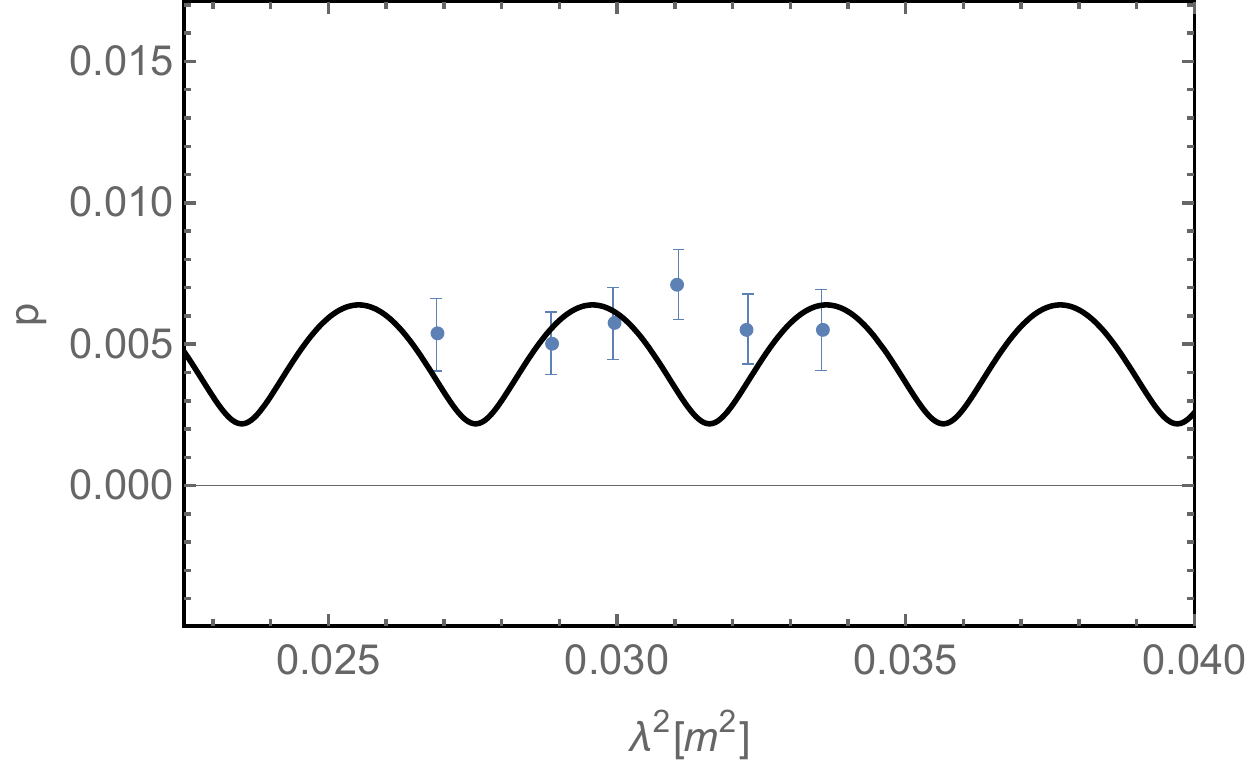}
        \includegraphics[width=0.4\textwidth]{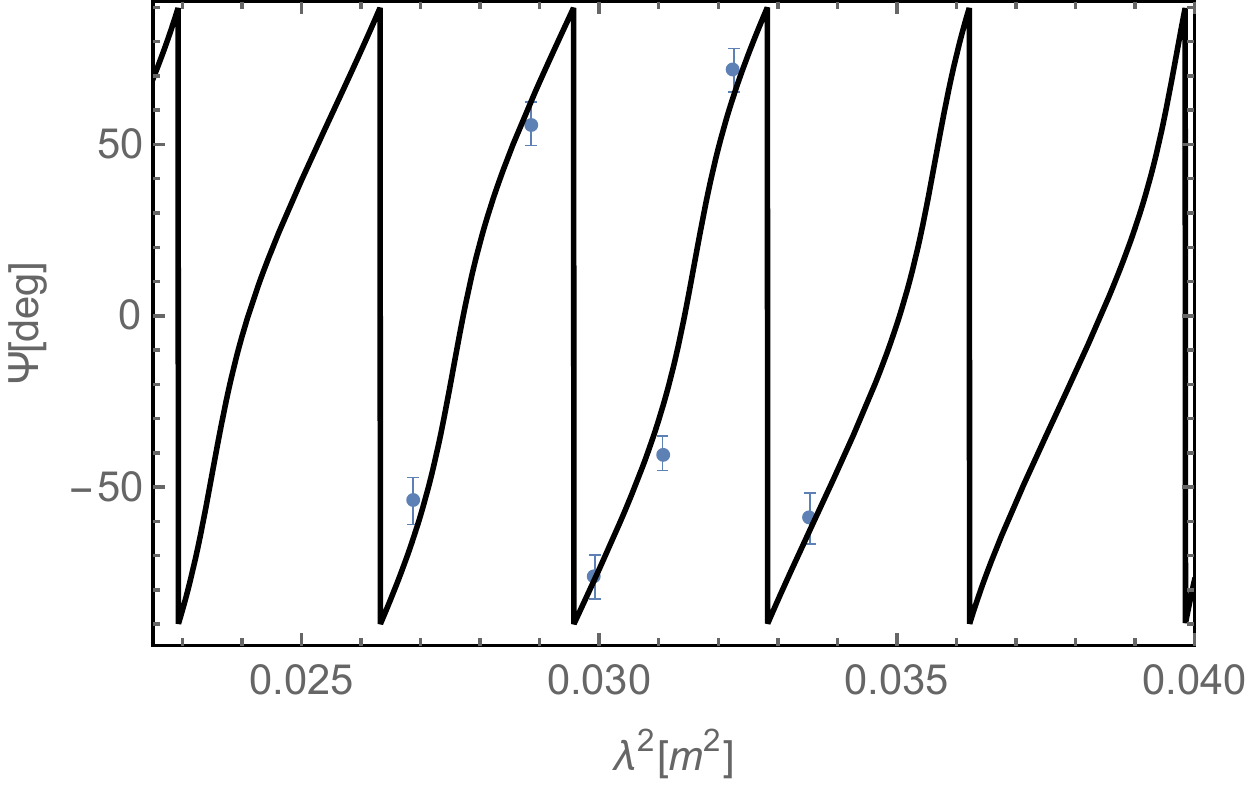}
        \includegraphics[width=0.4\textwidth]{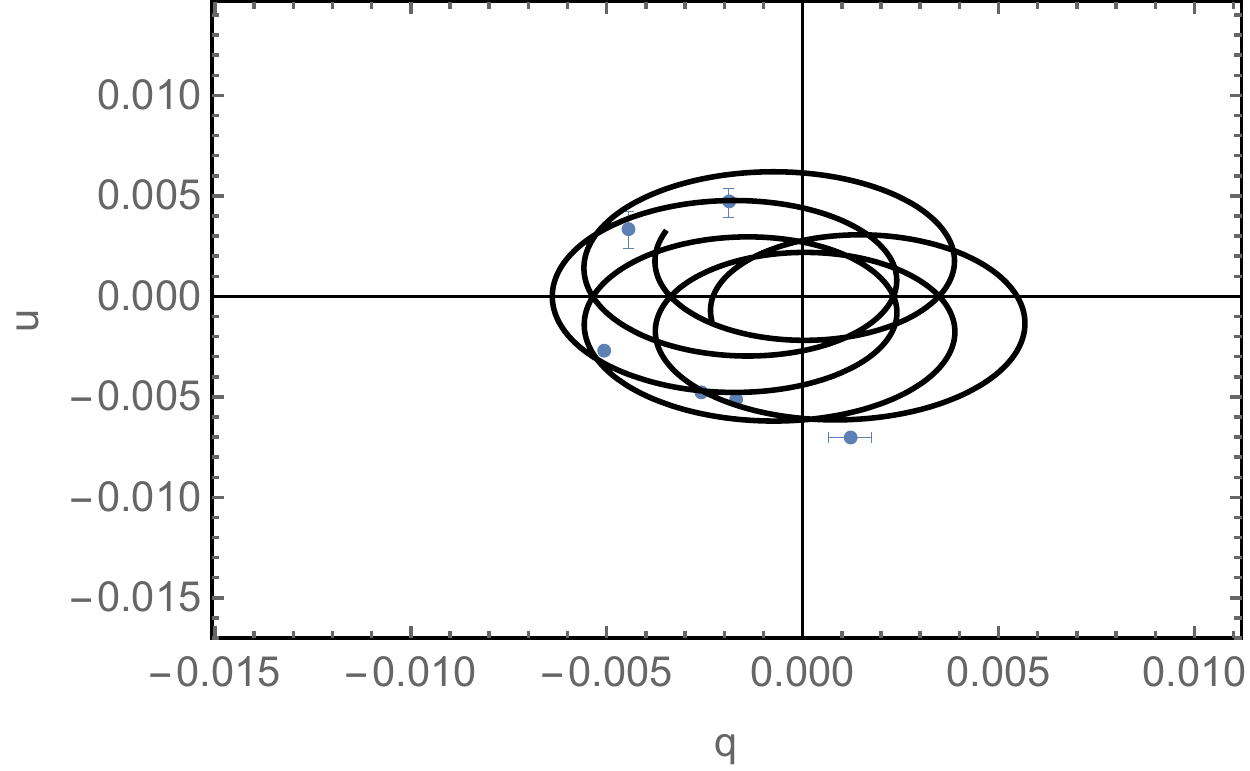}
\label{0243-0550Lbandmodel}
\end{center}

\end{figure*}


 \begin{figure*}[h]
\begin{center}
    \caption{Depolarization model for the source 1246--0730 at C and X bands: 2 components model}
        \includegraphics[width=0.4\textwidth]{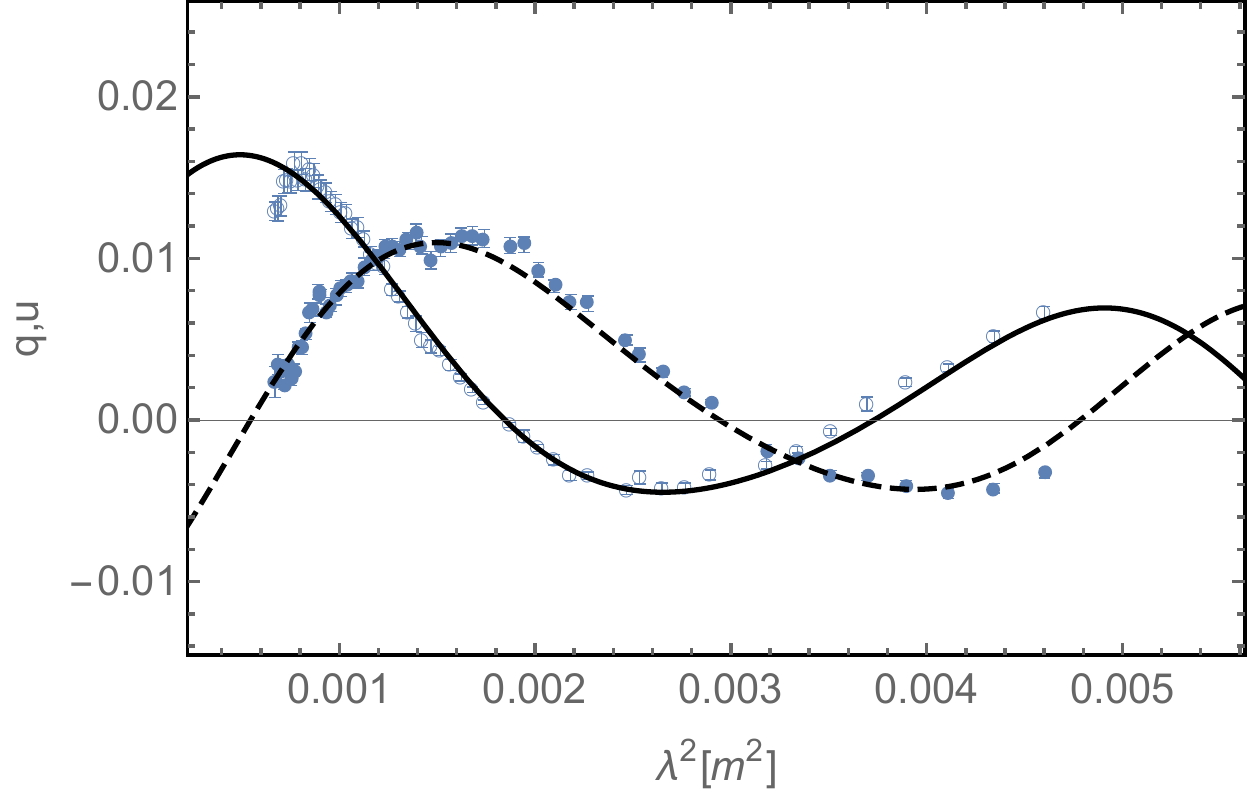}
        \includegraphics[width=0.4\textwidth]{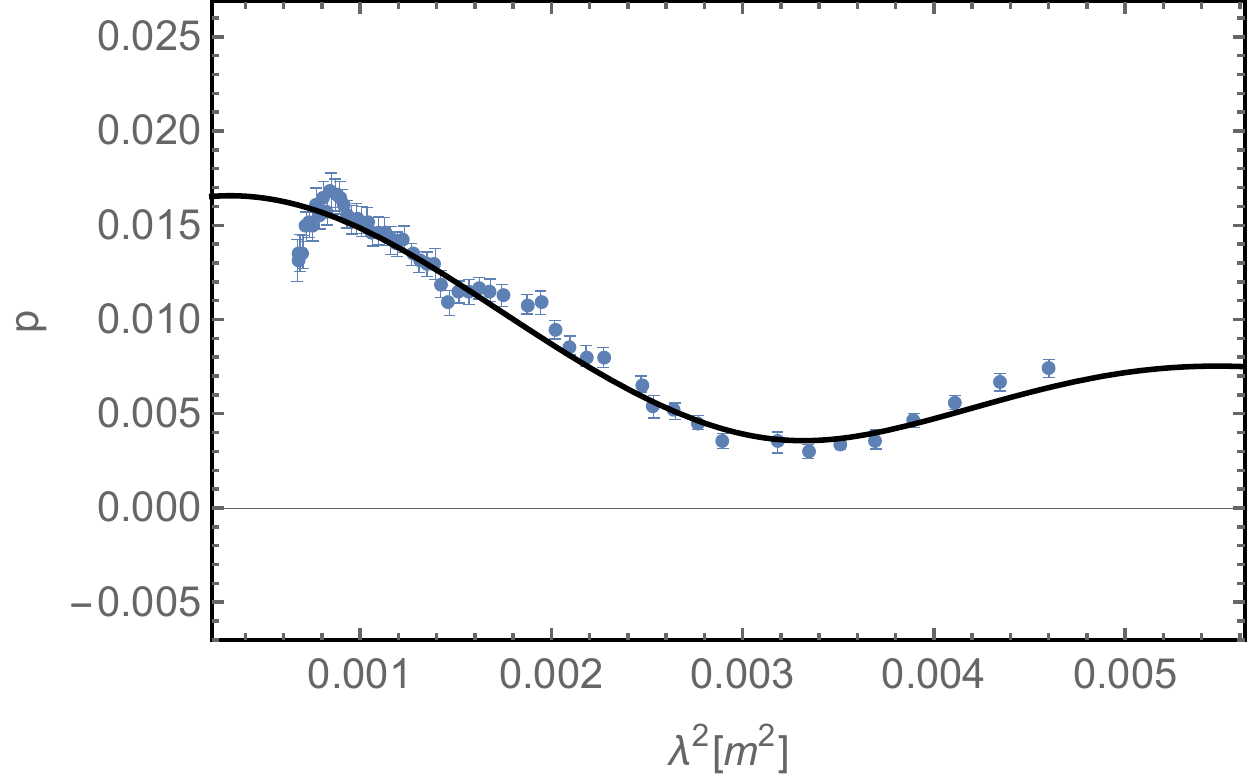}
        \includegraphics[width=0.4\textwidth]{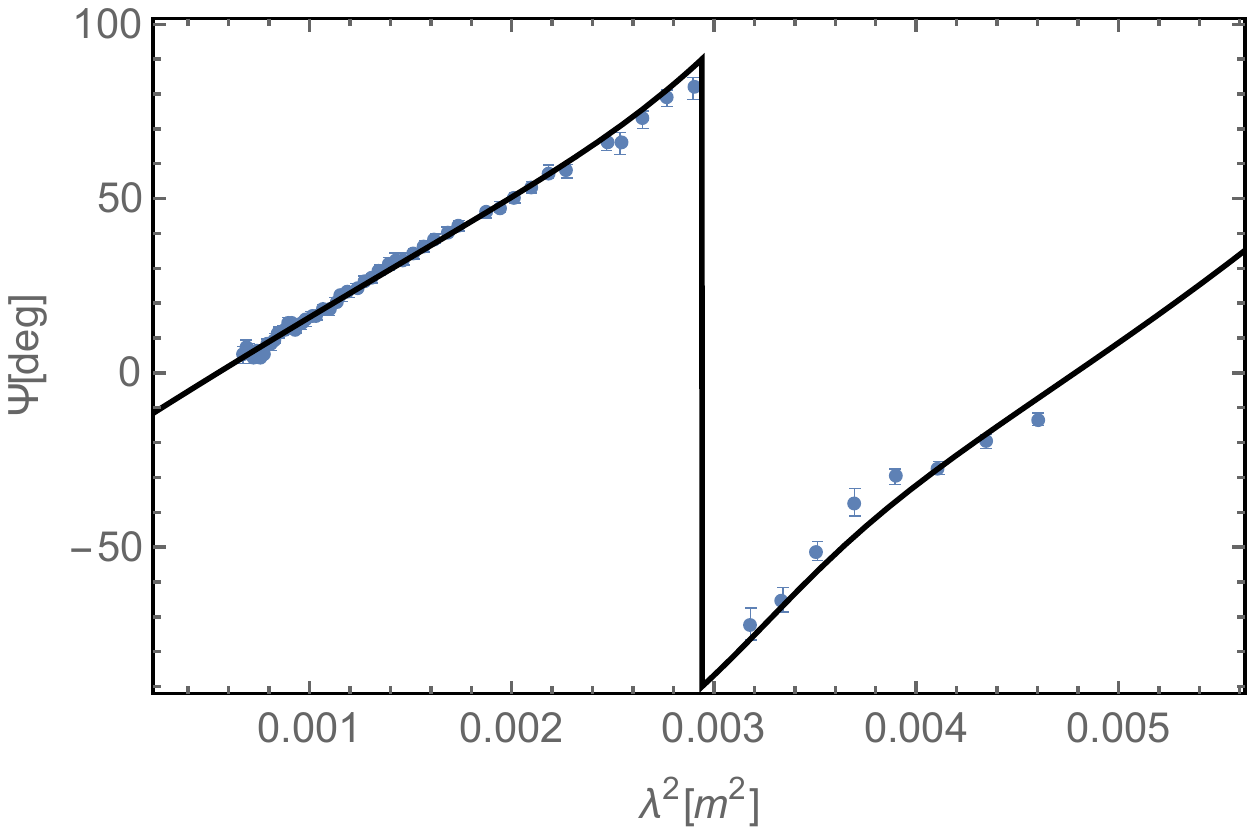}
        \includegraphics[width=0.4\textwidth]{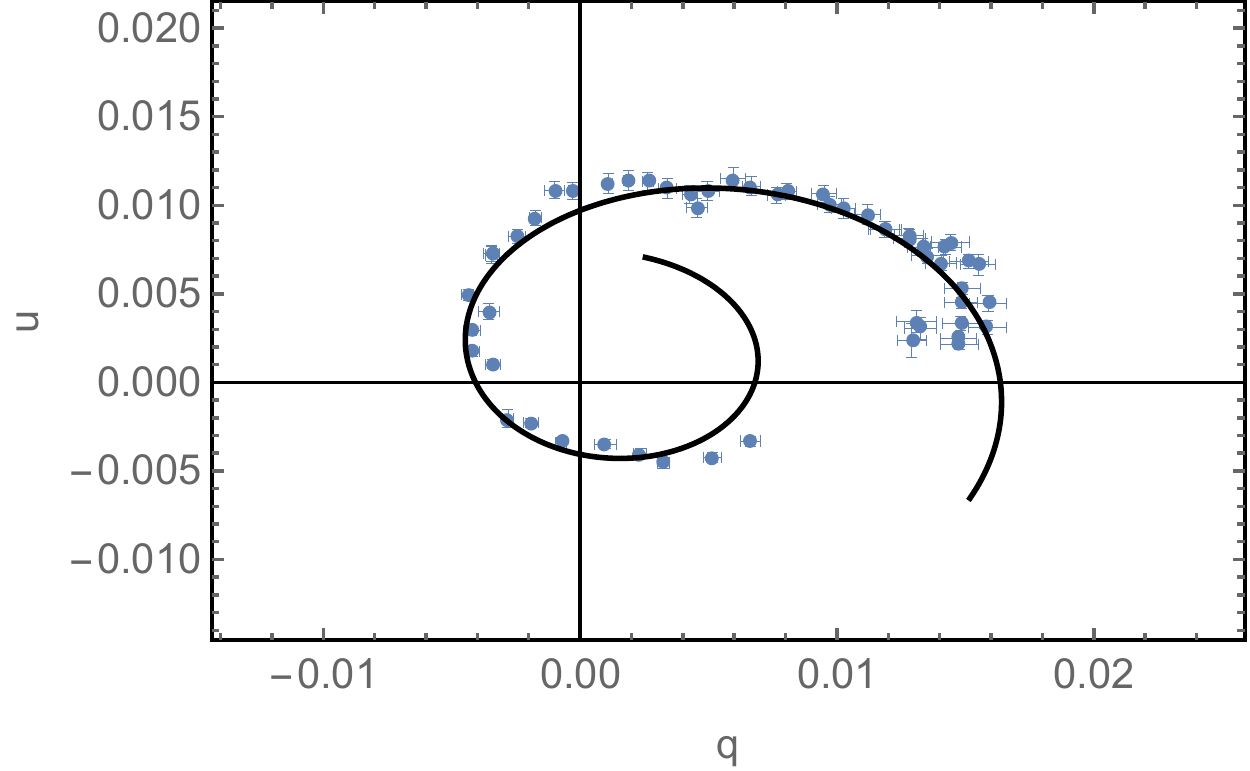}
\label{1246-0730CXbandmodel}
\end{center}
\end{figure*}

\begin{figure*}[h]
\begin{center}
    \caption{Depolarization model for the source 1246--0730 at L band: 2 components model}
        \includegraphics[width=0.4\textwidth]{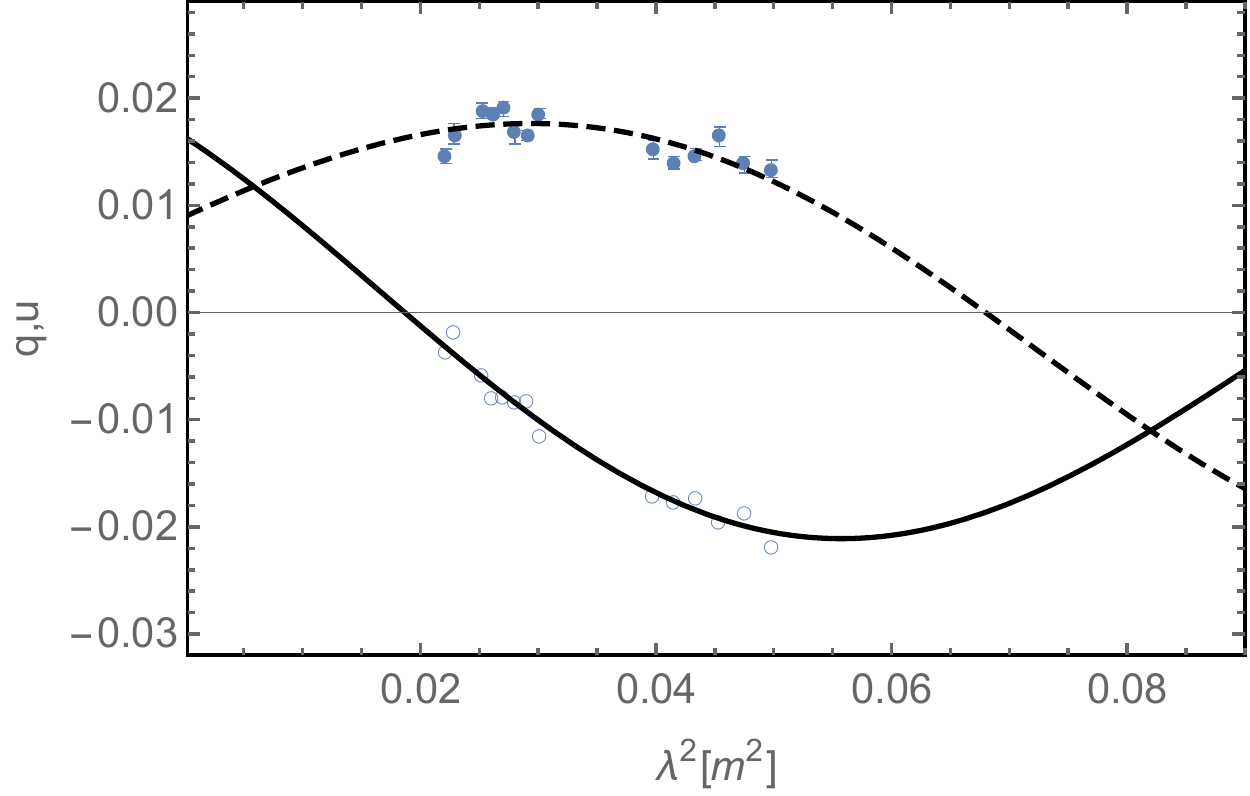}
        \includegraphics[width=0.4\textwidth]{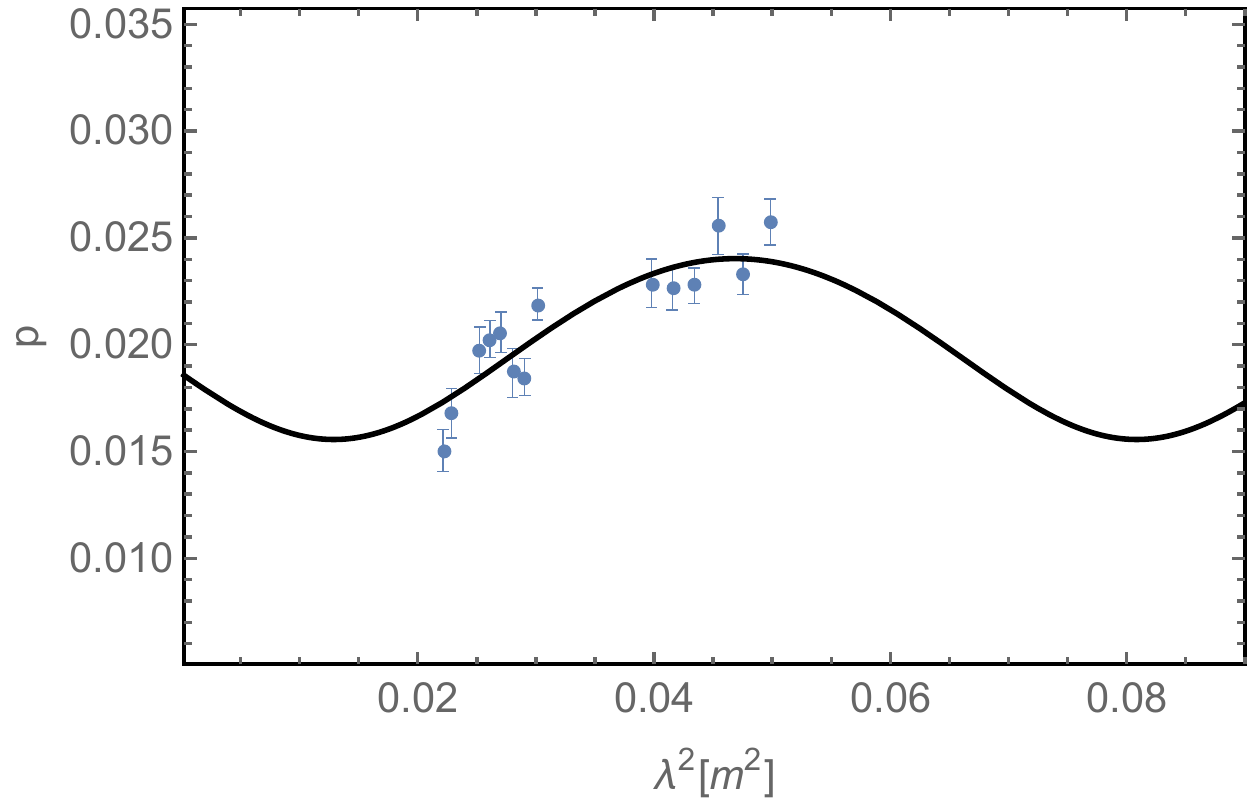}
        \includegraphics[width=0.4\textwidth]{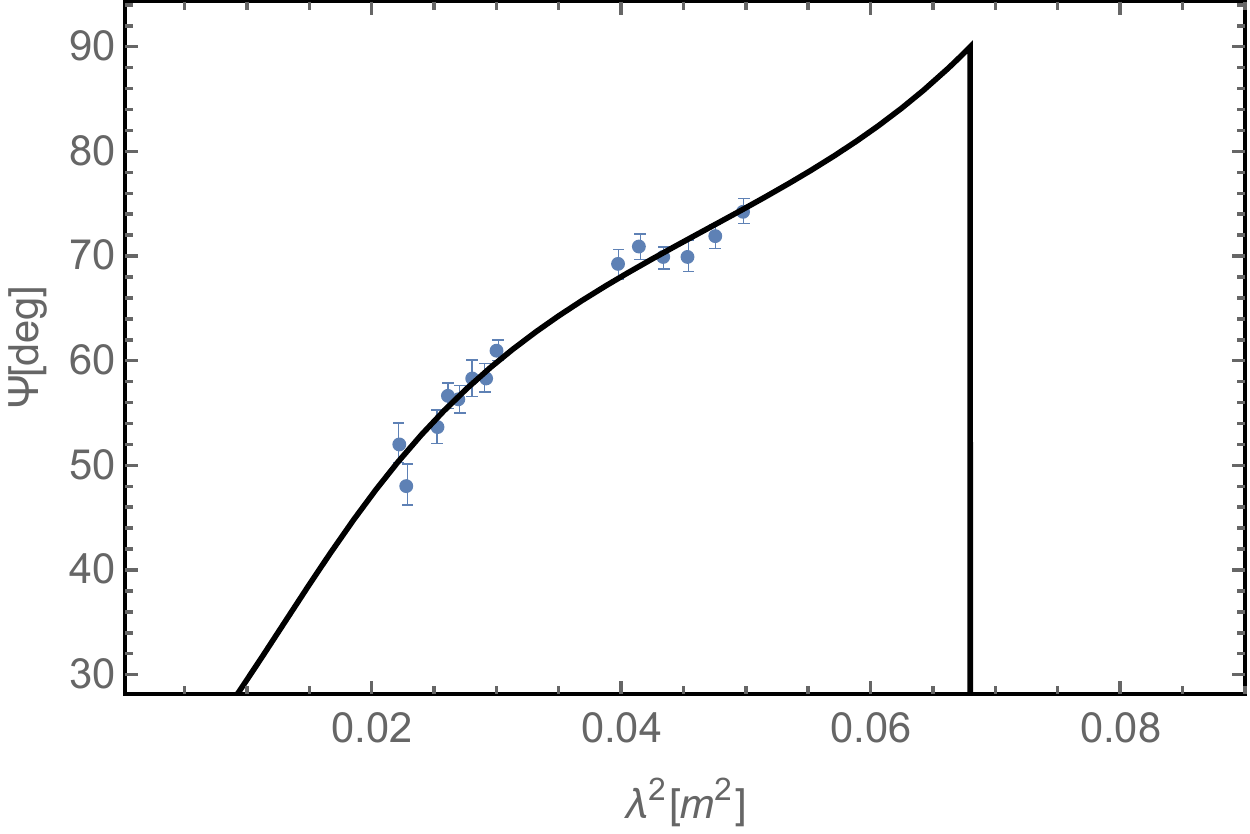}
        \includegraphics[width=0.4\textwidth]{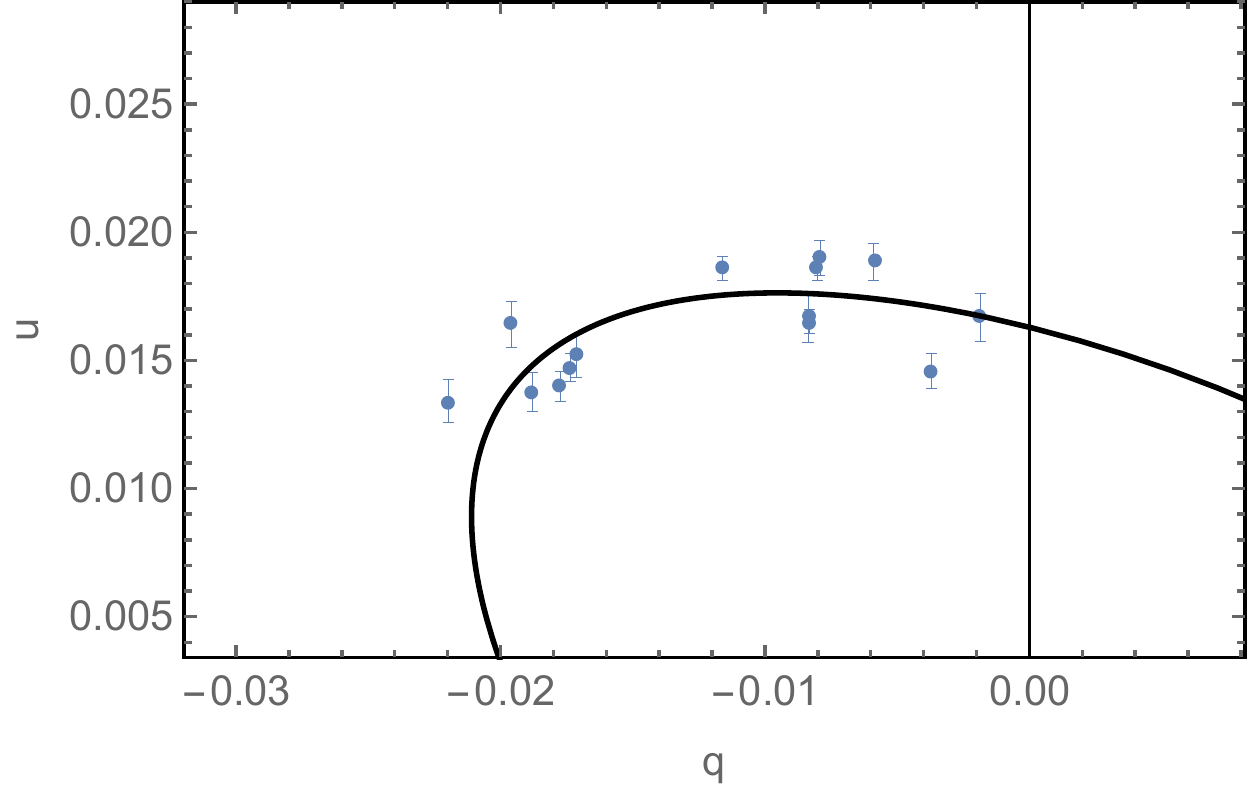}
\label{1246-0730Lbandmodel}
\end{center}
\end{figure*}


\begin{figure*}[h]
\begin{center}
    \caption{Depolarization model for the source 1405+0415 at C and X bands: 2 components model}
        \includegraphics[width=0.4\textwidth]{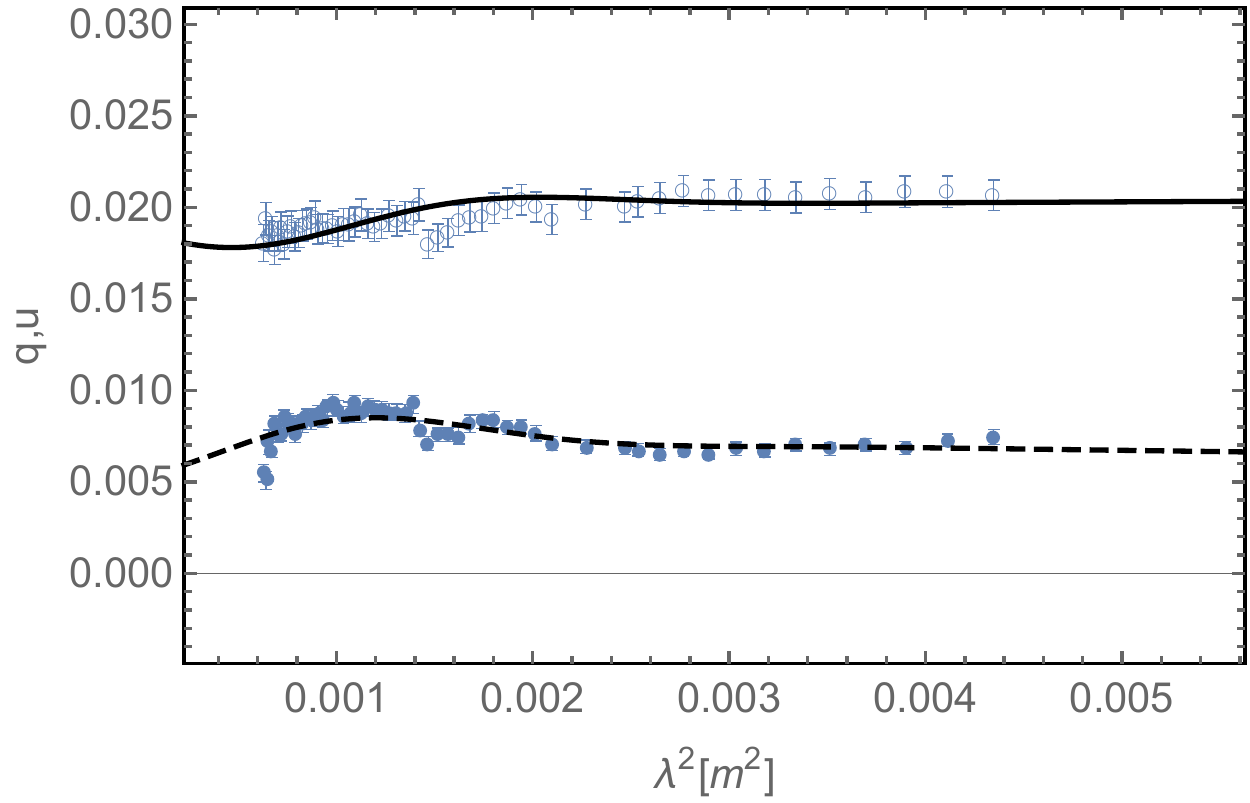}
        \includegraphics[width=0.4\textwidth]{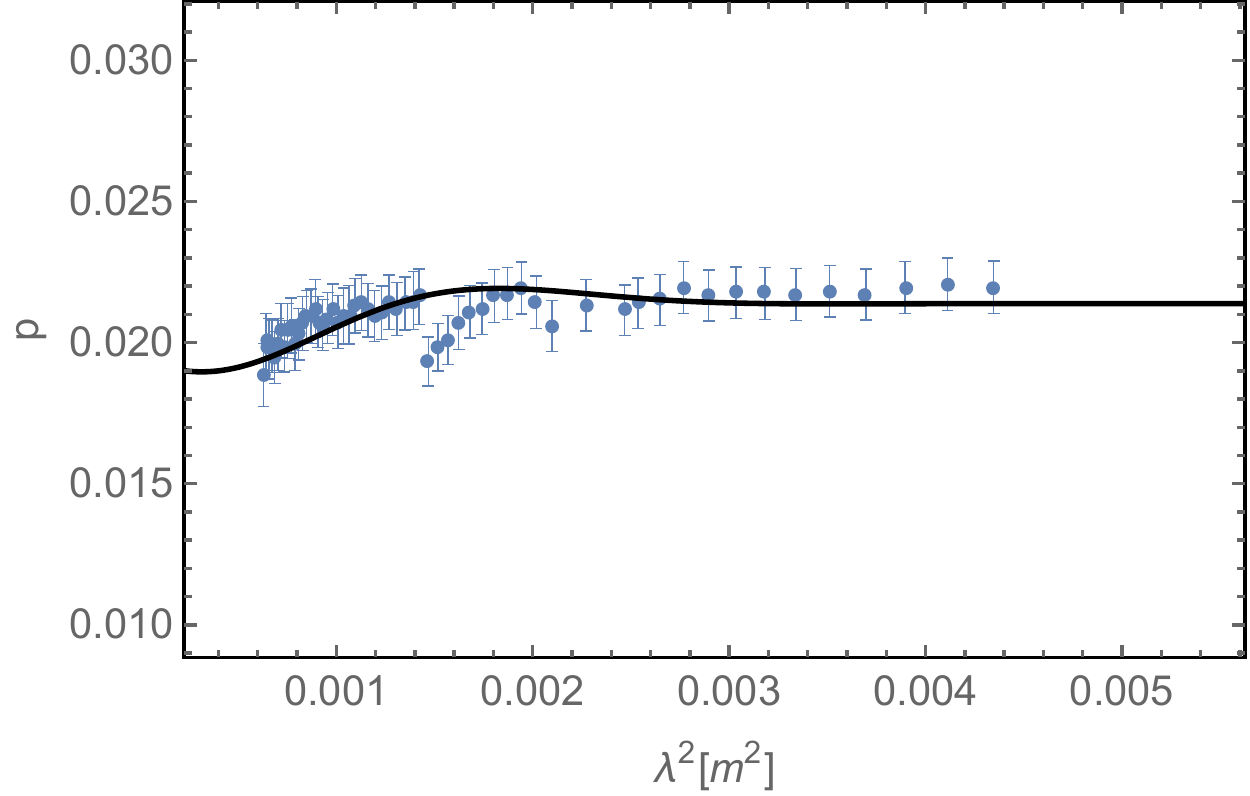}
        \includegraphics[width=0.4\textwidth]{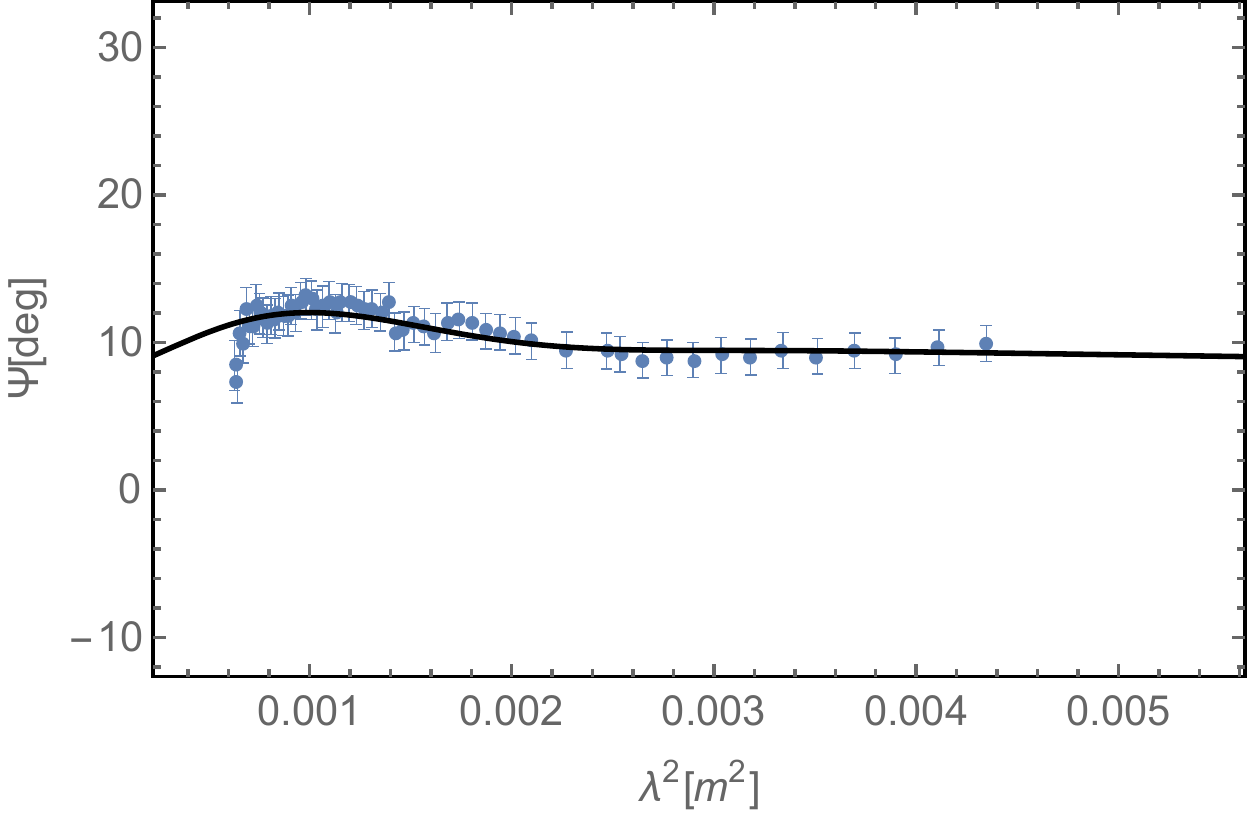}
        \includegraphics[width=0.4\textwidth]{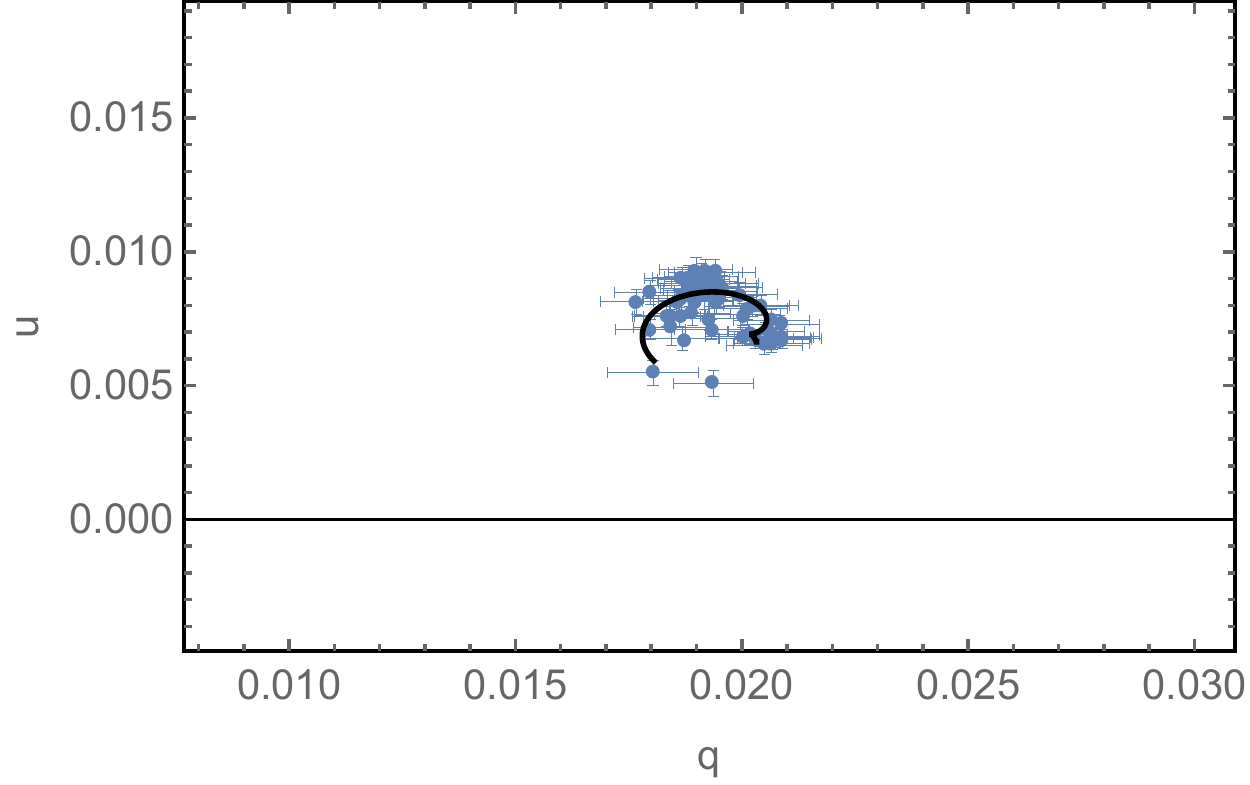}
\label{1405+0415CXbandmodel}

\end{center}
\end{figure*}

\begin{figure*}[h]
\begin{center}
    \caption{Depolarization model for the source 1405+0415 at L band: 2 components model}
        \includegraphics[width=0.4\textwidth]{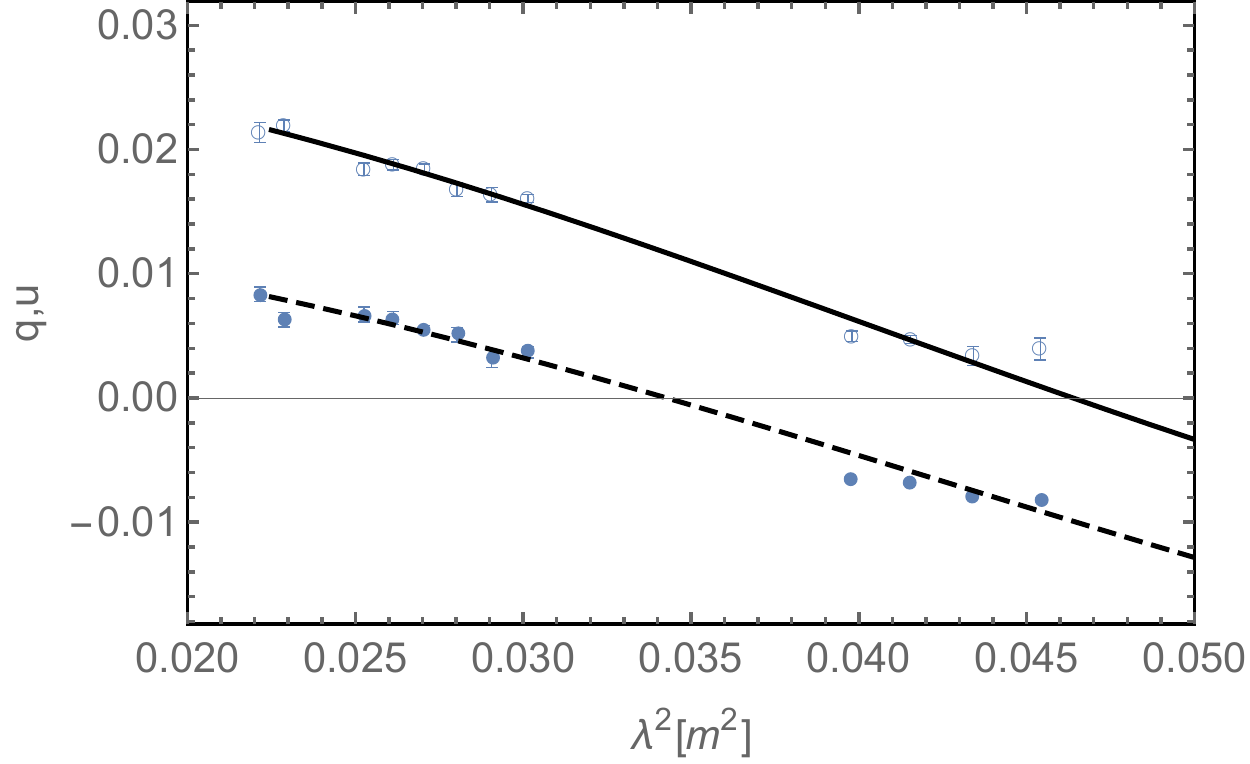}
        \includegraphics[width=0.4\textwidth]{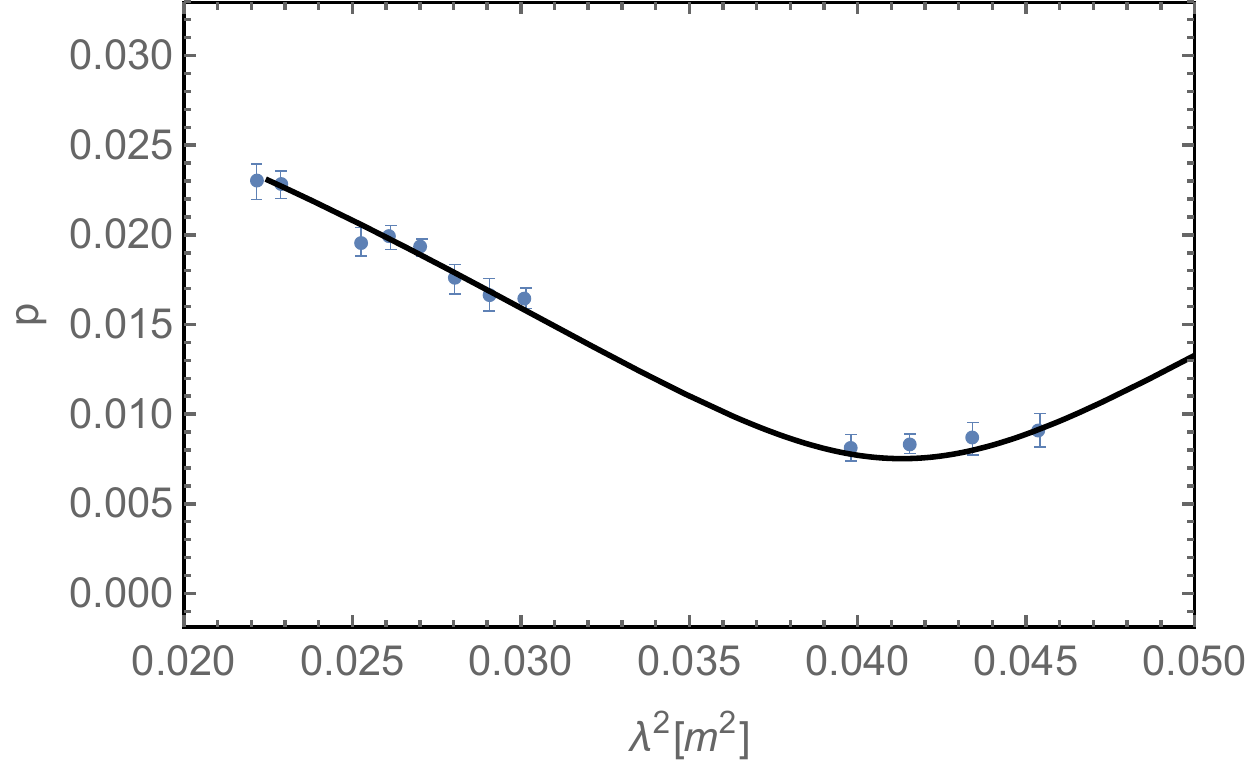}
        \includegraphics[width=0.4\textwidth]{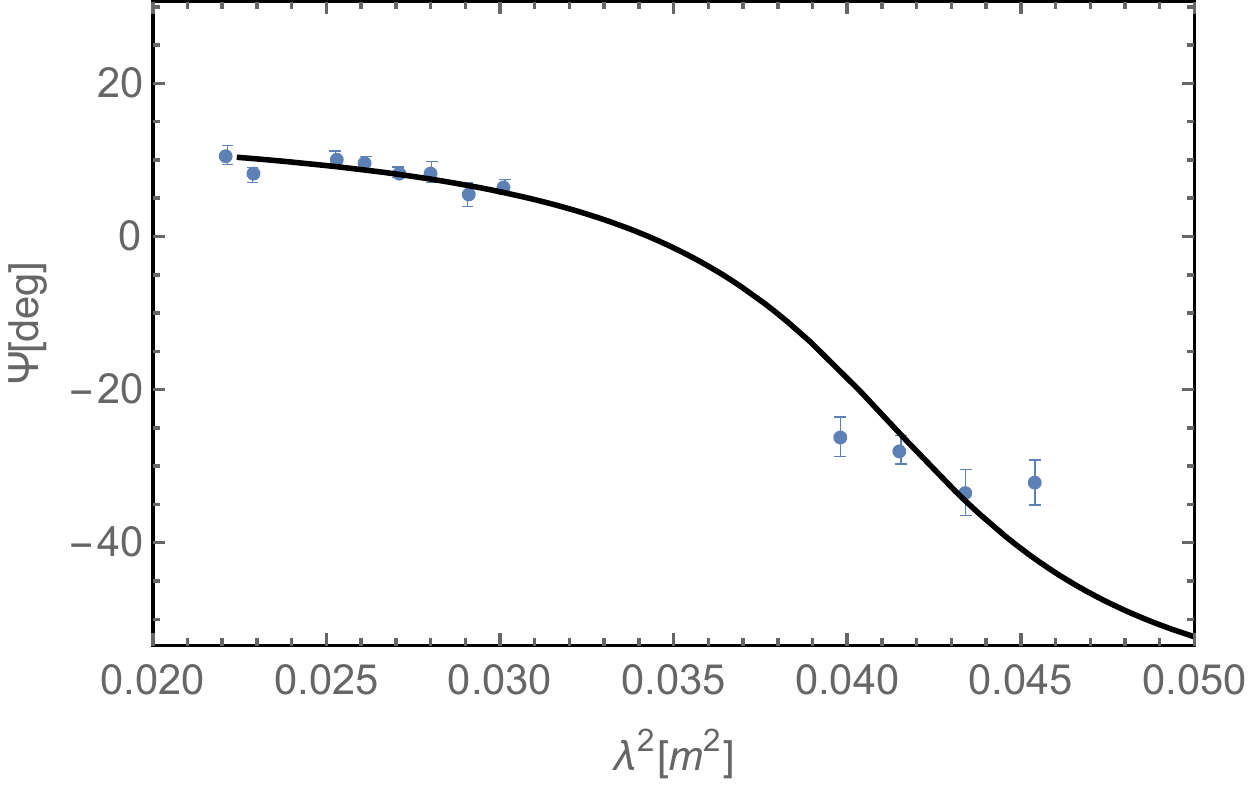}
        \includegraphics[width=0.4\textwidth]{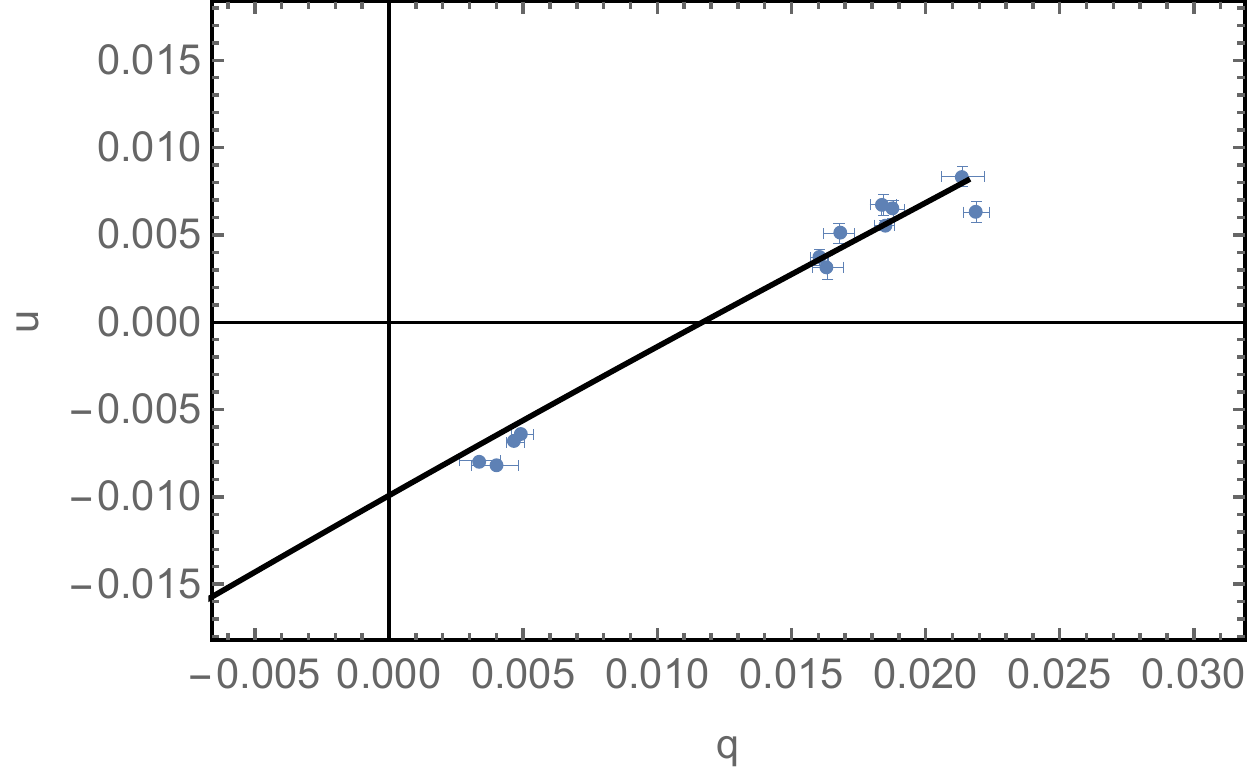}
\label{1405+0415Lbandmodel}

\end{center}
\end{figure*}

\end{appendix}

\end{document}